\documentclass[prd,preprint,tightenlines,floatfix,showpacs,preprintnumbers,nofootinbib,eqsecnum,superscriptaddress]{revtex4-1}

 \usepackage[dvips,final]{graphicx}
  \usepackage{amssymb}
   \usepackage{amsmath}
    \usepackage{amsfonts}
     \usepackage{epsfig}
      \usepackage{bm}
      
     \newcommand{\bp}{\mbox{\boldmath $p$}}
     
     \newcommand{\bk}{\mbox{\boldmath $k$}}

\begin{document}

\title{Exclusive \mbox{\boldmath $p p \to p p \pi^{0}$} reaction at high energies}

\author{Piotr Lebiedowicz}
\email{Piotr.Lebiedowicz@ifj.edu.pl}
\affiliation{Institute of Nuclear Physics PAN, PL-31-342 Cracow, Poland}

\author{Antoni Szczurek}
\email{Antoni.Szczurek@ifj.edu.pl}
\affiliation{Institute of Nuclear Physics PAN, PL-31-342 Cracow, Poland}
\affiliation{University of Rzesz\'ow, PL-35-959 Rzesz\'ow, Poland}

\begin{abstract}
The amplitudes for the $p p \to p p \pi^{0}$ process applicable at high energy are discussed in detail.
Both diffractive bremsstrahlung (Drell-Hiida-Deck--type model),
photon-photon and photon-omega exchange mechanisms are included in the calculation.
Large cross sections of the order of mb are predicted.
The corresponding differential cross sections 
in rapidities and transverse momenta of outgoing protons and pions
as well as relative azimuthal angle between outgoing protons 
are calculated for RHIC and LHC energies.
The hadronic bremsstrahlung contributions dominate at large 
(forward, backward) pion rapidities.
The diffractive nonresonant background contributes at small $\pi^0 p$ invariant mass
and could be therefore misinterpreted as the Roper resonance.
We predict strong dependence of the slope in $t$ 
(squared four-momentum transfer between ingoing and outgoing proton)
on the mass of the supplementary excited $\pi^{0}p$ system.
At high energies and midrapidities, the photon-photon contribution 
dominates over the diffractive components, 
however, the corresponding cross section is rather small.
The photon-odderon and odderon-photon contributions are included
in addition and first estimates (upper limits) of their contributions 
to rapidity and transverse momentum distribution of neutral pions are presented.
We suggest a search for the odderon contribution at midrapidity and
at $p_{\perp,\pi^{0}} \sim$ 0.5 GeV.
Our predictions are ready for verification at RHIC and LHC.
The bremsstrahlung mechanisms discussed here contribute also to the $pp \to p(n \pi^{+})$ reaction. 
Both channels give a sizable contribution to the low-mass 
single diffractive cross section and must be included in extrapolating
the measured experimental single diffractive cross section.
\end{abstract}

\pacs{13.60.Le, 13.85.-t, 12.40.Nn, 11.55.Jy}
\maketitle

\section{Introduction}

In the past we have studied exclusive production of pairs of charged pions in
the $pp \to pp \pi^{+}\pi^{-}$ \cite{LS10, SLTCS11}
and $pp \to nn \pi^{+}\pi^{+}$ \cite{LS11} processes.
There pomeron-pomeron or pomeron-reggeon exchanges are the dominant mechanisms.
The corresponding cross sections for high energies are rather large,
of the order of 100~$\mu$b.
Such processes could be measured with the help of the main ATLAS or CMS
detectors (for charged pions detection), ALFA \cite{ALFA}
or TOTEM \cite{TOTEM} detectors (for protons tagging),
and the zero degree calorimeters (ZDCs) (for neutrons detection).
Furthermore, the proposed forward shower counters,
to detect and trigger on rapidity gaps in diffractive events,
would improve the measurements at the LHC significantly~\cite{efficiency}.

Here we consider another simple final state with one pion only.
The exclusive process $p p \to p p \pi^0$ was measured
in detail only near to the pion threshold at the IUCF (Bloomington) 
\cite{IUCF}, CELSIUS (Uppsala) \cite{Celsius} and 
the COSY (J\"ulich) \cite{COSY}.
The total cross section for single pion production
grows from threshold to about 10 $\mu$b at the c.m. energy $\sqrt{s} \approx 3$~GeV.
Although only a few partial waves are involved close to threshold, 
the theoretical description is not easy (see e.g.~\cite{Juelich} and references therein).
For a summary on near-to-threshold meson production experiments see~\cite{Moskal}.

What happens when the energy increases?
In the range of center-of-mass energies $\sqrt{s} = 3-10$~GeV 
the nucleon resonances can be excited via meson exchange processes.
Evidence of proton excitation can be observed in 
the $p\pi^{0}$ mass spectrum ($\Delta^{+}$ or $N^{*+}$).
A nice summary of the intermediate energy data for $p p \to p p \pi^0$
can be found in Ref.\cite{Dahl_Jensen}.
In this region of energy the corresponding cross section systematically
decreases which is consistent with the meson exchange picture.
When energy increases further the role of many of the nucleon resonances 
diminishes and the mechanism becomes simpler.

In Refs~\cite{WA102_PLB427, kirk00} a study of pseudoscalar mesons produced centrally
by the CERN-WA102 Collaboration at $\sqrt{s} = 29.1$~GeV was performed.
The results show that the $\eta$ and $\eta'$ mesons appear to have a similar
production mechanism which considerably differs from that for the 
$\pi^{0}$ production \cite{WA102_PLB427}. To our knowledge this was
never explained theoretically. The WA102 Collaboration concentrated
on very central production of mesons and therefore measured protons with
large Feynman $x_{F} = 2 p_{\parallel} / \sqrt{s}$.
This condition eliminates contribution of the diffractive mechanisms discussed in our paper.
Reactions of this type $pp \to p M p$ are expected to be mediated by
double exchange processes, with a mixture of pomeron-pomeron,
reggeon-pomeron, and reggeon-reggeon exchange.
For instance, the $\eta$ and $\eta'$ mesons are produced dominantly
by double pomeron exchange (see \cite{LNS13} and references therein).
For the central exclusive $\pi^{0}$ production at intermediate energies 
the $\rho$-$\omega$ exchange may be the dominant mechanism.
The $\rho$-$a_{2}$ exchange could be another potential candidate.
The validity of these exchanges could be justified experimentally 
by the COMPASS Collaboration (see e.g., Ref.\cite{COMPASS}).

In the present paper we wish to concentrate
on the production of single neutral pions 
in the $p p \to p p \pi^0$ reaction at large energies (RHIC, LHC).
We hope that this process could be measured, 
at least in some corners of the phase space, at the LHC.
We shall refer also to $p p \to p (n \pi^+)$ and
$n p \to (p \pi^-) p$ reactions measured at lower energies
at Intersecting Storage Rings (ISR) and Fermilab in the 1970's \cite{Humble1975,Kerret1976,Mantovani1976,Babaev1976,Biel} 
(for a nice review we refer to \cite{Alberi_Goggi}). 
The mechanism of these reactions
is closely related to the $p p \to p p \pi^0$ reaction discussed here
and will be therefore a good reference point for our calculation.
As discussed in the past, the dominant hadronic bremsstrahlung-type 
mechanism is the Drell-Hiida-Deck (DHD) mechanism
for diffractive production of $\pi N$ final states in $NN$ collisions 
\cite{Deck}; for a review, see e.g.~\cite{Kaidalov, Alberi_Goggi}.
The exclusive pion production mechanism
is similar to $p p \to p p \omega$ \cite{CLSS} 
and $p p \to p p \gamma$ \cite{LS13} processes.


The $\pi^{0}$ can be also produced by $\gamma\gamma$, $\gamma\omega$, and $\gamma \mathcal{O}$ exchanges.
The soft odderon ($\mathcal{O}$) couples very weakly to the nucleon.
In Refs.~\cite{KN, BDDKNR} the authors discussed some results 
of exclusive pseudoscalar meson production in high energy $ep$ scattering.
It was shown in \cite{BDDKNR} (see also \cite{BN}) that odderon exchange
leads to a much larger inelastic than elastic $\pi^{0}$ production cross section.
As shown in Ref.\cite{BDDKNR}, the photon exchange is larger than the odderon exchange 
only at very small transverse momenta of $\pi^0$.
In this paper we shall consider the odderon contribution in proton-proton collisions
using a simple phenomenological approach for the odderon exchange.
We shall discuss how it can be separated
from the contribution of photon-photon fusion.

\section{The amplitudes for the \boldmath $p p \to p p \pi^0$ reaction}

\subsection{Diffractive bremsstrahlung mechanisms}
\begin{figure}[!ht]
(a)\includegraphics[width=3.7cm]{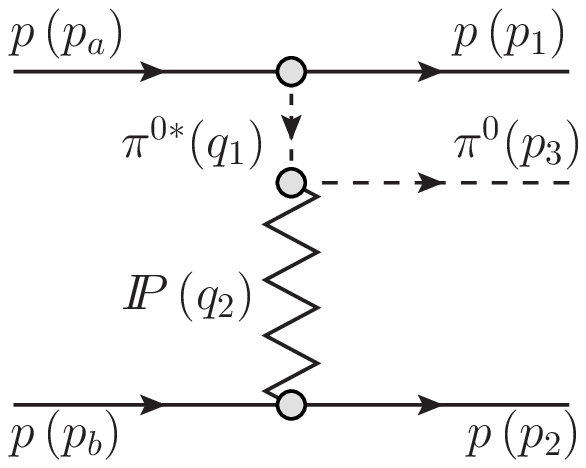}\quad
(b)\includegraphics[width=3.7cm]{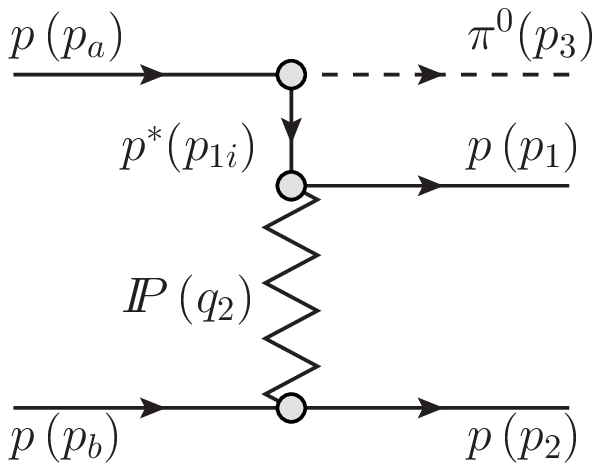}\quad
(c)\includegraphics[width=3.9cm]{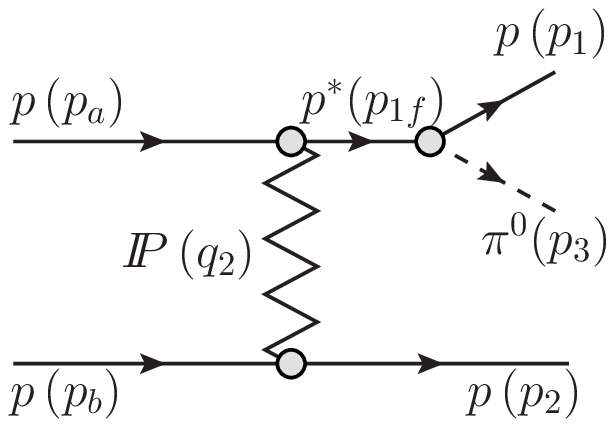}
\caption{\label{fig:diagrams_deck}
\small 
Diagrams of the $\pi^0$-bremsstrahlung amplitudes driven by
the pomeron exchange in proton-proton collisions:
(a) pion exchange, (b) proton exchange, and (c) direct production.
The direct-channel $p^{*}$ in (c) is an off-shell proton, not a proton resonance.
Some kinematical variables are shown in addition.}
\end{figure}
The diffractive bremsstrahlung mechanisms for the exclusive $p p \to p p \pi^0$ reaction are
driven by pomeron ($I\!\!P$) as depicted in Fig.\ref{fig:diagrams_deck} and reggeon ($I\!\!R$) exchanges.
At high c.m.~energies $\sqrt{s}$ the dominant contribution comes from the pomeron exchange.
There are two processes when the $\pi^{0}$ meson emitted by one of
the protons interacts with the second proton (diffractive $\pi^{0}$
rescattering), \footnote{ Discussed here diffractive mechanisms of
exclusive $\pi^{0}$ production are similar to the diffractive mechanism
of $p p \to p p \omega$ \cite{CLSS} and $pp \to pp \gamma$
\cite{LS13} processes.} as depicted in Fig.\ref{fig:diagrams_deck}(a),
and four processes in which protons interact and the $\pi^{0}$ emission
may occur, see Figs.\ref{fig:diagrams_deck}(b) and \ref{fig:diagrams_deck}(c).
In general, the amplitudes of these processes may interfere
but in the present case the interference is negligible
as the two processes are well separated in rapidity
as will be discussed in Section \ref{section:Results}.
The Born amplitudes for these processes, see Fig.\ref{fig:diagrams_deck},
can be written as
\begin{eqnarray}
{\cal M}^{(\pi-\mathrm{exchange})}_{\lambda_{a}\lambda_{b} \to \lambda_{1}\lambda_{2} \pi^{0}} &=&
\bar{u}(p_{1},\lambda_{1}) 
i\gamma_{5} 
u(p_{a},\lambda_{a}) 
\,S_{\pi}(t_{1}) 
\,g_{\pi NN}\, F_{\pi^{*}NN}(t_{1}) F_{I\!\!P \pi^{*}\pi}(t_{1})\, \nonumber \\
&\times &
\left( {A}_{I\!\!P}^{\pi N}(s_{23},t_{2}) + {A}_{I\!\!R}^{\pi N}(s_{23},t_{2}) \right) \,/(2s_{23})\nonumber \\
&\times & 
(q_{1} + p_{3})_{\mu}\,
\bar{u}(p_{2},\lambda_{2}) 
\gamma^{\mu} 
u(p_{b},\lambda_{b})\,,
\label{brem_e}\\
{\cal M}^{(p-\mathrm{exchange})}_{\lambda_{a}\lambda_{b} \to \lambda_{1}\lambda_{2} \pi^{0}} &=&
g_{\pi NN}\,
\bar{u}(p_{1},\lambda_{1})
\gamma^{\mu} 
S_{N}(p_{1i}^{2}) 
i\gamma_{5}
u(p_{a},\lambda_{a})\,
F_{\pi NN^{*}}(p_{1i}^{2}) \,
F_{I\!\!P N^{*}N}(p_{1i}^{2}) \nonumber \\
&\times &
\left( {A}_{I\!\!P}^{NN}(s_{12},t_{2}) + {A}_{I\!\!R}^{NN}(s_{12},t_{2}) \right) \,/(2s_{12})\nonumber \\
&\times &
\bar{u}(p_{2},\lambda_{2}) 
\gamma_{\mu} 
u(p_{b},\lambda_{b})\,,
\label{brem_c}\\
{\cal M}^{(\mathrm{direct\,production})}_{\lambda_{a}\lambda_{b} \to \lambda_{1}\lambda_{2} \pi^{0}} &=&
g_{\pi NN}\,
\bar{u}(p_{1},\lambda_{1}) 
i\gamma_{5} 
S_{N}(p_{1f}^{2})
\gamma^{\mu} 
u(p_{a},\lambda_{a})\,
F_{\pi N^{*}N}(p_{1f}^{2}) \,
F_{I\!\!P NN^{*}}(p_{1f}^{2}) \nonumber \\
&\times &
\left( {A}_{I\!\!P}^{NN}(s,t_{2}) + {A}_{I\!\!R}^{NN}(s,t_{2}) \right)\,/(2s)\nonumber \\
&\times &
\bar{u}(p_{2},\lambda_{2}) 
\gamma_{\mu} 
u(p_{b},\lambda_{b})\,,
\label{brem_a}
\end{eqnarray}
where $u(p,\lambda)$, $\bar{u}(p',\lambda')=u^{\dagger}(p',\lambda')\gamma^{0}$
are the Dirac spinors of the incident and outgoing protons 
with the four-momentum~$p$ and the helicities~$\lambda$;
normalized as $\bar{u}(p') u(p) = 2 m_{p}$.
The factors $\frac{1}{2s_{ij}}$ or $\frac{1}{2s}$ appear here
as a consequence of using spinors.
The four-momenta squared of intermediate particles
are defined in Fig.\ref{fig:diagrams_deck} and
$p_{1i,2i}^{2}=(p_{a,b}-p_{3})^{2}$,
$p_{1f,2f}^{2}=(p_{1,2}+p_{3})^{2}$,
$q_{1,2}^{2}=(p_{a,b}-p_{1,2})^{2}$,
the four-momentum transfers along the pomeron line
$t_{1,2}=q_{1,2}^{2}$
and $s_{ij}=(p_{i}+p_{j})^{2}$ are squared invariant masses of the $(i,j)$ system.
The propagators of the off-shell pion and proton are
\begin{eqnarray}
S_{\pi}(t)= {\frac{i}{t - m_{\pi}^{2}}} \,, \qquad
S_{N}(p^{2})= {\frac{i(p\!\!\!/ + m_{p})}{p^{2} - m_{p}^{2}}} \,,
\label{propagator_nucleon}
\end{eqnarray}
where $p\!\!\!/ = p_{\mu} \gamma^{\mu}$.

The energy dependence of the elastic scattering ${A}(s,t)$ was parametrized in the Regge-like form
with pomeron ($i = I\!\!P$) and reggeon 
($i = I\!\!R$ = $f_{2}$, $\rho$, $a_{2}$, $\omega$) exchanges,
\begin{eqnarray}
{A}_{i}^{el}(s,t) = 
\eta_{i} \, C_{i} \,s\,\left( \frac{s}{s_{0}}\right)^{\alpha_{i}(t)-1} 
\exp\left(\frac{B_{i}^{el} t}{2}\right) \,,
\label{amp_regge_NN}
\end{eqnarray}
where we use the scale parameter $s_{0} = 1$~GeV$^2$.
In writing the above amplitudes (\ref{amp_regge_NN}) we have omitted indices
related to helicities as we have assumed helicity conservation
in the rescattering process.
If the energy in the $\pi p$ or the $pp$ system is small,
then the secondary trajectories are also important, e.g. we have in (\ref{brem_e}) term
${A}^{\pi^{0} p} = A_{I\!\!P} + A_{f_{2}}$.
The strength parameters $C_{i}$, the values of signature factors $\eta_{i}$,
and the (linear) Regge trajectories
%
$\alpha_{i}(t) = \alpha_{i}(0)+\alpha'_{i}\,t$
%
are taken from the Donnachie-Landshoff analysis \cite{DL92}
of the total $NN$ and $\pi p$ cross sections
and are listed, e.g., in Ref.\cite{LS10}.
From the optical theorem we have 
$\sigma^{tot}(s) \cong 1/s\,\mathrm{Im} A^{el}(s,t=0)$.
%
The running slope for elastic scattering can be written as
%
$B(s) = B_{i}^{el} + 2 \alpha'_{i} \ln\left(s/s_{0}\right)$,
where $B_{I\!\!P}^{NN} = 9$~GeV$^{-2}$, $B^{\pi N}_{I\!\!P} = 5.5$~GeV$^{-2}$
and $B_{I\!\!R}^{NN} = 6$~GeV$^{-2}$, $B^{\pi N}_{I\!\!R} = 4$~GeV$^{-2}$ \cite{LS10, CLSS}
for pomeron and reggeon exchange, respectively.

Usually a high-energy approximate formula is used in the literature
in calculating differential cross sections.
We use a precise calculation of the phase space (see e.g., \cite{SL09}). 
This is important if one wants to go to lower energies and/or to large rapidities.
As will be discussed in the next section, for this particular reaction 
the cross section has maximum just at large rapidities, where
the often used formula is too approximate.
In the high-energy limit we obtain
\begin{eqnarray}
(q_{1} + p_{3})_{\mu}\,
\bar{u}(p_{2},\lambda_{2}) 
\gamma^{\mu} 
u(p_{b},\lambda_{b}) 
\cong
(q_{1} + p_{3})_{\mu}\,(p_{2} + p_{b})^{\mu} \delta_{\lambda_{2}\lambda_{b}}
\cong 
2s_{23}\,\delta_{\lambda_{2}\lambda_{b}}\,.
\label{amp_regge_piN_approx}
\end{eqnarray}
%

In the bremsstrahlung processes discussed here the intermediate nucleons
are off-mass shell. In the above equations the off-shell effects related
to the non-point-like protons in the intermediate state are included by 
the following form factors:
\begin{eqnarray}
F(p^{2}) = \frac{\Lambda_{N}^{4}}{(p^{2}-m_{p}^{2})^{2} + \Lambda_{N}^{4}}\,.
\label{extra_ff}
\end{eqnarray}
Such a form was used e.g. in Ref.\cite{OTL01} for $\omega$ photoproduction.
In general, the cutoff parameters in the form factors are not known
but could be fitted in the future to the (normalized) experimental data.
From our general experience in hadronic physics we expect $\Lambda_{N} \sim 1$~GeV.
We shall discuss how the uncertainties of the form
factors influence our final results (see Section \ref{section:Results}).

The pion-nucleon coupling constant $g_{\pi NN}$ 
is relatively well known \cite{Ericson-Weise,ELT02}.
In our calculations we take $g_{\pi NN}^{2}/4\pi = 13.5$.
$F_{\pi^{*} N N}(t)$ is a vertex form factor due to the extended nature of particles involved. 
Unfortunately, the off-shell form factor
is not well known as it is due to nonperturbative effects 
related to the internal structure of the respective objects.
This discussion of form factors applies also to the other $I\!\!P \pi^{*}\pi$ vertices.
We parametrize these form factors in the simple exponential form,
\begin{eqnarray}
F_{\pi^{*} N N}(t) = 
F_{I\!\!P \pi^{*}\pi}(t) = \exp \left( \frac{t - m_{\pi}^2}{\Lambda_{\pi}^2}\right)\,,
\label{formfactors_pi0_rescattering}
\end{eqnarray}
which is conventionally normalized to unity on the pion-mass shell and 
$\Lambda_{\pi} = 1$~GeV is a reasonable choice.

The pion exchange as a meson exchange is 
a correct description at rather low $\pi p$ energies while
a reggezation of pion is required at higher energies.
We propose to use a generalized pion propagator (see e.g.~\cite{LPS11})
at an appropriate subsystem energy and $t$,
\begin{eqnarray}
S_{\pi}(t) \rightarrow
\beta_{M}(s) S_{\pi}(t) + \beta_{R}(s) \mathcal{P}^{\pi}(t,s) \,,
\label{generalized_pion_propagator}
\end{eqnarray}
where the pion Regge propagator with the Euler's gamma function,
%
\begin{eqnarray}
\mathcal{P}^{\pi}(t,s) =
\frac{\pi \alpha'_{\pi}}{2\Gamma(\alpha_{\pi}(t)+1)}
\frac{1+\exp(-i\pi \alpha_{\pi}(t))}{\sin (\pi \alpha_{\pi}(t))}
\left(\frac{s}{s_{0}}\right)^{\alpha_{\pi}(t)} \,,
\label{pion_propagator}
\end{eqnarray}
gives a suppression for large values of $t$.
We have introduced extra phenomenological functions $\beta_M(s)$
and $\beta_R(s)$ to interpolate between meson and reggeon exchanges.
We parametrize them as
\begin{eqnarray}
\beta_M(s) = \exp\left( -(s-s_{thr})
/\Lambda_{int}^2 \right) \,, \quad \beta_R(s) = 1 - \beta_M(s)\,.
\label{low_high_matching}
\end{eqnarray}
%
From our general experience in hadronic physics the parameter 
$\Lambda_{int} \cong 1$~GeV.
The pion trajectory is taken as
$\alpha_{\pi}(t) = \alpha'_{\pi}(t - m_{\pi}^2)$ 
with a slope parameter $\alpha'_{\pi} = 0.7$~GeV$^{-2}$.

We improve the parametrization of $p$-exchange amplitude (\ref{brem_c})
to reproduce the high-energy Regge dependence by the factor
$\left(s_{13}/s_{thr} \right)^{\alpha_{N}(p_{1i}^{2})-\frac{1}{2}}$
or
by the factor
$\left(s_{23}/s_{thr} \right)^{\alpha_{N}(p_{2i}^{2})-\frac{1}{2}}$,
where the threshold factor $s_{thr} = (m_{p}+m_{\pi^{0}})^{2}$ and
the nucleon trajectory is
$\alpha_{N}(p_{1i,2i}^{2})=-0.3 + \alpha'_{N} \, p_{1i,2i}^{2}$ with $\alpha'_{N}=0.9$ GeV$^{-2}$.
\subsection{Absorptive corrections}
\begin{figure}[!ht]
(a)\includegraphics[width=3.7cm]{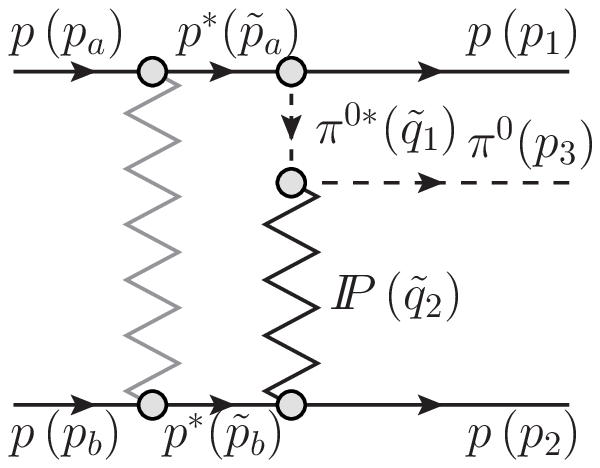}
  \includegraphics[width=3.7cm]{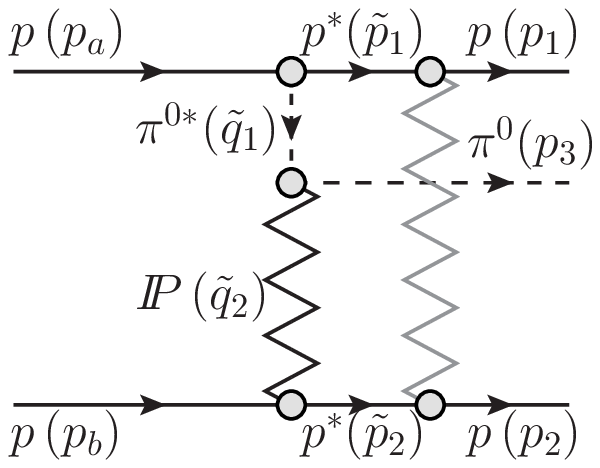}\quad
(b)\includegraphics[width=3.7cm]{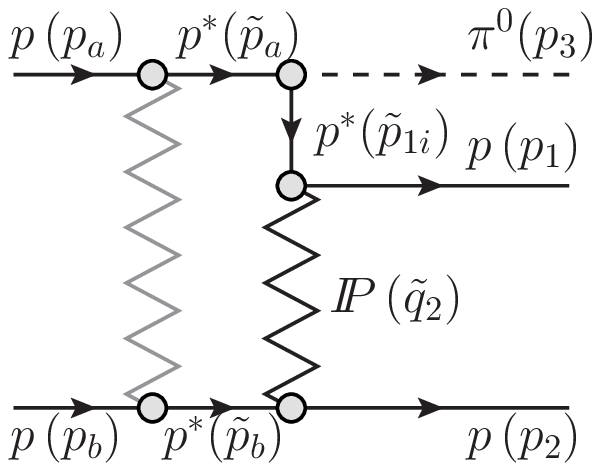}
  \includegraphics[width=3.7cm]{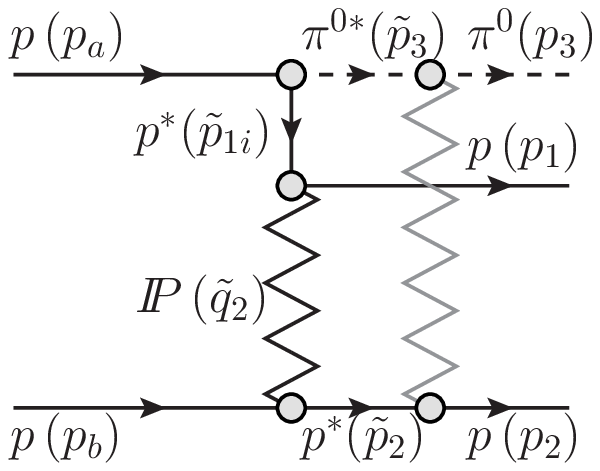}
\caption{\label{fig:diagrams_abs}
\small 
Typical absorptive correction diagrams to: 
(a) pion exchange and (b) proton exchange.
The stars attached to protons and $\pi^{0}$ meson denote the fact that they are off-mass shell.
}
\end{figure}
Let us estimate absorptive corrections $\delta {\cal M}$ 
shown in Fig.\ref{fig:diagrams_abs}.
Diagrams which involve the elastic scattering of the incident protons
are termed ``initial-state" absorption.
From physical reasons discussed in \cite{Berger,Alberi_Goggi} the diagrams,
when the transition of excited proton $p^{*} \to p \pi^{0}$ occurs inside,
do not contribute significantly at high energies.
Diagrams with the ``final-state" absorption corrections
provide the dominant absorptive effect \cite{Berger}.
In the quasieikonal approximation 
which takes into account contribution of elastic rescatterings
the absorbed amplitudes can be expressed as
\begin{eqnarray}
{\cal M}_{\mathrm{abs}}(-\bp_{1\perp},-\bp_{2\perp})=
{\cal M}(-\bp_{1\perp},-\bp_{2\perp}) - 
\delta{\cal M}(-\bp_{1\perp},-\bp_{2\perp})\,,
\label{absorptive_corrections}
\end{eqnarray}
where 
$\delta{\cal M}$ for the diagrams with ``initial-state" absorption 
is the sum of convolution integral
\begin{eqnarray}
&&\delta {\cal M}_{\lambda_a \lambda_b \to \lambda_1 \lambda_2 \pi^{0}}^{\mathrm{initial\,state\,abs}}
(-\bp_{1\perp},-\bp_{2\perp}) =
\frac{i}{8 \pi^2 s} \int d^2 k_{\perp} \,
A^{NN}_{\lambda_a \lambda_b \to \lambda_a' \lambda_b'}(s,\bk_{\perp}) \nonumber\\
&&
\times \left[ 
{\cal M}_{\lambda_a' \lambda_b' \to \lambda_1 \lambda_2 \pi^{0}}^{(\pi-\mathrm{exchange})} 
(-\tilde{\bp}_{1\perp},-\tilde{\bp}_{2\perp})+
{\cal M}_{\lambda_a' \lambda_b' \to \lambda_1 \lambda_2 \pi^{0}}^{(p-\mathrm{exchange})}
(-\tilde{\bp}_{1\perp},-\tilde{\bp}_{2\perp}) 
\right]
\,
\label{absorptive_corrections1}
\end{eqnarray}
and in the case of diagrams with ``final-state" absorption we have
\begin{eqnarray}
&&\delta {\cal M}_{\lambda_a \lambda_b \to \lambda_1 \lambda_2 \pi^{0}}^{\mathrm{final\,state\,abs}}
(-\bp_{1\perp},-\bp_{2\perp}) =\nonumber\\
&&\frac{i}{8 \pi^2} \int d^2 k_{\perp} 
\frac{1}{s_{12}}
{\cal M}_{\lambda_a \lambda_b \to \lambda_1' \lambda_2' \pi^{0}}^{(\pi-\mathrm{exchange})}
(-\tilde{\bp}_{1\perp},-\tilde{\bp}_{2\perp})\,
A^{NN}_{\lambda_1' \lambda_2' \to \lambda_1 \lambda_2}(s_{12},\bk_{\perp}) \nonumber\\
+ &&\frac{i}{8 \pi^2} \int d^2 k_{\perp} 
\frac{1}{s_{23}}
{\cal M}_{\lambda_a \lambda_b \to \lambda_1 \lambda_2' \pi^{0}}^{(p-\mathrm{exchange})}
(-\tilde{\bp}_{1\perp},-\tilde{\bp}_{2\perp})\,
A^{\pi N}_{\lambda_2' \to \lambda_2}(s_{23},\bk_{\perp}) 
\,,
\label{absorptive_corrections2}
\end{eqnarray}
where the two-dimensional transverse vectors
$-\tilde{\bp}_{1\perp} = -\bp_{1\perp} + \bk_{\perp}$
and $-\tilde{\bp}_{2\perp} = -\bp_{2\perp} - \bk_{\perp}$
are the transverse components of the momenta
of final state protons and $\bk_{\perp}$ is the momentum transfer.
$A^{el}(s,\bk_{\perp})$ is an elastic scattering amplitude given by Eq.(\ref{amp_regge_NN})
at an appropriate energy and for the momentum transfer $\bk_{\perp}$.
Since in our calculations we include effective pomeron and reggeon exchanges,
i.e. pomerons and reggeons describing approximately 
nucleon-nucleon or pion-nucleon elastic scattering,
no explicit absorption corrections have to be included in addition.

Experience from hadronic phenomenology
(for several analyses of two-body reactions see \cite{Kane_Seidl}) 
suggest that the purely elastic rescattering taken into account by 
Eq.~(\ref{absorptive_corrections}) are insufficient,
and inelastic intermediate states (screening corrections)
lead to an enhancement of absorptive corrections.
This is sometimes included in a phenomenological way
by a factor $\lambda_{sc}$~($\lambda_{sc} > 1$).
%
%
Taking into account absorption corrections,
the DHD mechanism was shown to give a reasonable explanation for the main properties 
of the low-mass diffractive dissociation \cite{Tarasiuk}.
The effect of the absorption in diffractive dissociation is also discussed in \cite{Tsarev}.

\subsection{\mbox{\boldmath $\gamma \gamma$} and \mbox{\boldmath $\gamma \omega$} exchange mechanisms}
\begin{figure}[!ht]    
(a)\includegraphics[width=4.0cm]{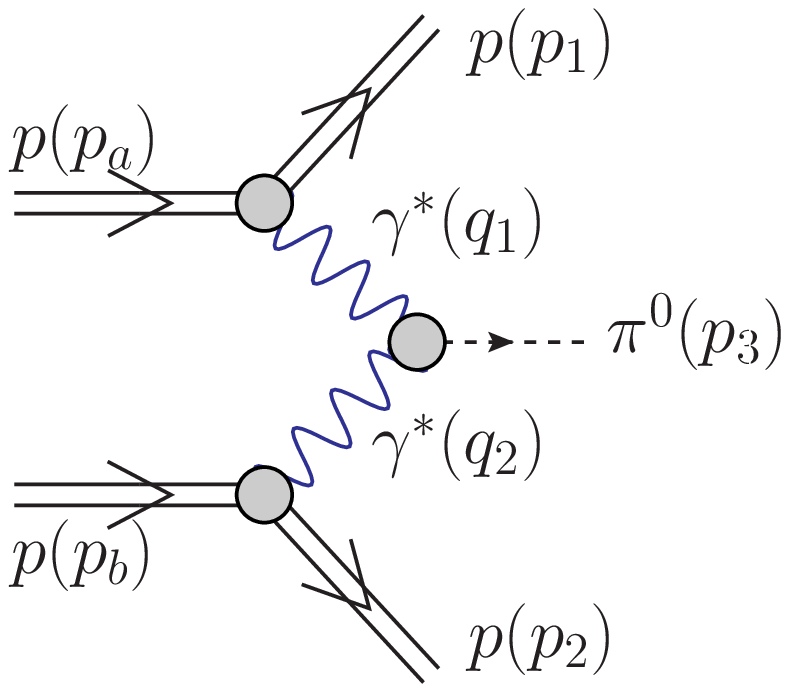} \quad
(b)\includegraphics[width=4.0cm]{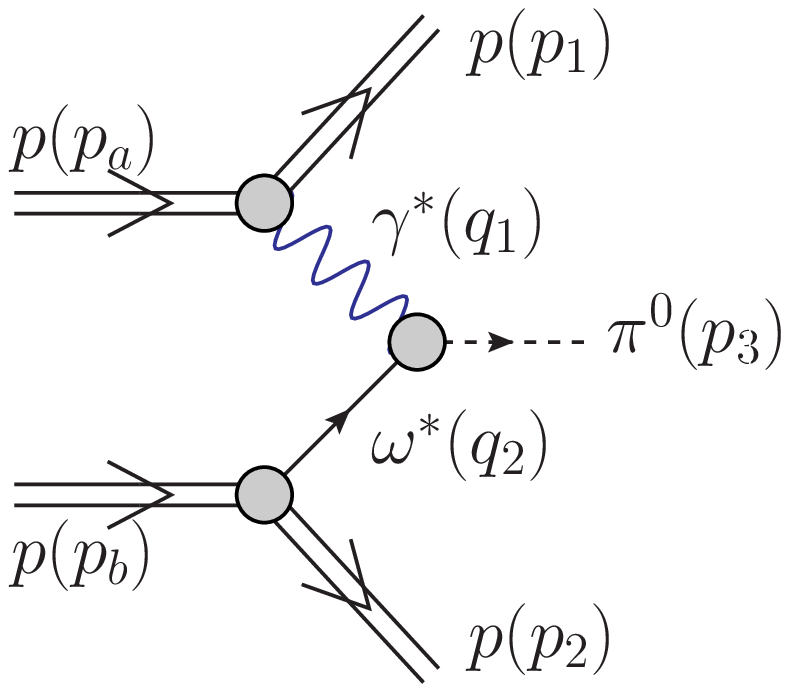} 
  \includegraphics[width=4.0cm]{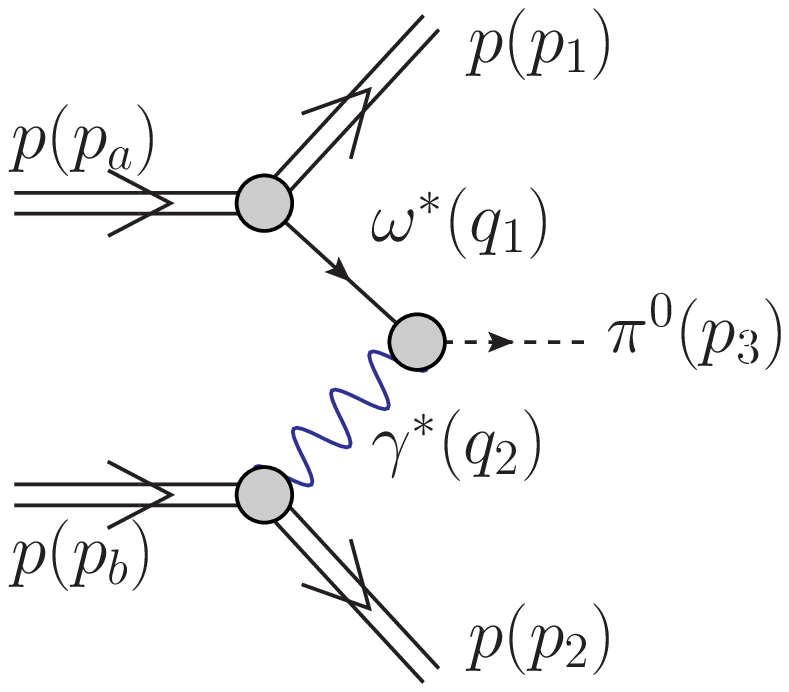}
\caption{\label{fig:diagram_gamgam_pi0}
\small
A sketch of the photon-photon (a) and photon-omega meson (b) exchanges 
induced production of $\pi^0$ in proton-proton collisions.}
\end{figure}

In the following we wish to investigate 
competitive mechanisms to the diffractive mechanisms discussed in the previous subsection.
The new mechanisms, never discussed so far in the literature,
are shown schematically in Fig.\ref{fig:diagram_gamgam_pi0}.
%
%
In the most general case the corresponding Born amplitudes read
\begin{eqnarray}
{\cal M}^{\gamma \gamma-\mathrm{exchange}}_{\lambda_{a}\lambda_{b} \to \lambda_{1}\lambda_{2} \pi^{0}}&=&
e\,\bar{u}(p_{1},\lambda_{1})
\gamma^{\mu}
u(p_{a},\lambda_{a})\,F_{1}(t_{1})
\nonumber \\ & \times &
\frac{g_{\mu \mu'}}{t_1} \,
(-i) \, e^{2}\,
\epsilon^{\mu' \nu' \rho \sigma} \, q_{1,\rho} q_{2,\sigma} \,
F_{\gamma^* \gamma^* \to \pi^0}(t_1,t_2)\,
\frac{g_{\nu \nu'}}{t_2} 
\nonumber \\ & \times &
e\,\bar{u}(p_{2},\lambda_{2})
\gamma^{\nu}
u(p_{b},\lambda_{b}) \,F_{1}(t_{2})\,, 
\label{QED_full_amplitude}\\
{\cal M}^{\gamma \omega-\mathrm{exchange}}_{\lambda_{a}\lambda_{b} \to \lambda_{1}\lambda_{2} \pi^{0}}&=&
e\,\bar{u}(p_{1},\lambda_{1})
\gamma^{\mu}
u(p_{a},\lambda_{a}) \,F_{1}(t_{1})
\nonumber \\ & \times &
\frac{g_{\mu \mu'}}{t_1} \,
(-i) \, g_{\gamma\omega\pi^{0}}\,
\epsilon^{\mu' \nu' \rho \sigma} \, q_{1,\rho} q_{2,\sigma} \,
F_{\gamma^* \omega^* \to \pi^0}(t_1,t_2)\,
\frac{-g_{\nu \nu'}+\frac{q_{\nu}q_{\nu'}}{m_{\omega}^{2}}} {t_{2} - m_{\omega}^{2}}\,
\nonumber \\ & \times &
g_{\omega NN} \, \bar{u}(p_{2},\lambda_{2})
\gamma^{\nu}
u(p_{b},\lambda_{b}) \, F_{\omega NN}(t_{2})\, {\cal F}(s_{23},t_{2})\,, 
\label{gamma_omega_amplitude}\\
{\cal M}^{\omega \gamma-\mathrm{exchange}}_{\lambda_{a}\lambda_{b} \to \lambda_{1}\lambda_{2}  \pi^{0}}&=&
g_{\omega NN} \, \bar{u}(p_{1},\lambda_{1})
\gamma^{\mu}
u(p_{a},\lambda_{a}) \, F_{\omega NN}(t_{1})\, {\cal F}(s_{13},t_{1})
\nonumber \\ & \times &
\frac{-g_{\mu \mu'}+\frac{q_{\mu}q_{\mu'}}{m_{\omega}^{2}}} {t_{1} - m_{\omega}^{2}}\,
(-i) \, g_{\gamma\omega\pi^{0}}\,
\epsilon^{\mu' \nu' \rho \sigma} \, q_{1,\rho} q_{2,\sigma} \,
F_{\gamma^* \omega^* \to \pi^0}(t_2,t_1)\,
\frac{g_{\nu \nu'}}{t_2} \,
\nonumber \\ & \times &
e\,\bar{u}(p_{2},\lambda_{2})
\gamma^{\nu}
u(p_{b},\lambda_{b}) \,F_{1}(t_{2})\,, 
\label{omega_gamma_amplitude}
\end{eqnarray}
%
where the $\gamma^{*} N N$ vertices are
parametrized by 
%
%
the proton's Dirac electromagnetic form factor 
%
\begin{eqnarray}
F_{1}(t)= \frac{4 m_{p}^{2}-2.79\,t}{(4 m_{p}^{2}-t)(1-t/m_{D}^{2})^{2}} \,,
\label{Fpomproton}
\end{eqnarray}
where $m_{D}^{2} = 0.71$~GeV$^{2}$ is a phenomenological parameter.

The central vertices involve off-shell particles. 
The $t$ dependences of $F_{\gamma^* \gamma^* \to \pi^0}(t_1,t_2)$
electromagnetic off-shell form factor
are the least known ingredients in formula (\ref{QED_full_amplitude}).
It is known experimentally only for one virtual photon $\gamma \gamma^* \to \pi^0$ (e.g., \cite{Belle}).
In the present calculation we use a vector meson dominance model inspired parametrization 
of the $\gamma^* \gamma^* \to \pi^0$ transition form factor,
\begin{eqnarray}
F_{\gamma^* \gamma^* \to \pi^0}(t_1,t_2) = 
\frac{F_{\gamma^* \gamma^*  \pi^0}(0,0)}
     {(1-t_1/m_{\rho}^2)(1-t_2/m_{\rho}^2)} \,,
\label{em_off_shell_formfactor1}
\end{eqnarray}
where $m_{\rho}$ is the $\rho$ meson mass.
The form factor is normalized to
%
$F_{\gamma^* \gamma^* \pi^0}(0,0) = \frac{N_{c}}{12 \pi^2 f_{\pi}}$\,,
%
where $N_{c}=3$ is the number of quark colors and
$f_{\pi} = 93$~MeV is the pion decay constant.

The coupling of the omega meson to the nucleon is described by the coupling constant
$g^2_{\omega N N}/4\pi = 10$ and the corresponding form factor is taken
in the exponential form:
\begin{equation}
F_{\omega NN}(t) = \exp \left( \frac{t - m_{\omega}^2}{\Lambda_{\omega NN}^2} \right) \,,
\label{omegaNN_ff}
\end{equation}
where $\Lambda_{\omega NN} = 1$~GeV.
The $g_{\omega \pi^0 \gamma} \simeq 0.7$~GeV$^{-1}$ constant was obtained
from the omega partial decay width as discussed in Ref.~\cite{CLSS}.
The $\gamma \omega$ and $\omega \gamma$ form factors are taken in the
following factorized form:
\begin{eqnarray}
F_{\gamma^{*} \omega^{*} \to \pi^{0}}(t_1, t_2) &=& 
\frac{m_{\rho}^2}{m_{\rho}^2 - t_1}
\exp \left( \frac{ t_2 - m_{\omega}^2 }{\Lambda_{\omega \pi \gamma}^2}
\right)
\, .
\label{omega_pi_gamma_ff}
\end{eqnarray}
%
In practical calculations we take $\Lambda_{\omega \pi \gamma} = 0.8$~GeV \cite{CLSS}
as found from the fit to the $\gamma p \to \omega p$ experimental data.

At larger subsystem energies, $s_{ij} \gg s_{thr}$,
one should rather use reggeons than mesons.
The ``reggezation'' is included here only approximately
by a factor assuring asymptotically correct high energy dependence,
\begin{eqnarray}
{\cal F}(s,t) = 
\left( \frac{s}{s_{thr}} \right)^{\frac{2}{\pi} 
\arctan((s-s_{thr})/\Lambda_{thr}^{2})(\alpha_{I\!\!R}(t)-1)}\,,
\label{aux_omeome}
\end{eqnarray}
where $\Lambda_{thr} \simeq 1$~GeV,
$\alpha_{I\!\!R}(0)$ = 0.5, and $\alpha'_{I\!\!R}$ = 0.9 GeV$^{-2}$.

%
%
In the high energy limit
%
%
%
we can write a relatively simple formula of two-photon fusion amplitude squared
and averaged over initial and summed over final spin polarizations (see \cite{SPT07}):
\begin{eqnarray}
\overline{|{\cal M}_{pp \to pp \pi^0}^{\gamma \gamma-\mathrm{exchange}}|^2}
\cong 4 s^2 e^8
\frac{F_1^2(t_1)}{t_1^2}
\frac{F_1^2(t_2)}{t_2^2} 
|F_{\gamma^* \gamma^* \to \pi^0}(t_1,t_2)|^2\, 
|{\bf q}_{1\perp}|^2 
|{\bf q}_{2\perp}|^2 \sin^2(\phi_{12}) \,,
\label{gamma_gamma_amplitude_squared}
\end{eqnarray}
where $\phi_{12} = \phi_{1} - \phi_{2}$ is the azimuthal angle between the two outgoing protons.

The amplitude for processes shown in Fig.\ref{fig:diagram_gamgam_pi0}
are calculated numerically for each point in the phase space.
In calculating cross section we perform integration in
$\log_{10}(p_{1\perp})$ 
and $\log_{10}(p_{2\perp})$ 
instead in $p_{1\perp}$ and $p_{2\perp}$, 
which is useful numerically because of photon propagators.

\subsection{\mbox{\boldmath $\gamma \mathcal{O}$} and \mbox{\boldmath $\mathcal{O} \gamma$} exchanges}
\begin{figure}[!ht]    
(a)\includegraphics[width=4.0cm]{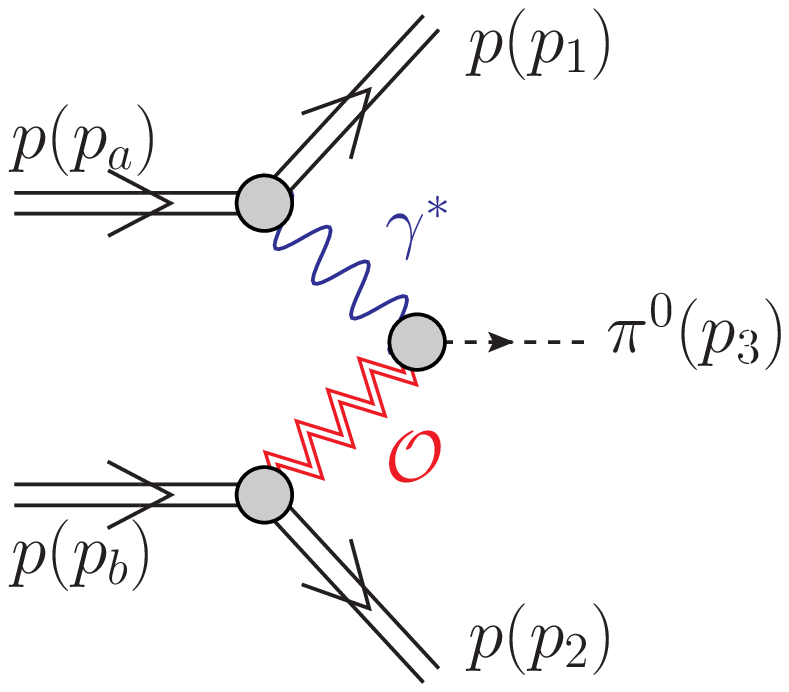} \quad
(b)\includegraphics[width=4.0cm]{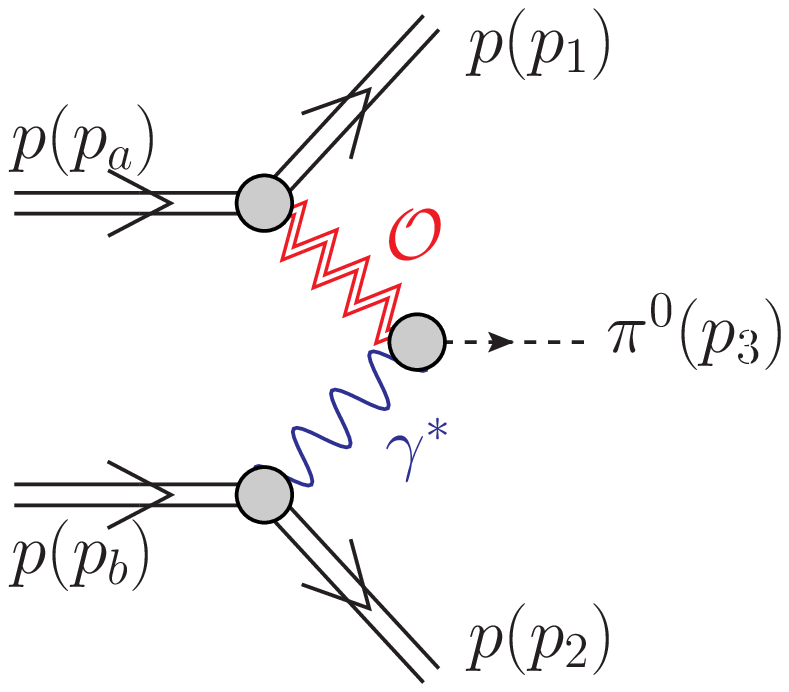} 
\caption{\label{fig:diagrams_oddereon}
\small
Diagrams with the photon-odderon (a) and odderon-photon (b) exchanges in the $pp \to pp\pi^0$ reaction.}
\end{figure}

As will be shown in Section \ref{section:Results}, at the $\pi^0$ midrapidity 
only the $\gamma \gamma \to \pi^0$, out of the mechanisms considered so far,
contributes, i.e. the corresponding cross section is rather small.
This gives a chance to search for $\gamma \mathcal{O}$ and $\mathcal{O} \gamma$ 
exchange processes shown in Fig.\ref{fig:diagrams_oddereon}. 
The $\gamma p \to \pi^0 p$ reaction was proposed some time ago as a good
candidate for identifying the odderon exchange, 
the $C = -1$ partner of the pomeron \cite{BDDKNR}. 
They have predicted cross section of about 341 nb at the HERA energy. 
However, the search performed at HERA \cite{H1} 
was negative and found only an upper limit for this process 
$\sigma_{\gamma p \to \pi^0 p} < 49$~nb.
Ewerz and Nachtmann \cite{EN2007} found an explanation 
of this discrepancy within a nonperturbative approach 
using approximate chiral symmetry and 
partially conserved axial vector current (PCAC). 
They have found that the amplitude for diffractive neutral pion
production is proportional to $m_{\pi}^{2}$
and vanishes in the chiral limit ($m_{\pi} \to 0$).
They have estimated that the cross section is probably damped by a
factor of 50 (see \cite{DDN06}) compared to the early estimate in \cite{BDDKNR}.
The exclusive production of neutral pions at midrapidity can be used
to search for odderon exchange as well as to test the predictions of 
Ref.\cite{EN2007}.

The cross section for photon-odderon and odderon-photon exchanges 
can be estimated in the equivalent photon approximation
similar as was done in Ref.\cite{LS13} for other photon induced processes.
In such an approach the distribution of the neutral pions can be written as
\begin{eqnarray}
\frac{d\sigma}{d y d p_{\perp}^2} &=&
z_1 f(z_1) \frac{d \sigma_{\gamma p \to \pi^0 p}}{dt_2}
\left(s_{23},t_2 \approx -p_{\perp}^2 \right)  \nonumber \\
&+& z_2 f(z_2) \frac{d \sigma_{\gamma p \to \pi^0 p}}{dt_1}
\left(s_{13},t_1 \approx -p_{\perp}^2 \right)\,,
\label{EPA}
\end{eqnarray}
where $f(z)$ is an elastic photon flux in the proton; an explicit
formula can be found e.g. in \cite{DZ89}. In the formula above,
$z_{1/2} = \frac{m_{\perp}}{\sqrt{s}}\exp (\pm y)$ with 
$m_{\perp} = \sqrt{m_{\pi}^2 + p_{\perp}^2}$.

The differential cross section $\gamma p \to \pi^0 p$ will be
parametrized in the present paper as:
\begin{equation}
\frac{d \sigma_{\gamma p \to \pi^0 p}}{d t} = B^2 (-t) \exp(B t) \sigma_{\gamma p \to \pi^0 p} \; .
\label{gammap_pi0p}
\end{equation}
The differential cross section vanishes at $t = 0$ which is due to helicity flip
in the $\gamma \to \pi^{0}$ transition.
The slope parameter can be expected to be typically as for other soft 
processes $B \sim 4-8$~GeV$^{-2}$. At the LHC and at midrapidities 
typical energies in the photon-proton subsystems are similar as at the HERA.
In the following we shall consider two scenarios: 
HERA upper limit ($\sigma_{\gamma p \to \pi^0 p} = 49$~nb) and
Ewerz-Nachtmann estimate ($\sigma_{\gamma p \to \pi^0 p} = 6$~nb).

\section{Results}
\label{section:Results}
\begin{figure}[!ht]    
\includegraphics[width=0.9\textwidth]{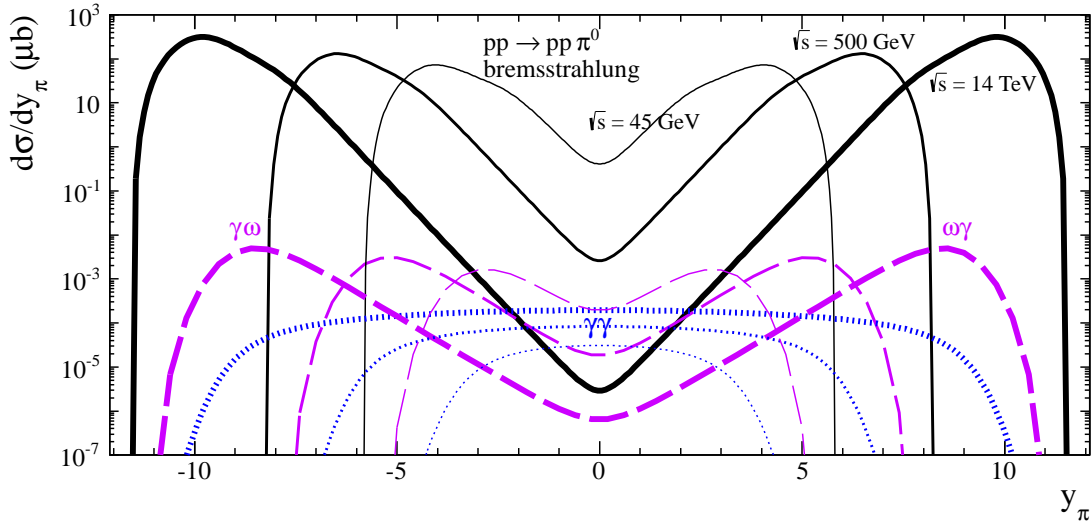}
\caption{\label{fig:dsig_dy_all}
\small The distribution of $\pi^0$ in rapidity 
at $\sqrt{s}=45$~GeV (ISR), 500~GeV (RHIC), and 14~TeV (LHC).
The $\pi^{0}$-bremsstrahlung contribution (black solid lines) and
$\omega \gamma$ ($\gamma \omega$) exchanges (violet dashed lines) 
peaks at forward (backward) region of $y_{\pi^{0}}$, respectively,
while $\gamma \gamma$ fusion (blue dotted lines) contributes at midrapidity.
In this calculation we have used 
$\Lambda_{N} = \Lambda_{\pi} = 1$~GeV of the hadronic form factors.
No absorption effects are included here.
}
\end{figure}

Now we present our calculations of cross sections and distributions
of the exclusive $\pi^{0}$ meson production in proton-proton collisions.
The rapidity distributions of $\pi^{0}$
are shown in Fig.\ref{fig:dsig_dy_all} at center-of-mass energies
$\sqrt{s}=45$~GeV (ISR), 500~GeV (RHIC) and 14~TeV (LHC)
for all processes considered in the present paper.
We present results for the diffractive $\pi^{0}$-bremsstrahlung mechanisms
as well as photon-photon fusion and
photon-omega (omega-photon) exchange processes not discussed so far in the literature.
The higher the energy, the two $\pi^{0}$-bremsstrahlung contributions become better separated.
The dotted line corresponds to the photon-photon fusion mechanism.
At the LHC energy and in the rapidity region $-2 < y_{\pi^{0}} < 2$
it even dominates over the diffractive mechanism.
The cross section for the $\pi^{0}$-bremsstrahlung contribution
at the LHC energy and at midrapidity is much smaller
than e.g. for the production of heavy quarkonia:
$J/\psi$ \cite{SS07}, $\Upsilon$ \cite{RSS08} or $\chi_{c0}$ \cite{PST07,LPS11}.
Clearly an experimental measurement there would be a challenge.

Let us look now how absorption effects discussed
in the theory section (see Fig.{\ref{fig:diagrams_abs})
can modify the results obtained with the bare amplitudes (see Fig.{\ref{fig:diagrams_deck}). 
In Fig.\ref{fig:dsig_dy_deco} we present, in addition,
individual contributions for the $\pi^{0}$-bremsstrahlung mechanism.
We observe a large cancellation between the two terms in the amplitude
[between the initial ($p$-exchange) and final state radiation (direct production)].
Because of destructive interference of bare and absorptive correction amplitudes,
the resulting cross section is by a factor 2 to 3 smaller 
than that for the bare amplitude.
The difference between the solid ($\Lambda_{N} = \Lambda_{\pi} = 1$~GeV)
and dashed ($\Lambda_{N} = 0.6$~GeV and $\Lambda_{\pi} = 1$~GeV)
curves represents the uncertainties on the form factors.
\begin{figure}[!ht]    
(a)\includegraphics[width=0.465\textwidth]{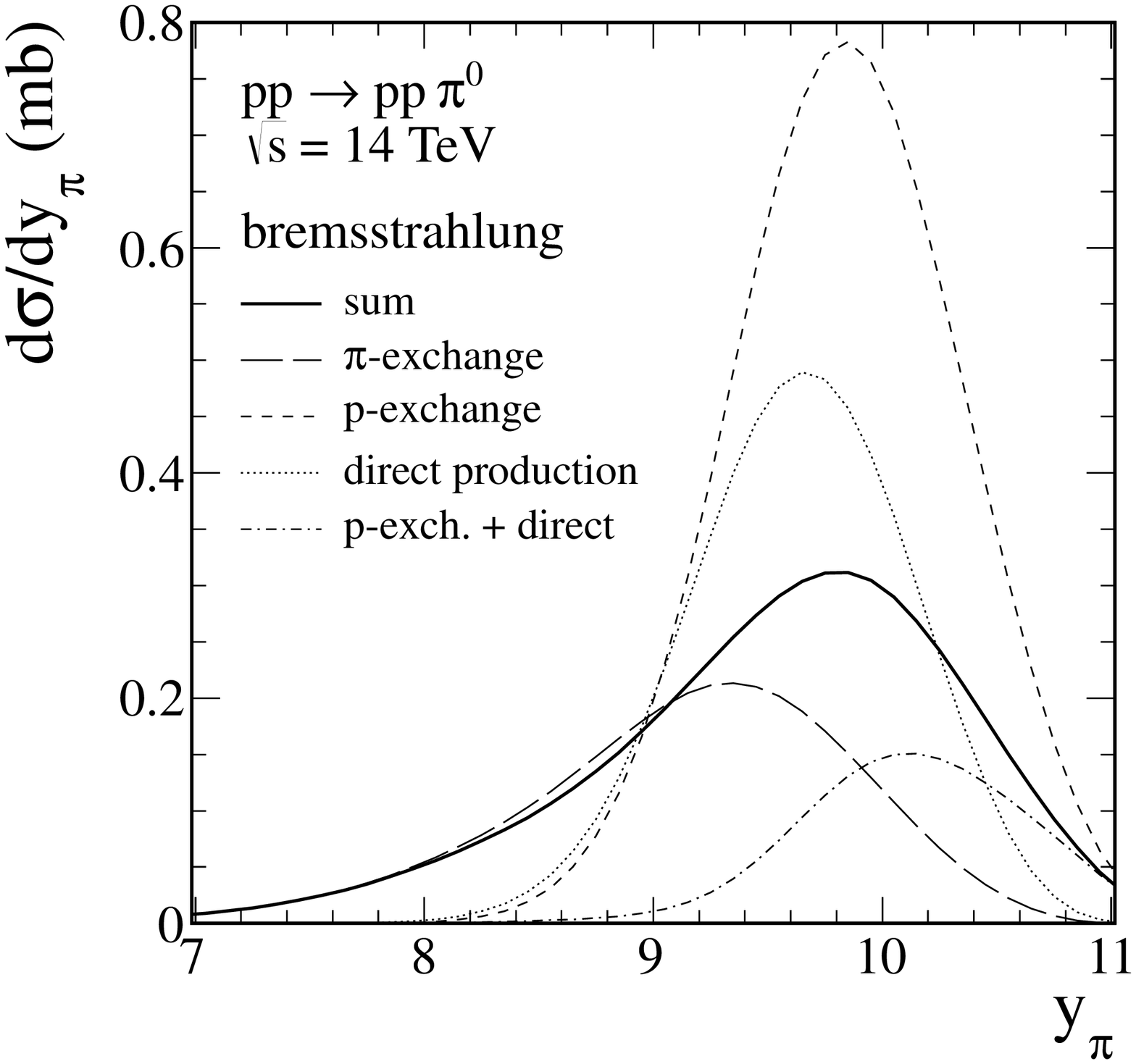}\\
(b)\includegraphics[width=0.465\textwidth]{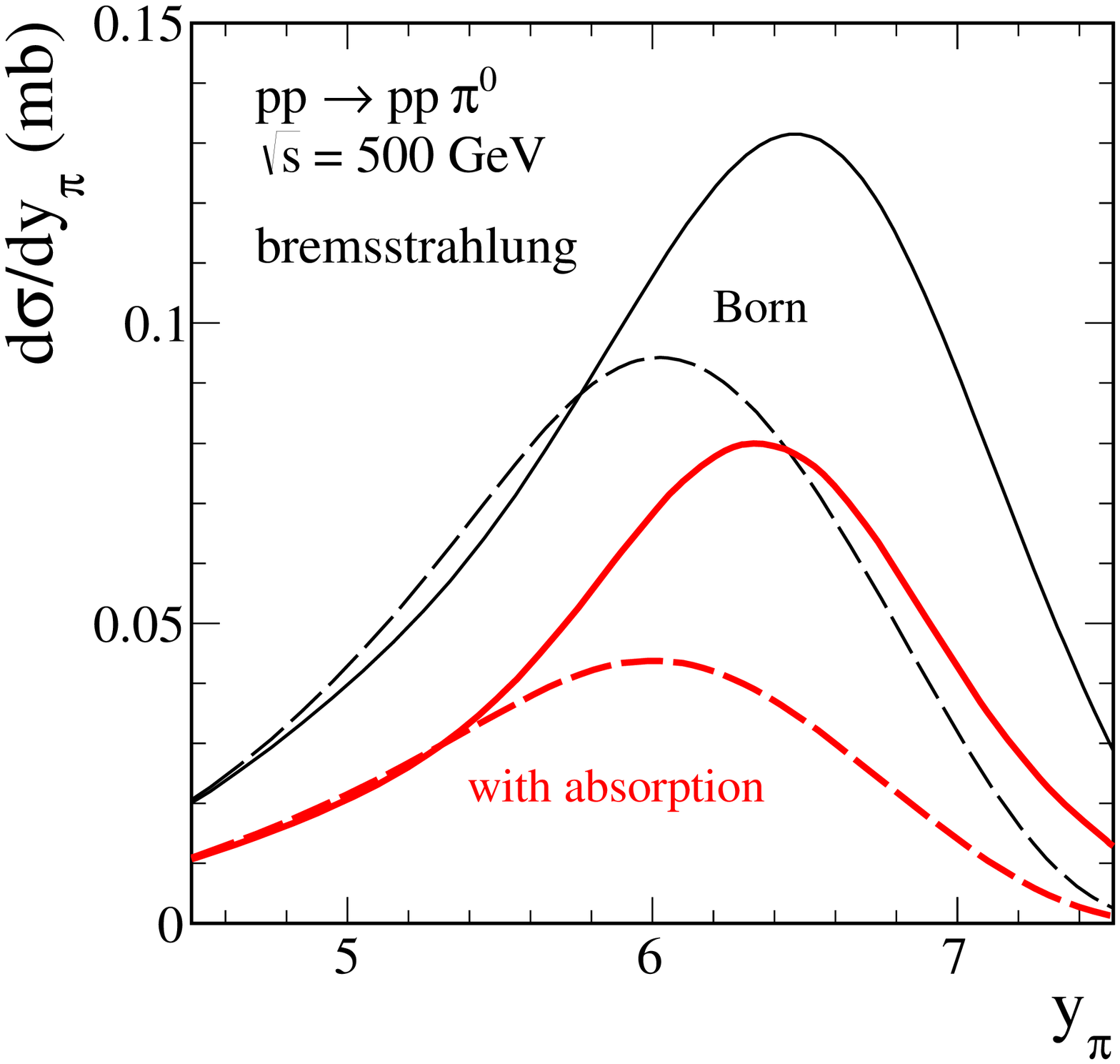}
(c)\includegraphics[width=0.465\textwidth]{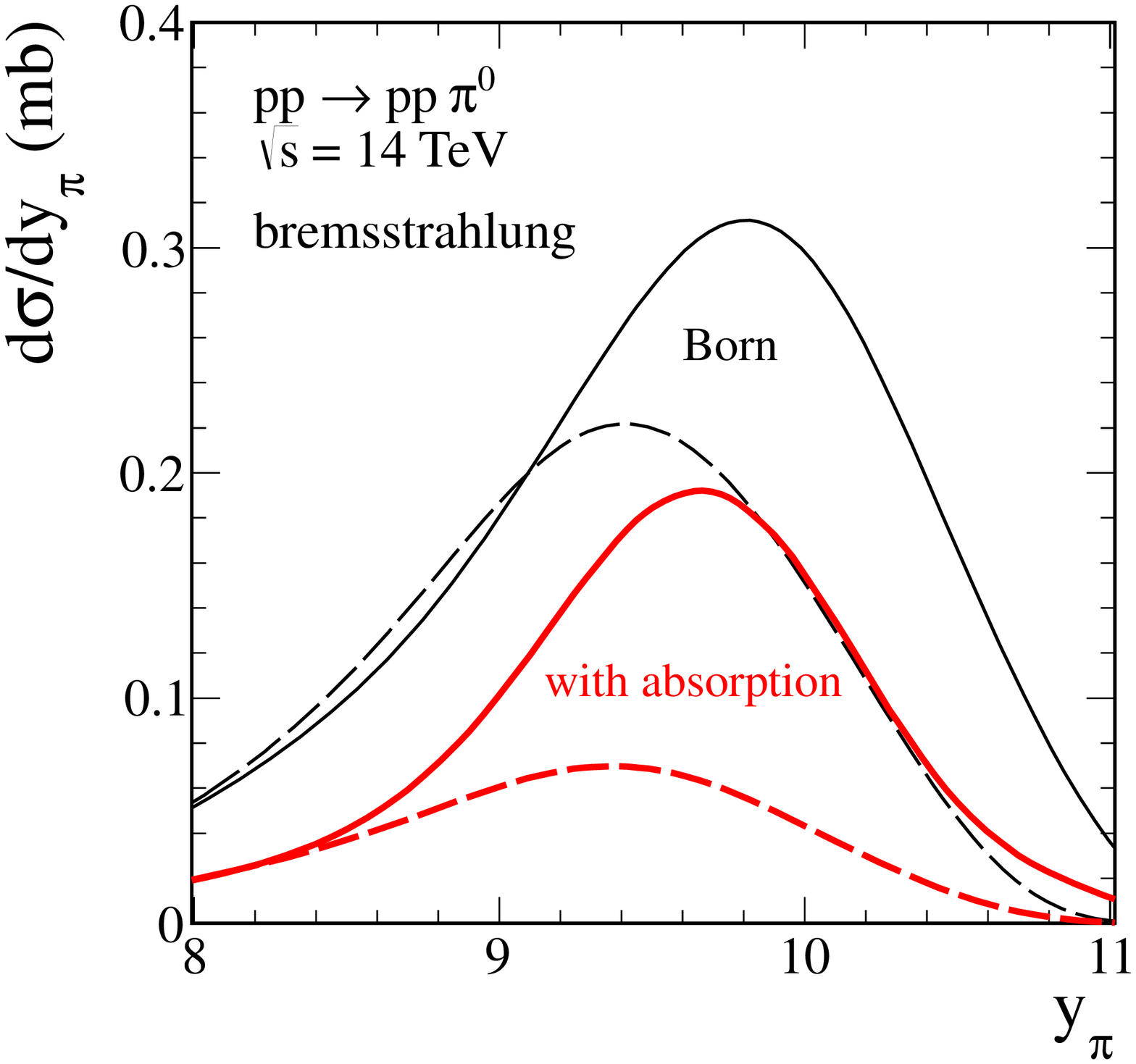}
\caption{\label{fig:dsig_dy_deco}
\small The distribution of $\pi^{0}$ in rapidity for $\sqrt{s} = 0.5, 14$~TeV.
In panel (a) we show individual contributions to the Born cross section.
A large cancellation between the initial ($p$-exchange) 
and final state radiation (direct production) can be observed.
In panels (b) and (c) the upper solid line corresponds to calculations without absorption effects,
the lower solid line with absorption effects.
The solid lines are for $\Lambda_{N} = \Lambda_{\pi} = 1$~GeV
while the dashed lines are for $\Lambda_{N} = 0.6$~GeV and $\Lambda_{\pi} = 1$~GeV.
}
\end{figure}

At large $y_{\pi^{0}}$ another mechanism may come into the game
-- diffractive excitation of nucleon resonances.
The resonances may occur when the energy
in the $\pi N$ subsystem $W_{\pi N} \in {\cal R}$,
where ${\cal R}$ is the nucleon resonance domain.
If $\left\langle W_{13} \right\rangle (y_{\pi^{0}}) \in {\cal R}$ 
or $\left\langle W_{23} \right\rangle (y_{\pi^{0}}) \in {\cal R}$ then
an extra strength due to resonance excitation may occur.
In Fig.\ref{fig:ave_w.eps} we present 
the average value of subsystem energies $\left\langle W_{13} \right\rangle$ 
and $\left\langle W_{23} \right\rangle$ 
as a function of $y_{\pi^{0}}$ at $\sqrt{s} = 0.5, 14$~TeV.
Only some nucleon resonances can be excited diffractively.
At the LHC they can occur for $8 < |y_{\pi^{0}}| < 11$
and at the RHIC for $4.5 < |y_{\pi^{0}}| < 7.5$.
One way to introduce resonances in the DHD model is to include
them as intermediate states in the direct production term in Eq.(\ref{brem_a}) 
[see also Fig.\ref{fig:diagrams_deck}(c)].
Calculating the contribution of diffractively produced resonances 
is more complicated and clearly goes beyond the scope of the present paper.
The reader can find some theoretical attempts in Ref.\cite{JKOS2012}.
\begin{figure}[!ht]    
\includegraphics[width=0.465\textwidth]{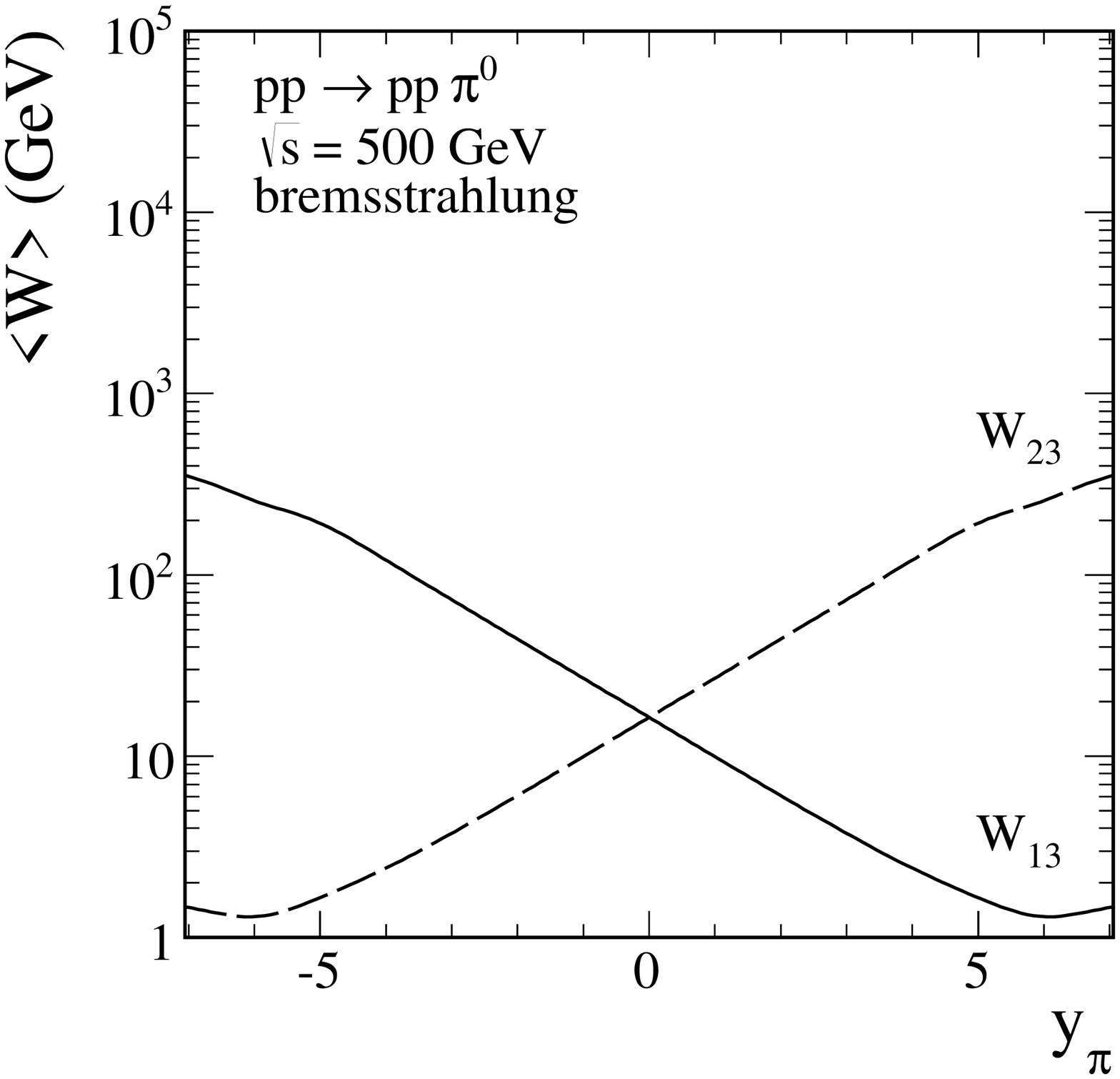}
\includegraphics[width=0.465\textwidth]{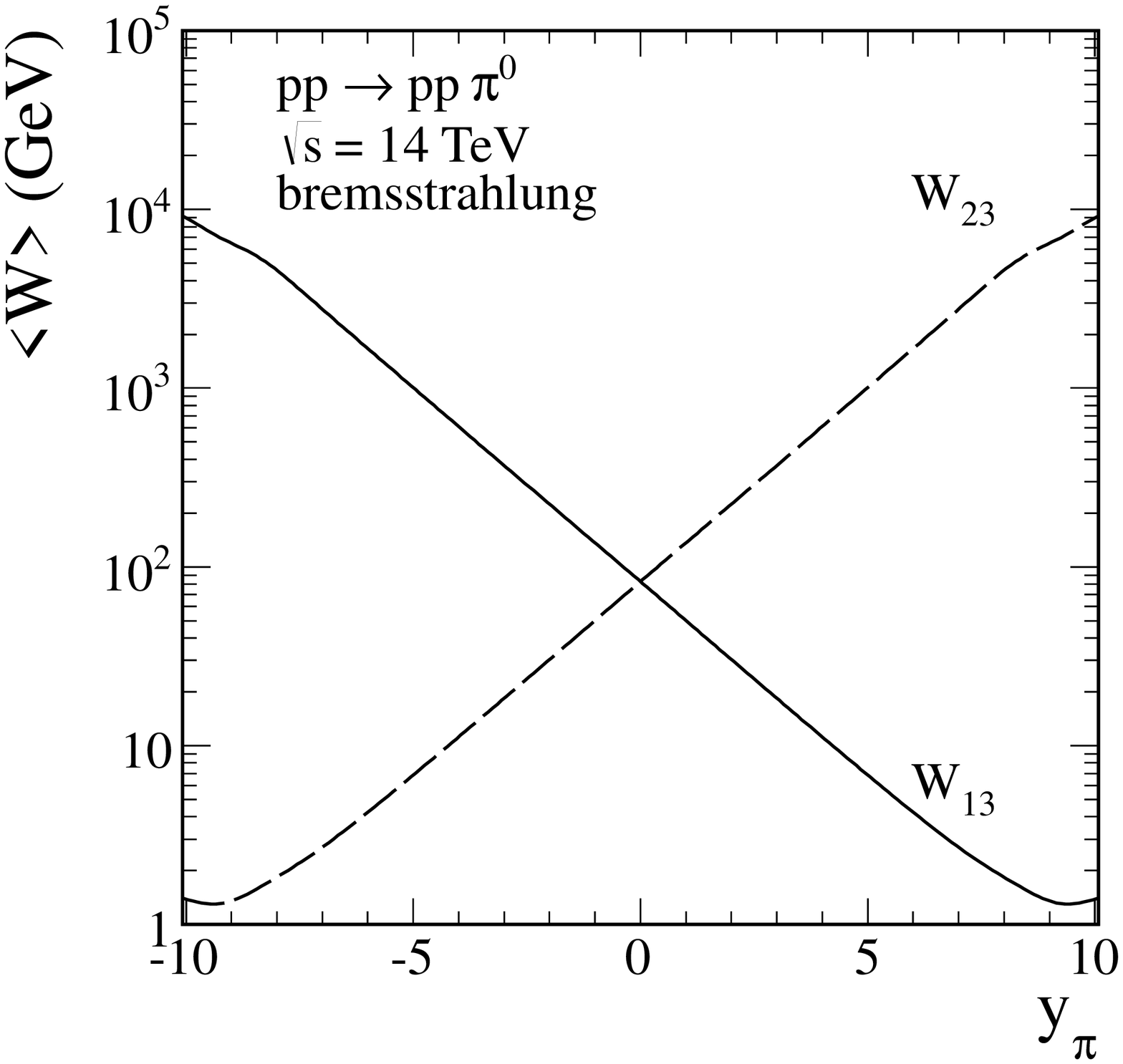}
\caption{\label{fig:ave_w.eps}
\small 
The average value of subsystem energies $\left\langle W_{13} \right\rangle (y_{\pi^{0}})$ (solid line) 
and $\left\langle W_{23} \right\rangle (y_{\pi^{0}})$ (dashed line) at $\sqrt{s} = 0.5, 14$~TeV.
Here $\Lambda_{N} = \Lambda_{\pi} = 1$~GeV.
}
\end{figure}

In Fig.\ref{fig:dsig_deta} we show corresponding distribution in proton pseudorapidity
again without (upper lines) and with (lower lines) absorption effects
and for two sets of $\Lambda_{\pi}$ and $\Lambda_{N}$ parameters.
At the LHC protons could be measured by the ALFA (ATLAS) or TOTEM (CMS) detectors.
\begin{figure}[!ht]    
\includegraphics[width=0.465\textwidth]{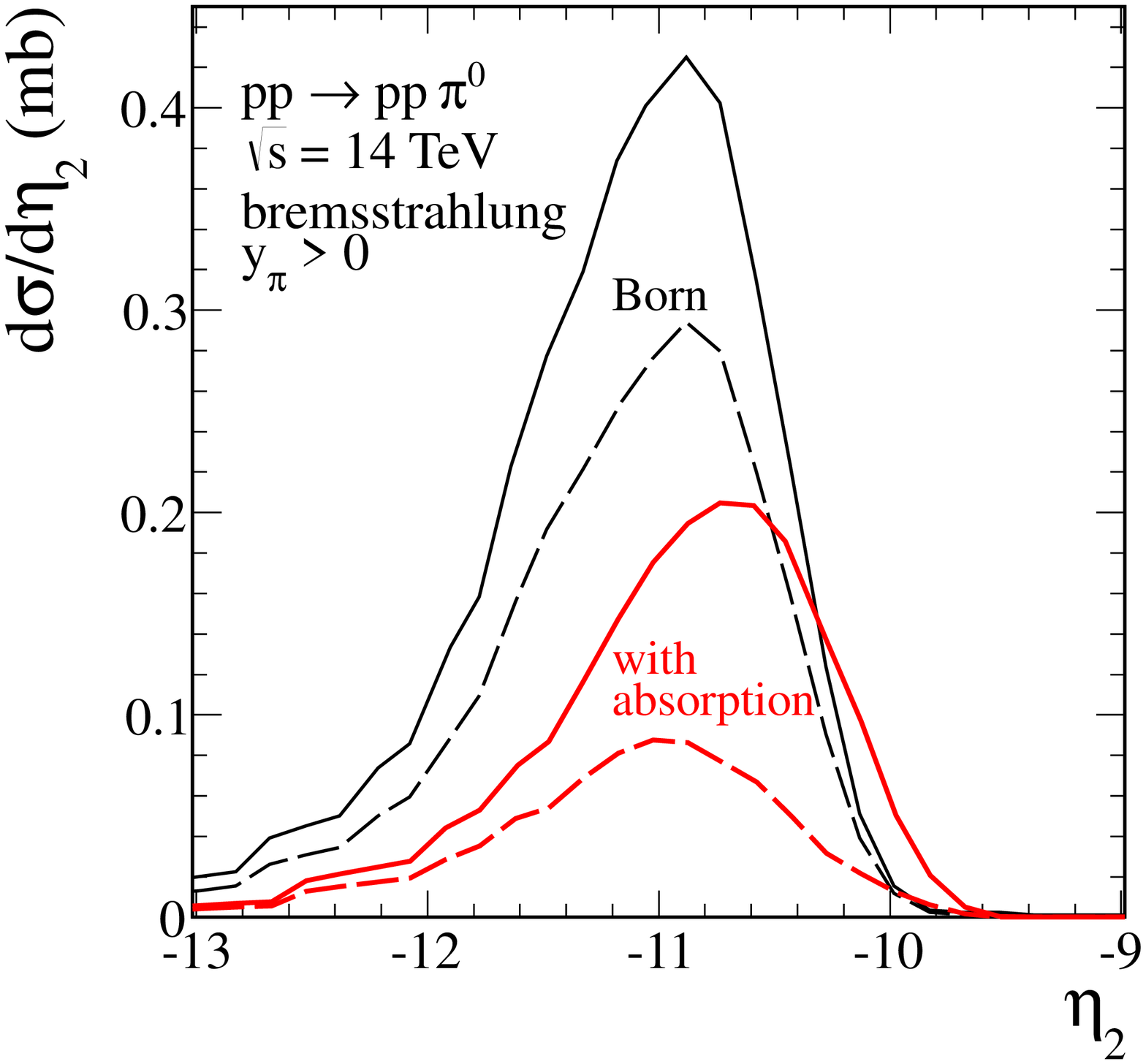}
\includegraphics[width=0.465\textwidth]{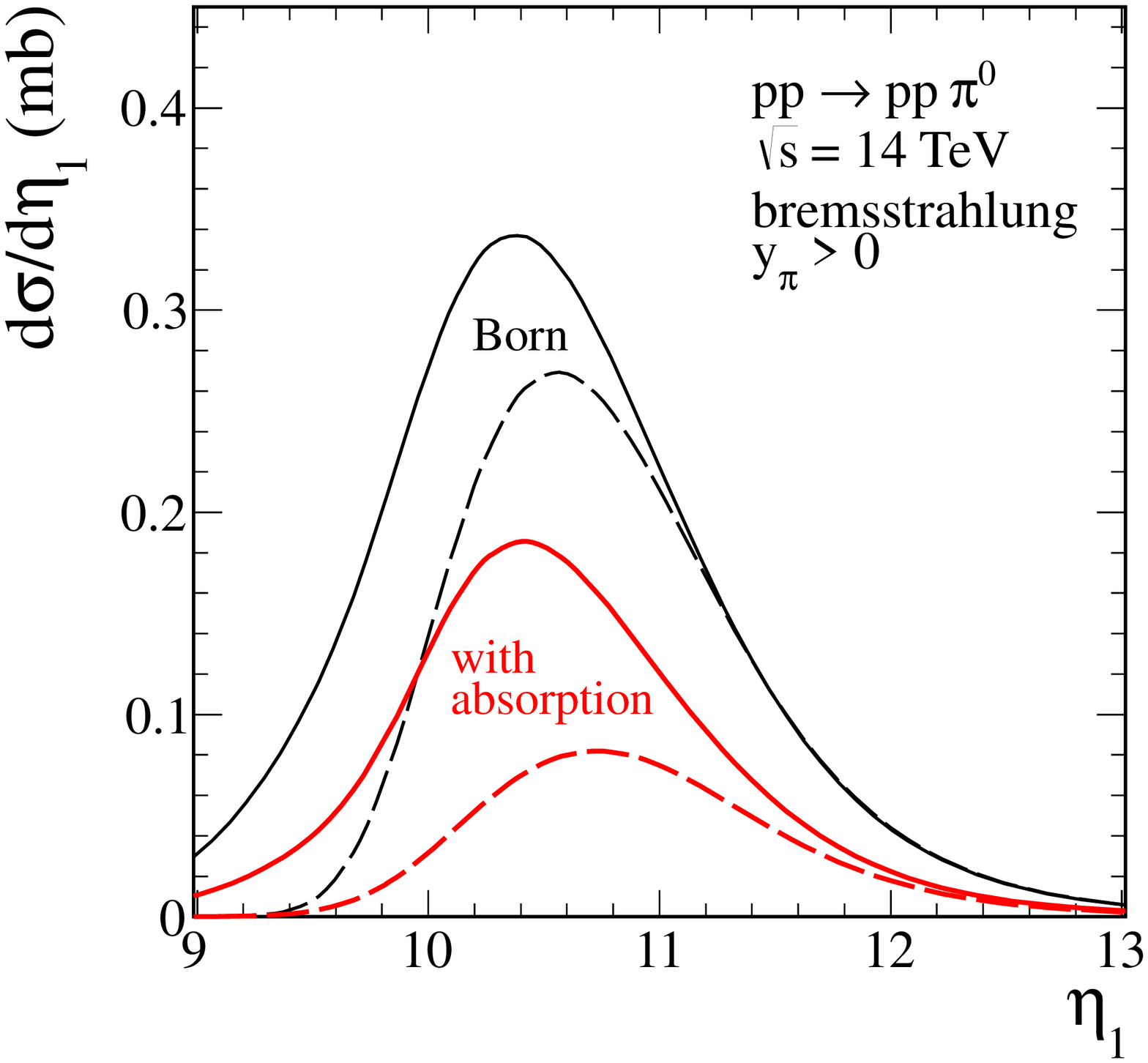}
\caption{\label{fig:dsig_deta}
\small 
The distribution in pseudorapidity of protons in the backward (left panel) 
and the forward (right panel) hemisphere at $\sqrt{s} = 14$~TeV and for $y_{\pi^{0}} > 0$.
Here $\Lambda_{N} = \Lambda_{\pi} = 1$~GeV (solid line) or
$\Lambda_{N} = 0.6$~GeV and $\Lambda_{\pi} = 1$~GeV (dashed line).
}
\end{figure}

The effect of absorption on transverse momentum spectra of protons 
and neutral pions is more complicated. 
In Fig.\ref{fig:dsig_dpt1} we show distribution in transverse momentum of outgoing protons.
Absorption causes a transverse momentum dependent 
damping of the cross section at small $p_{\perp,p}$
and an enhancement at large $p_{\perp,p}$ (compare upper and lower solid line).
\begin{figure}[!ht]    
\includegraphics[width=0.465\textwidth]{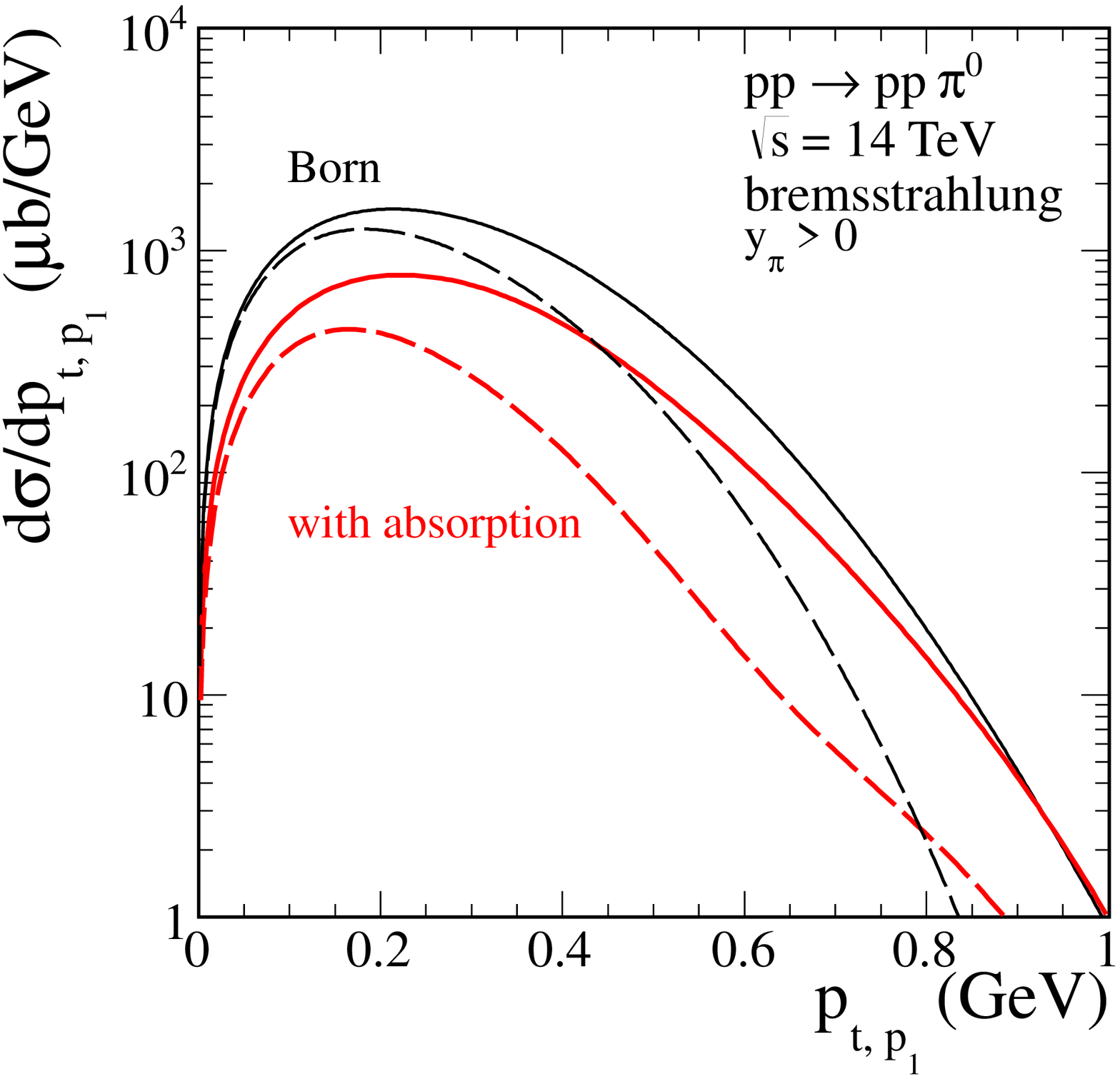}
\includegraphics[width=0.465\textwidth]{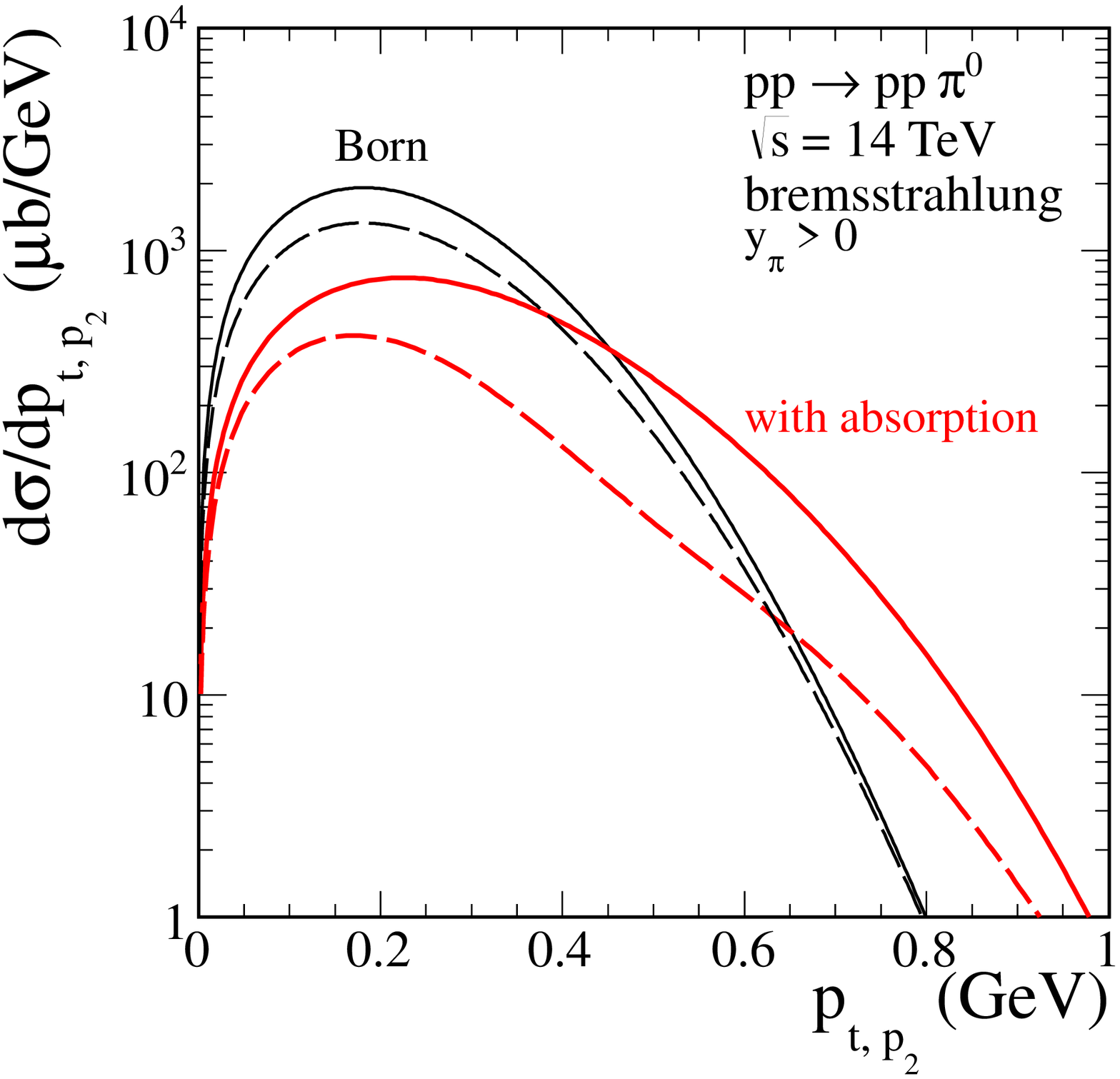}
\caption{\label{fig:dsig_dpt1}
\small  
The distribution of outgoing protons in transverse momentum for $\sqrt{s}=14$~TeV and for $y_{\pi^{0}} > 0$.
As in the previous figure we show results without and with absorption effects.
Here $\Lambda_{N} = \Lambda_{\pi} = 1$~GeV (solid line) or
$\Lambda_{N} = 0.6$~GeV and $\Lambda_{\pi} = 1$~GeV (dashed line).
}
\end{figure}

In Fig.\ref{fig:dsig_dpt3} we show distribution in transverse momentum of $\pi^0$ meson.
As in the previous figure we show results without and with absorption effects.
The distributions are peaked at $p_{\perp,\pi} \sim 0.2$~GeV.
\begin{figure}[!ht]    
(a)\includegraphics[width=0.465\textwidth]{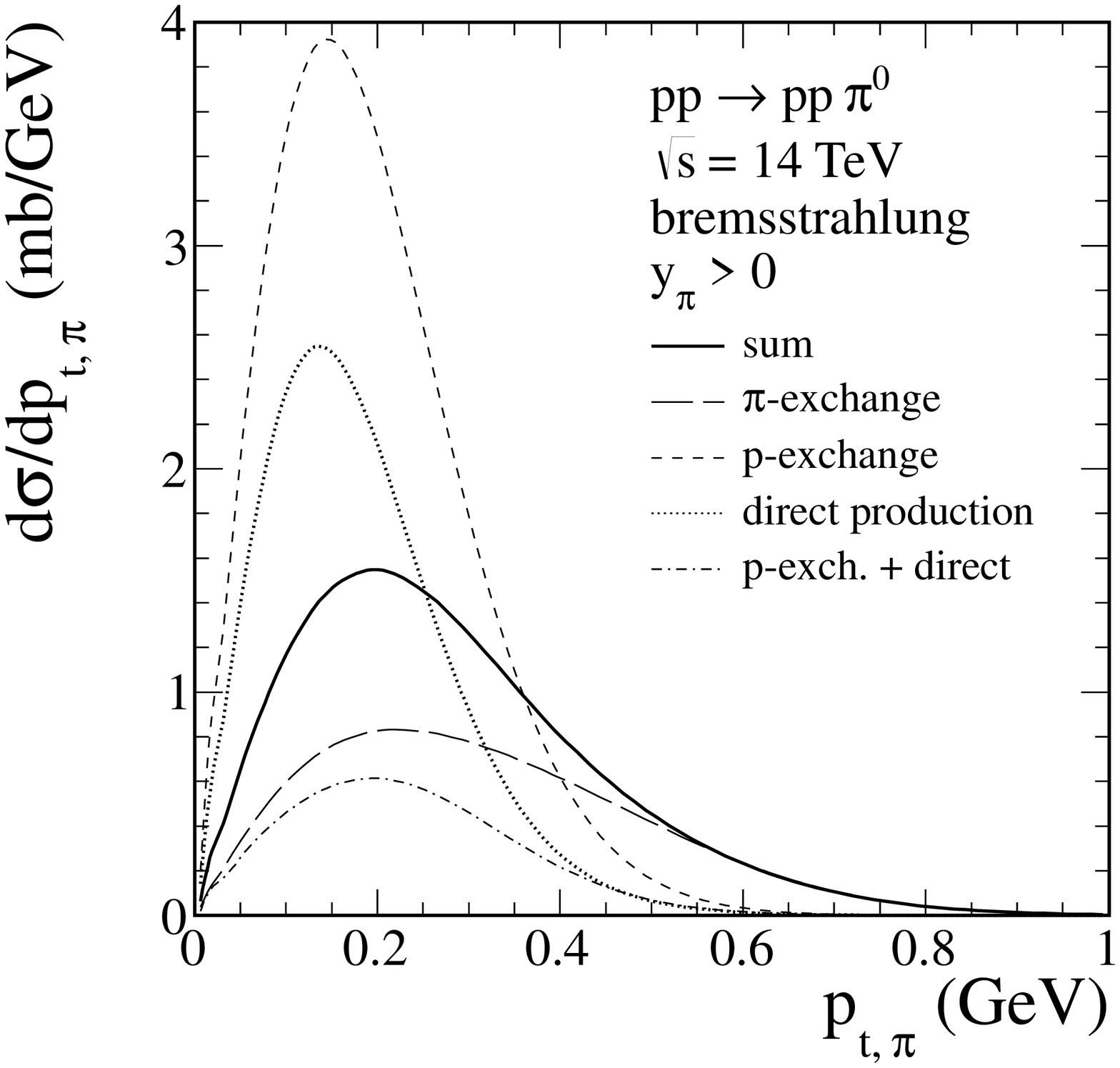}\\
(b)\includegraphics[width=0.465\textwidth]{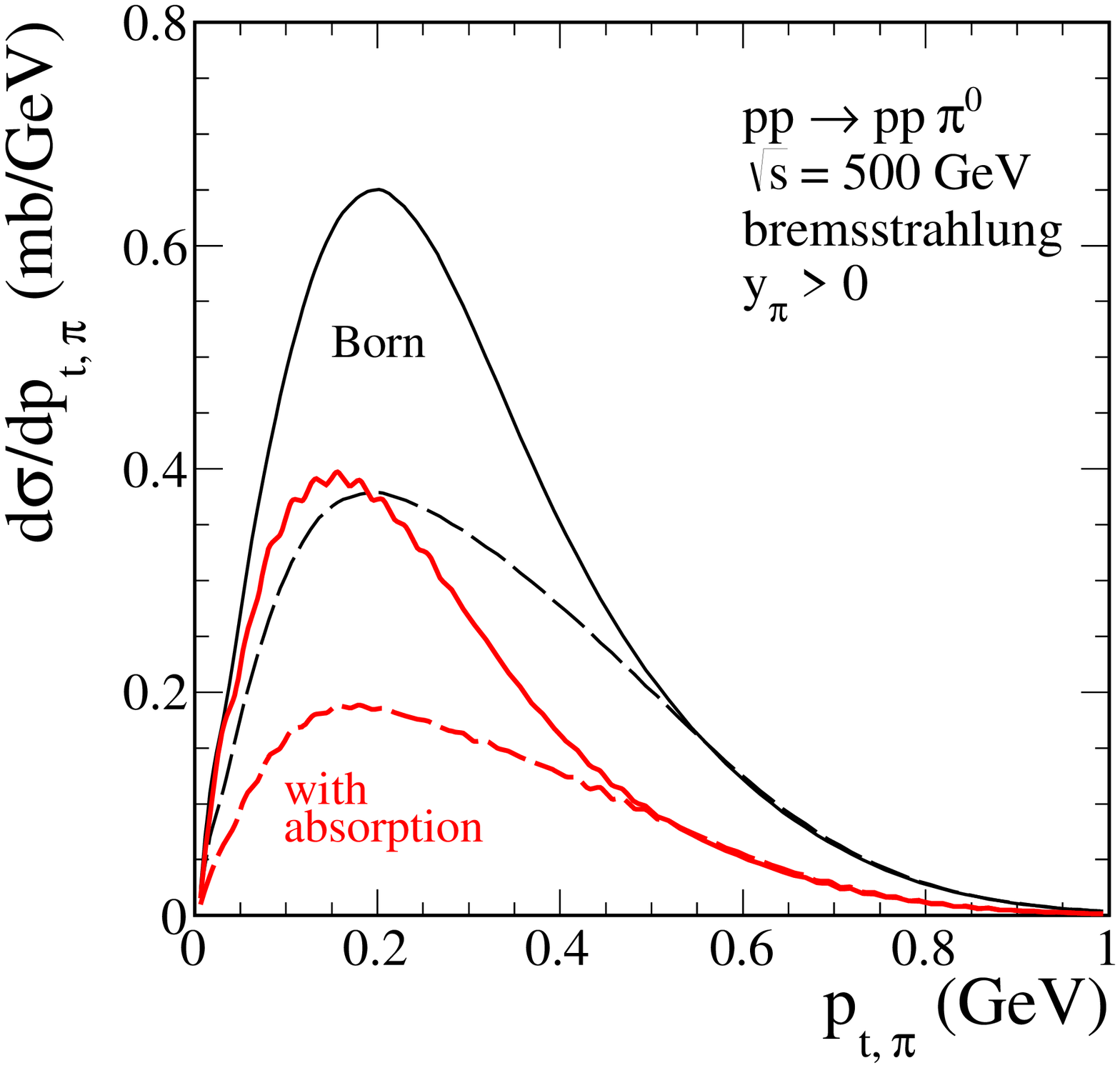}
(c)\includegraphics[width=0.465\textwidth]{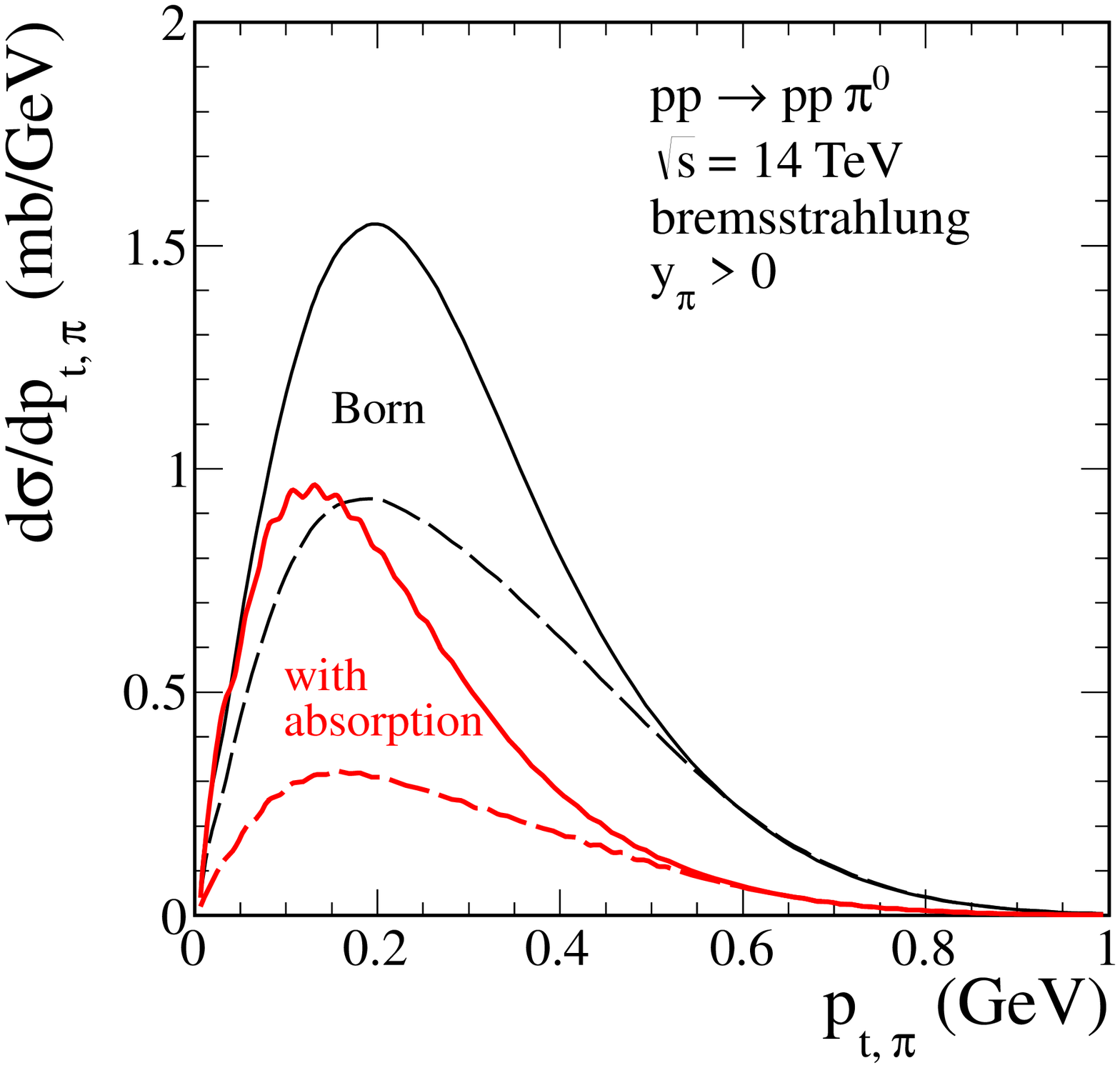}
\caption{\label{fig:dsig_dpt3}
\small
The distribution of $\pi^0$ mesons in transverse momentum for $\sqrt{s}=0.5, 14$~TeV and for $y_{\pi^{0}} > 0$.
In panel (a) we show individual contributions to the Born cross section. 
In the other panels as in the previous figures we show theoretical uncertainties 
(in e.g. form factors).
Here $\Lambda_{N} = \Lambda_{\pi} = 1$~GeV (solid line) or
$\Lambda_{N} = 0.6$~GeV and $\Lambda_{\pi} = 1$~GeV (dashed line).
}
\end{figure}

In Fig.\ref{fig:dsig_dt1_14000_absf} we show distribution
in the square of four-momentum transfer between initial and final protons.
In panels (a) and (b) we show the separate contributions of different exchange terms.
As in the previous figure we show results without and with absorption effects.
One can observe much large tails of distributions in $t_{1}$ than in $t_{2}$
($y_{\pi^{0}} > 0$ was assumed).
\begin{figure}[!ht]
(a)\includegraphics[width = 0.465\textwidth]{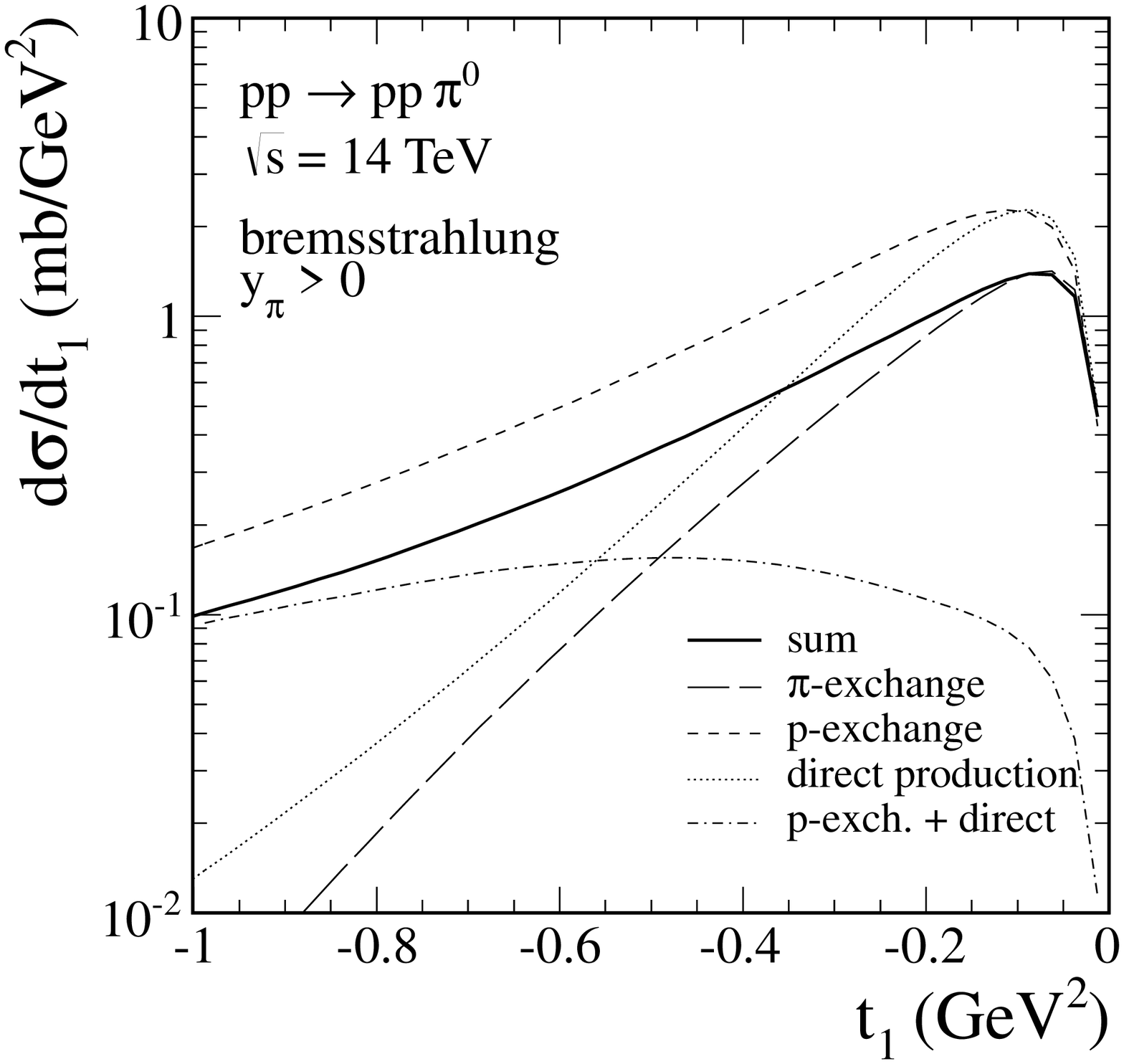}
(b)\includegraphics[width = 0.465\textwidth]{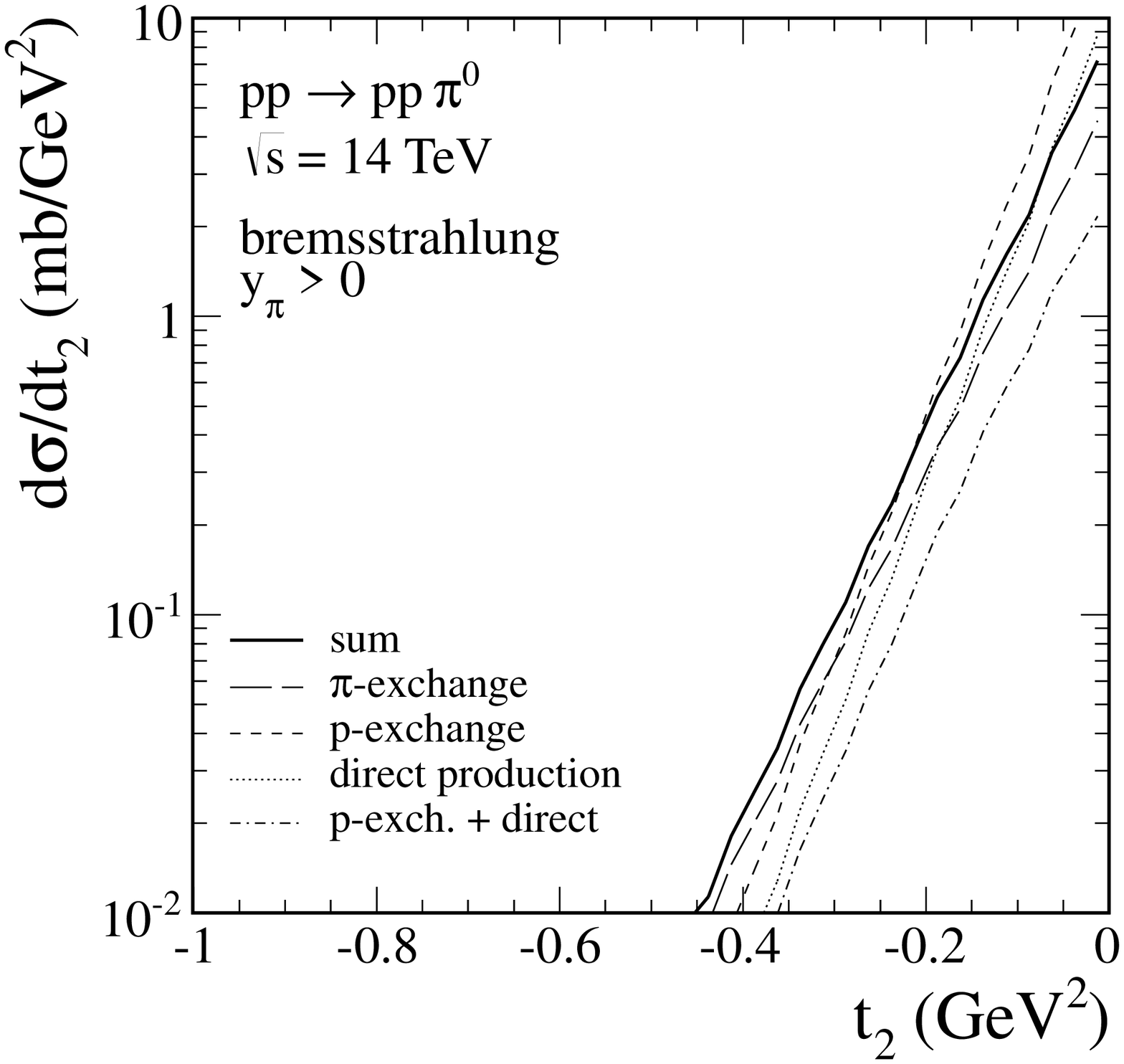}\\
(c)\includegraphics[width = 0.465\textwidth]{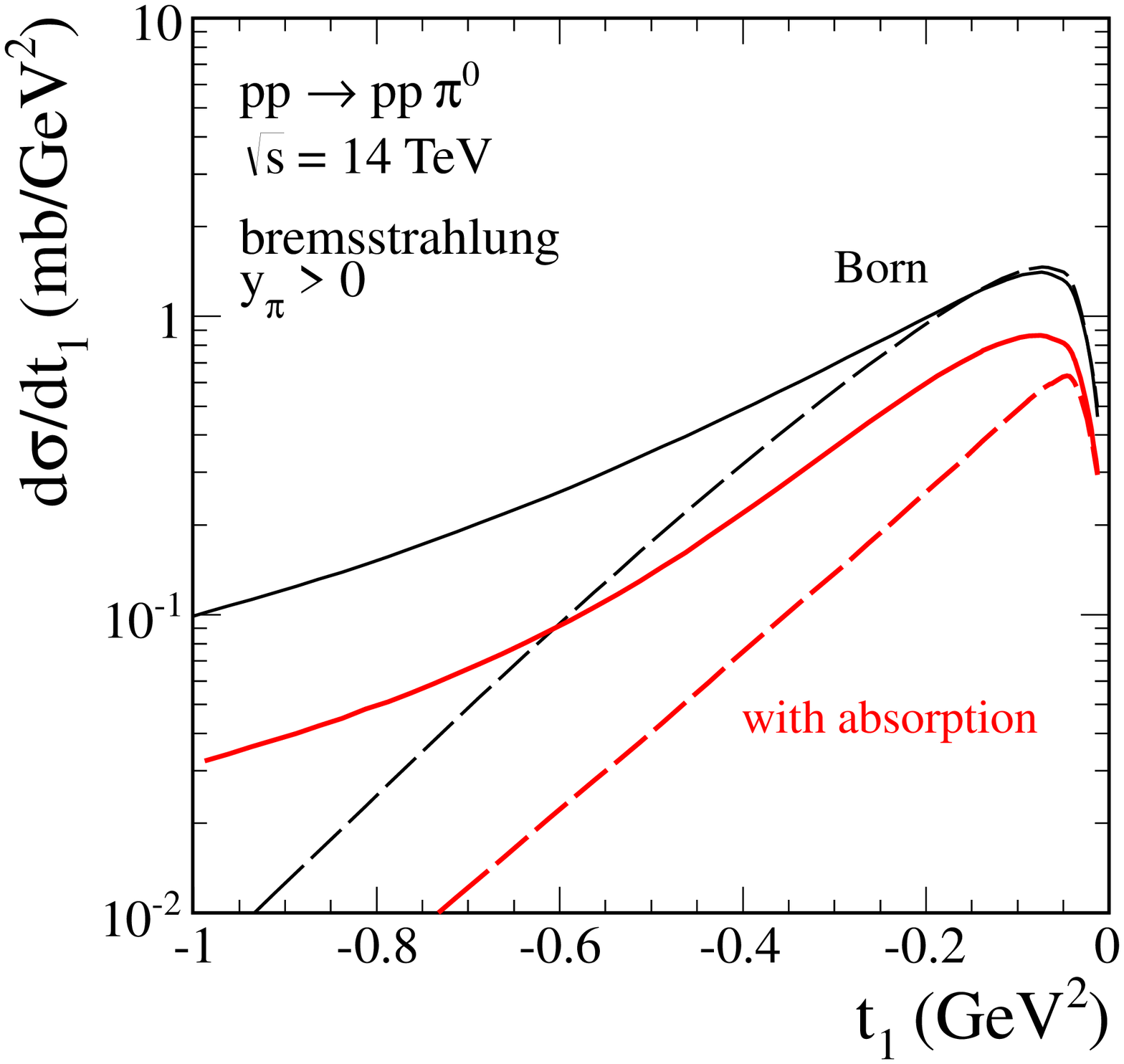}
(d)\includegraphics[width = 0.465\textwidth]{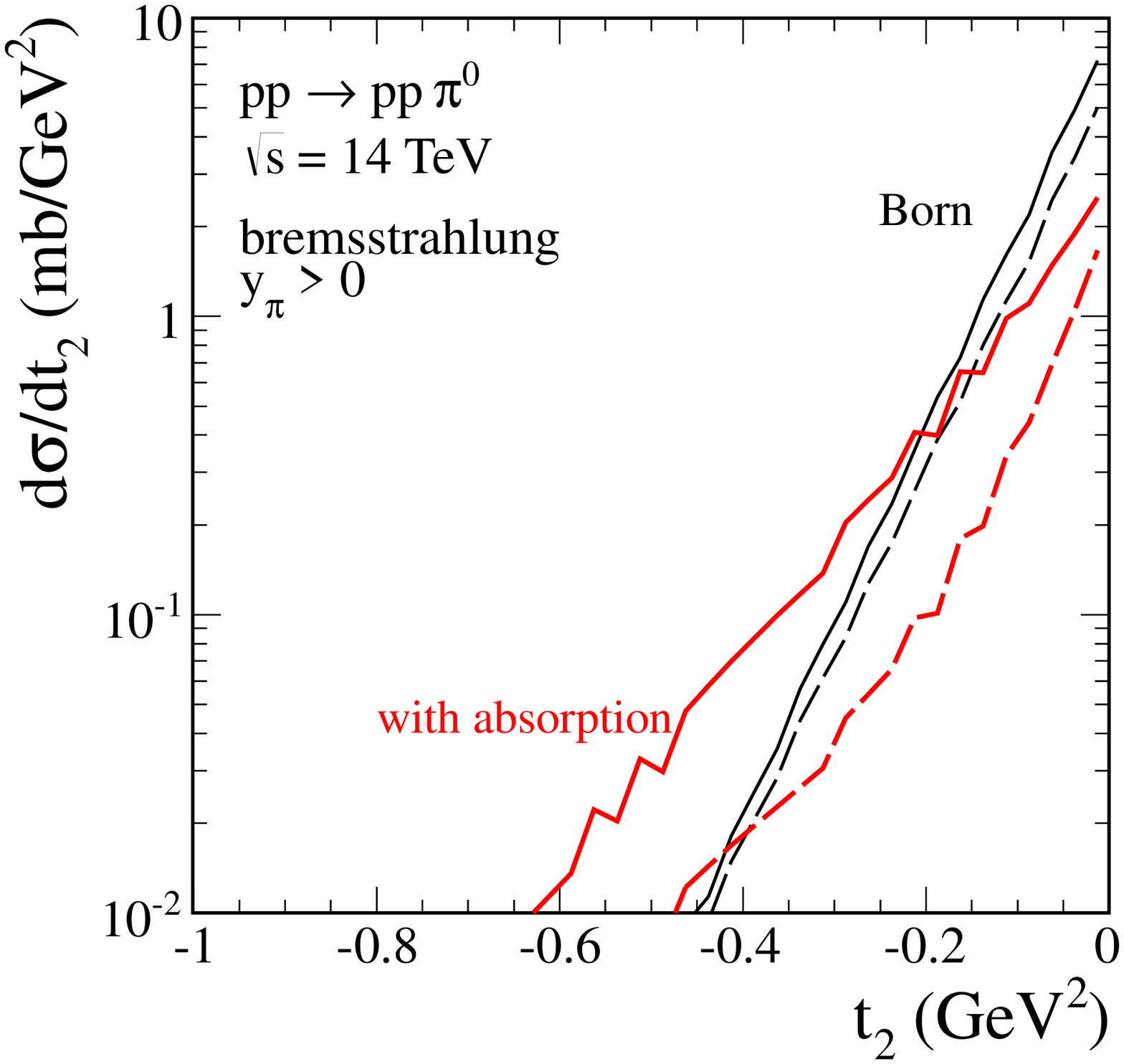}
  \caption{\label{fig:dsig_dt1_14000_absf}
  \small
Distribution in the four-momentum transfer squared between initial and final protons
at $\sqrt{s} = 14$~TeV and for $y_{\pi^{0}} > 0$. 
In panels (a) and (b) we show individual contributions to the Born cross section.
The theoretical uncertainties are shown in panels (c) and (d).
Here $\Lambda_{N} = \Lambda_{\pi} = 1$~GeV (solid line) or
$\Lambda_{N} = 0.6$~GeV and $\Lambda_{\pi} = 1$~GeV (dashed line).
}
 \end{figure}

In Fig.\ref{fig:map_t1t2} we show distribution in two-dimensional space $(t_1,t_2)$
for the $\pi^{0}$-bremsstrahlung contribution 
at $\sqrt{s} = 14$~TeV (top panels) and $\sqrt{s} = 500$~GeV (bottom panels)
without (left panel) and with (right panel) absorption effects.
The distributions in $t_{1}$ or $t_{2}$ are different because we have limited
to the case of $y_{\pi^{0}} > 0$ only.
The distributions discussed here could in principle be obtained with 
the TOTEM detector at CMS to supplement the ZDC detector 
for the measurement of neutral pions.
Similar analysis could be done by the ALFA detector for proton tagging at ATLAS.
\begin{figure}[!ht]
\includegraphics[width = 7.cm]{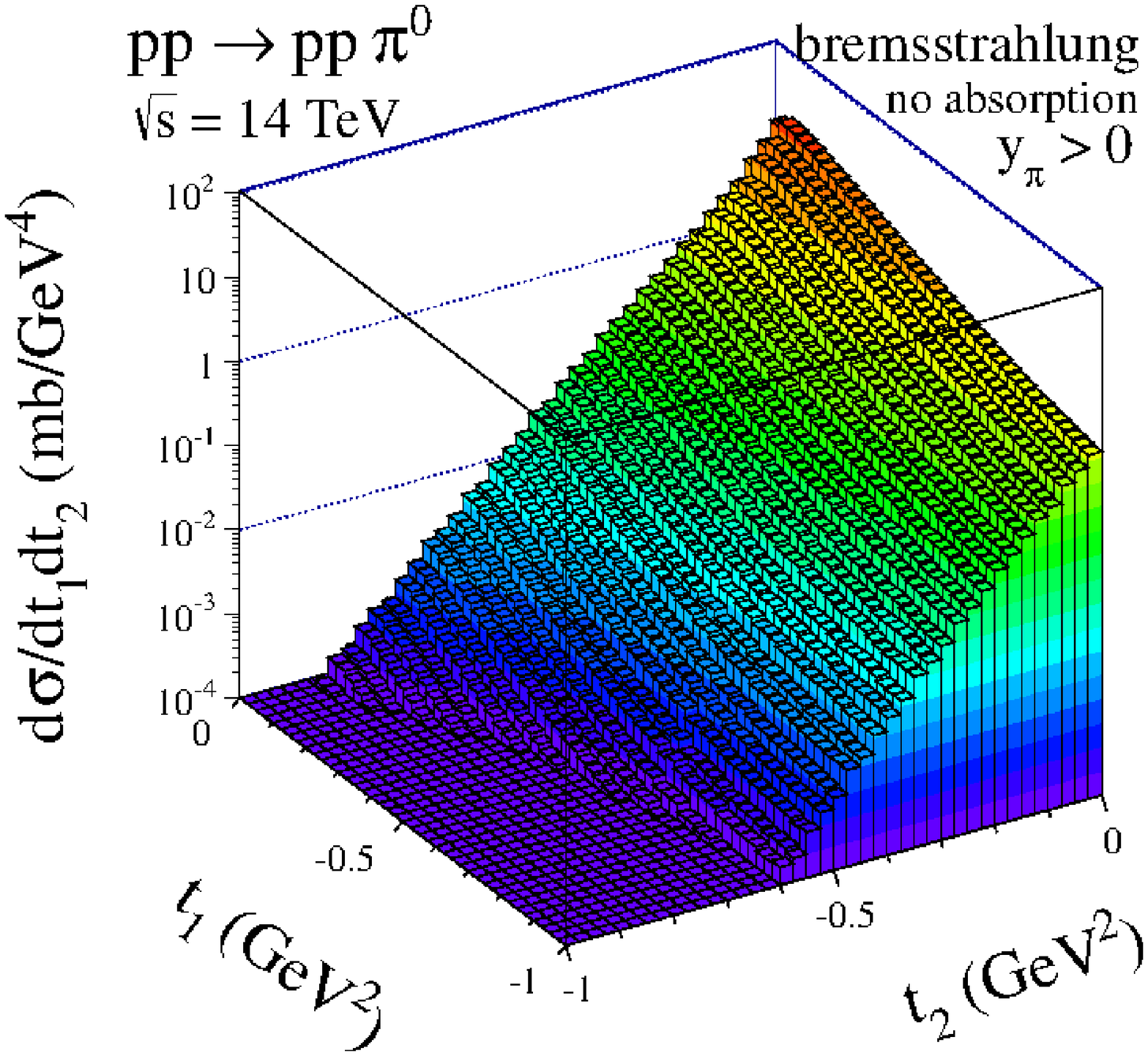}
\includegraphics[width = 7.cm]{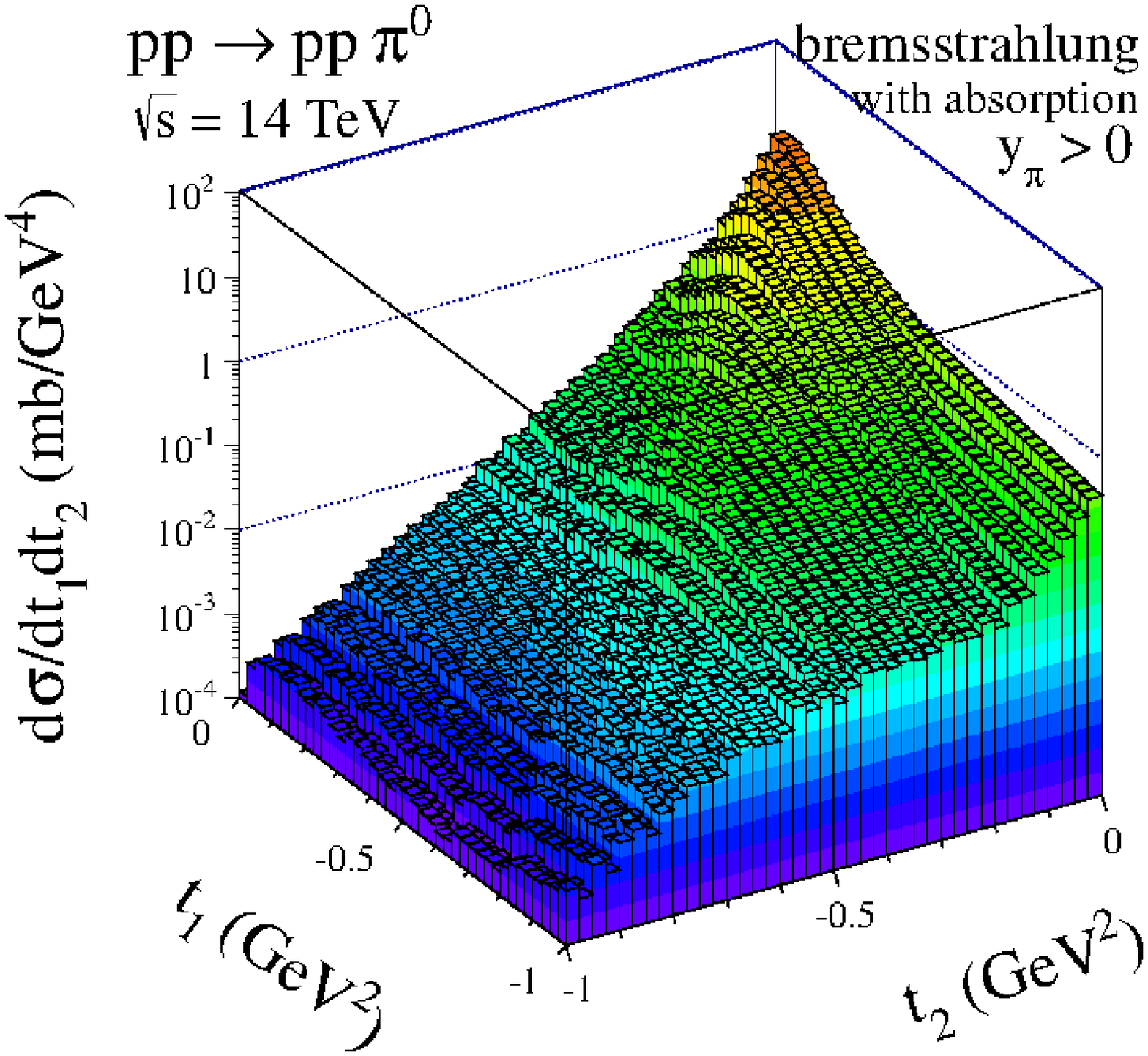}\\
\includegraphics[width = 7.cm]{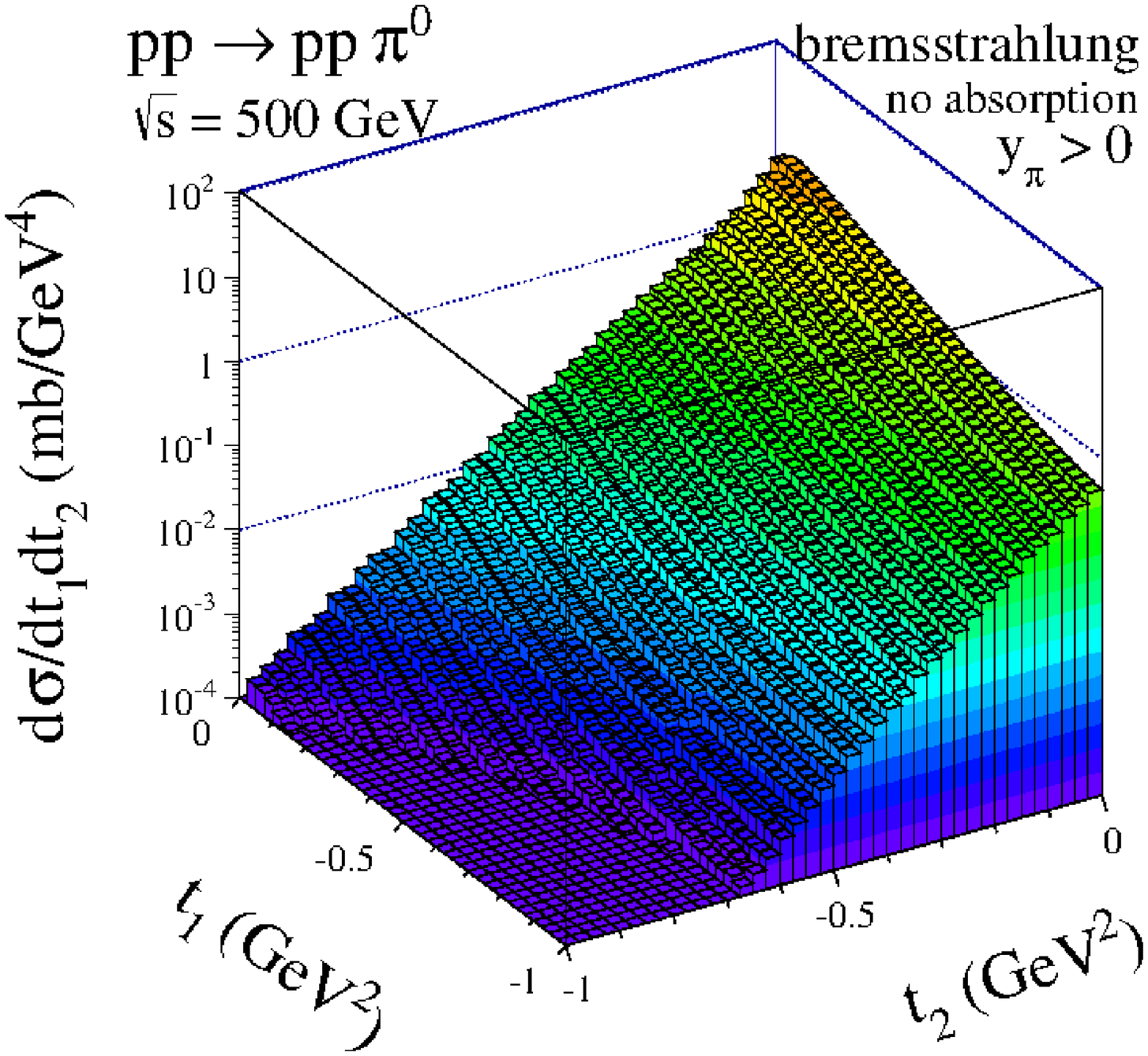}
\includegraphics[width = 7.cm]{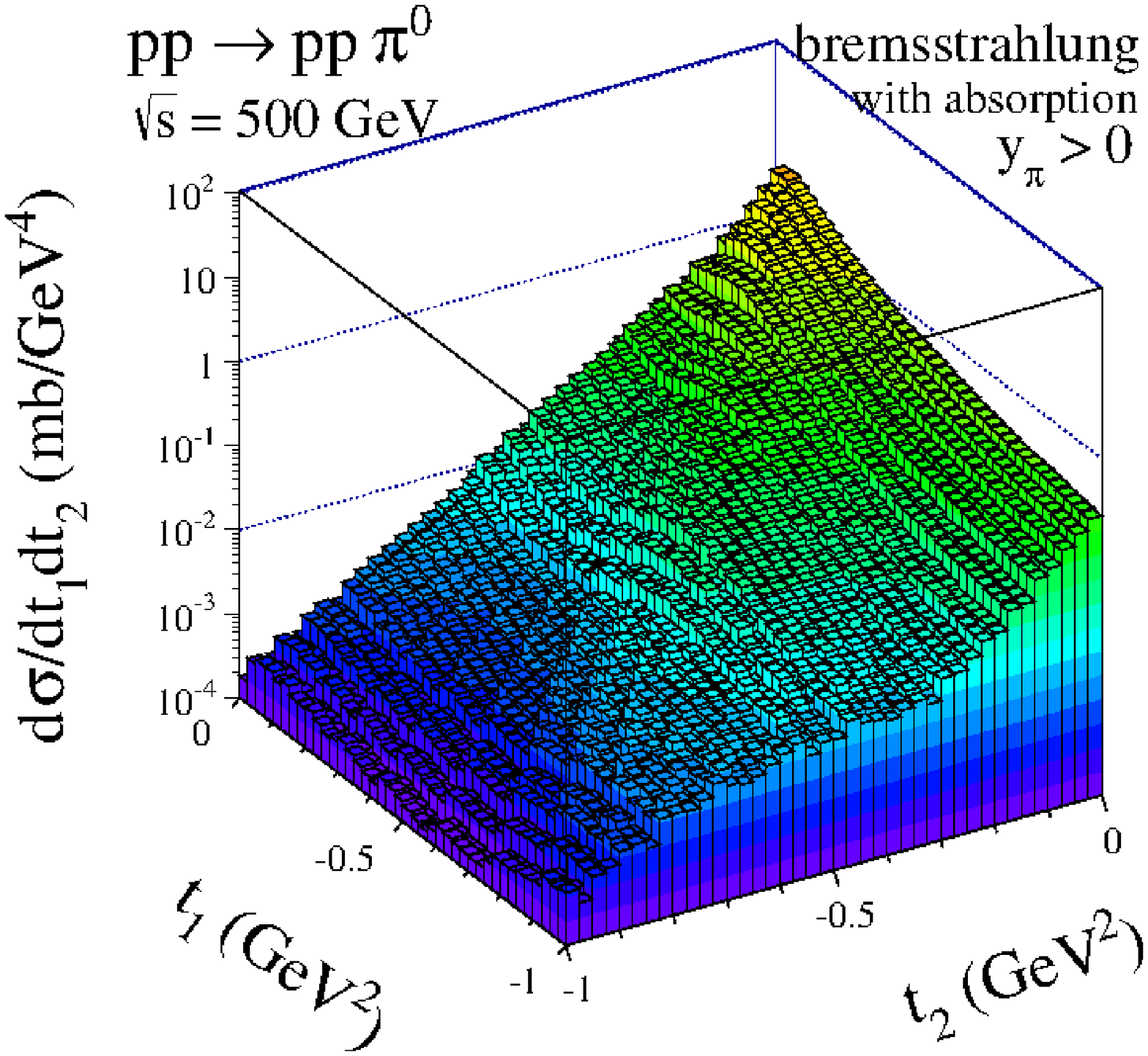}
  \caption{\label{fig:map_t1t2}
  \small
Distribution in ($t_{1}, t_{2}$) for the $\pi^{0}$-bremsstrahlung contribution
at $\sqrt{s} = 14$~TeV (top panels) and
$\sqrt{s} = 500$~GeV (bottom panels) and for $y_{\pi^{0}}>0$
without (left panels) and with (right panels) absorption effects.
Here $\Lambda_{N} = \Lambda_{\pi} = 1$~GeV.
}
\end{figure}

The pion energy spectrum for $y_{\pi^{0}} > 0$ drops relatively slowly with pion energy
which is shown in Fig.\ref{fig:dsig_dE3}.
We show results without and with absorption effects.
\begin{figure}[!ht]
(a)\includegraphics[width = 0.465\textwidth]{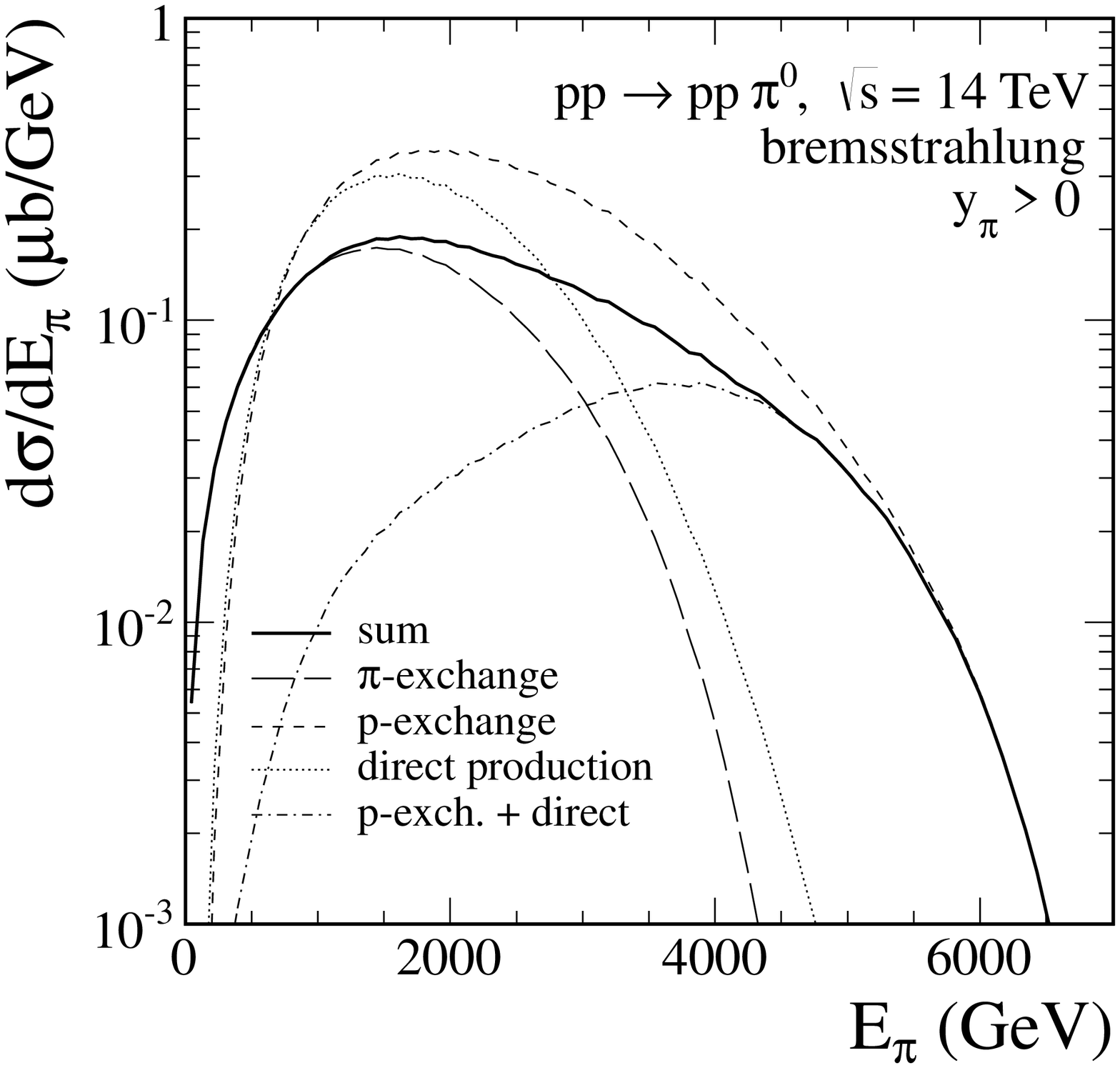}\\
(b)\includegraphics[width = 0.465\textwidth]{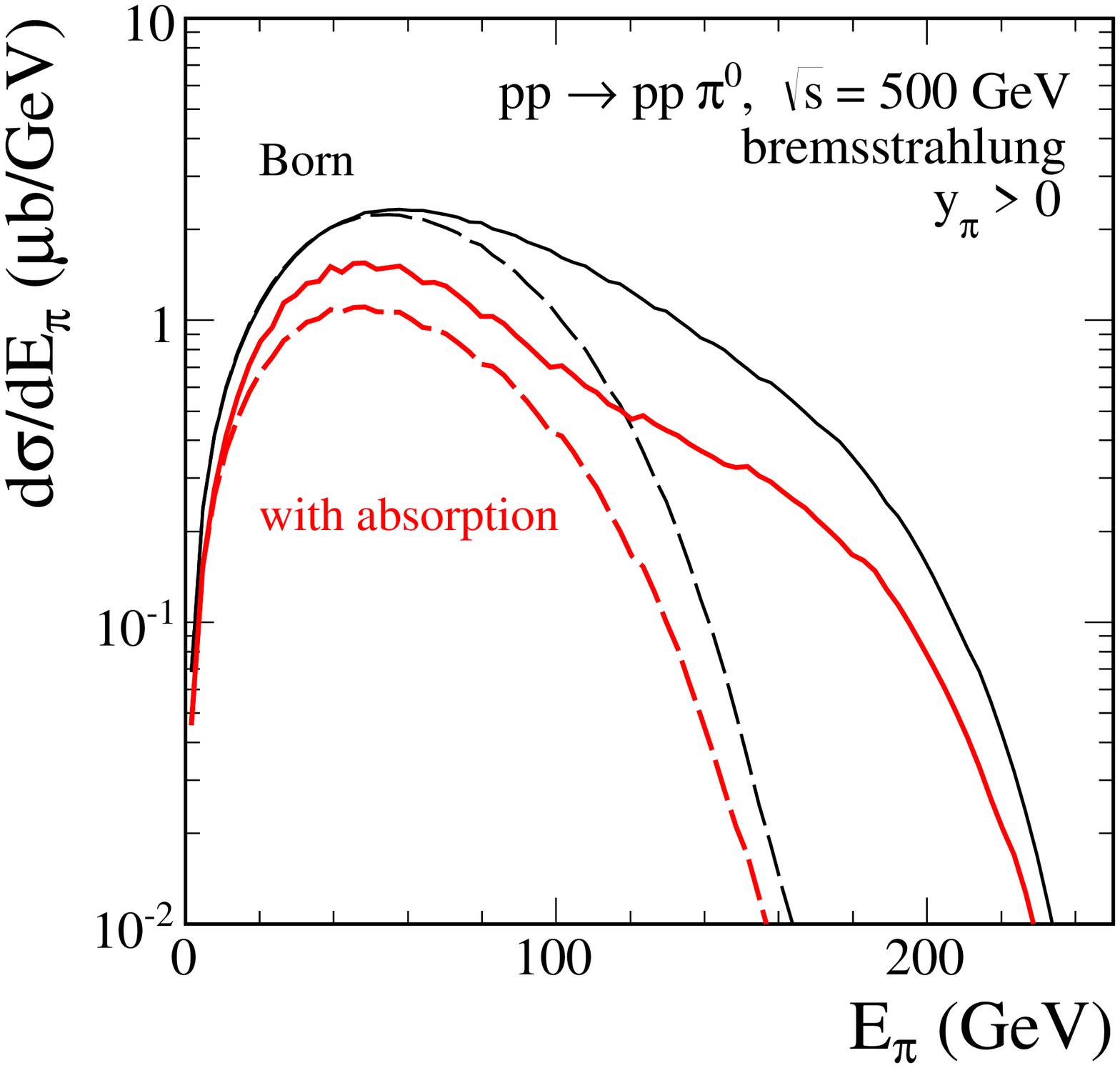}
(c)\includegraphics[width = 0.465\textwidth]{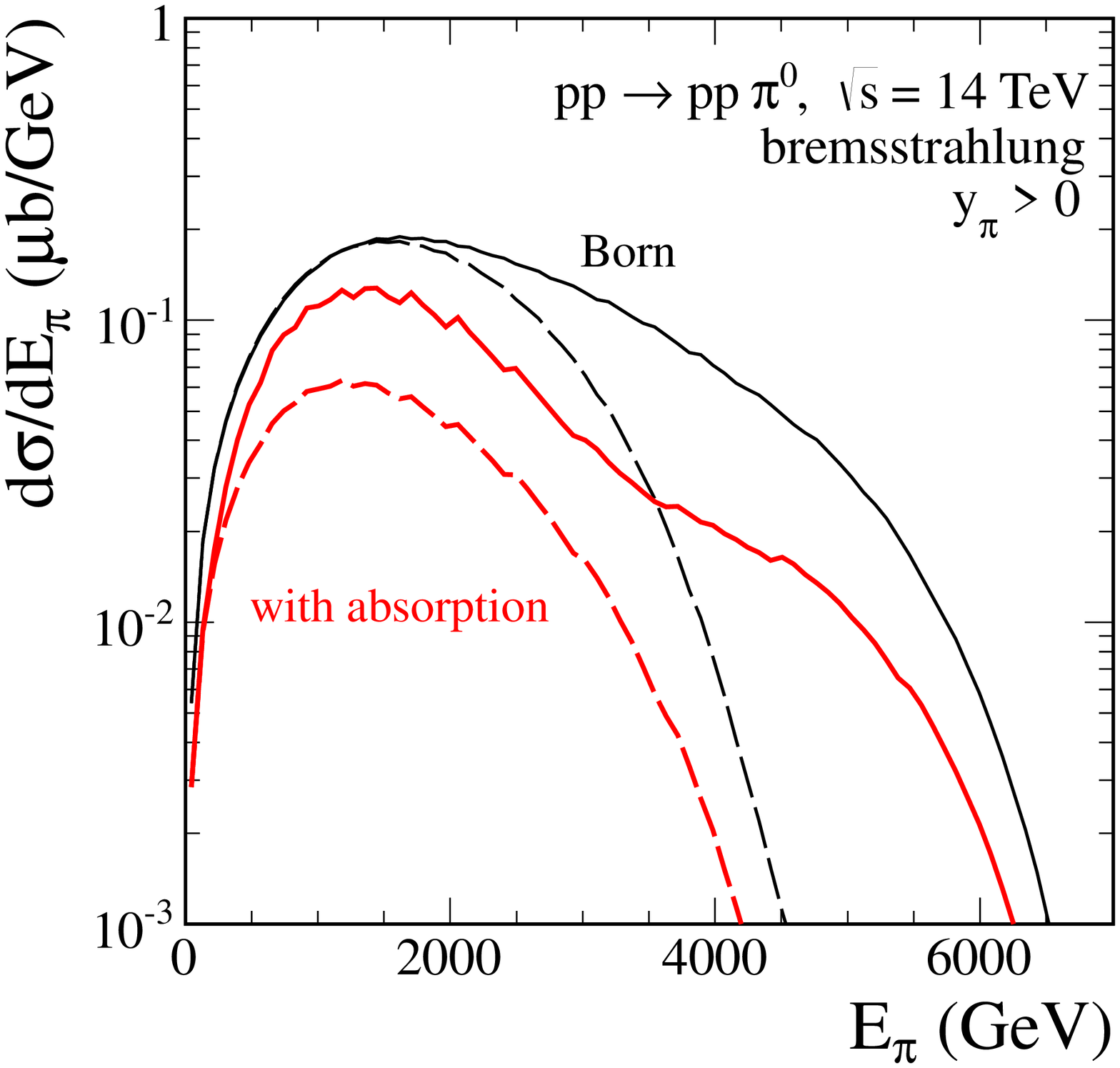}
  \caption{\label{fig:dsig_dE3}
  \small
Energy spectrum of pions at $\sqrt{s} = 0.5, 14$~TeV and for $y_{\pi^{0}} > 0$.
In panel (a) we show individual contributions to the Born cross section.
Theoretical uncertainties are shown in panels (b) and (c).
Here $\Lambda_{N} = \Lambda_{\pi} = 1$~GeV (solid line)
or $\Lambda_{N} = 0.6$~GeV and $\Lambda_{\pi} = 1$~GeV (dashed line).
}
\end{figure}

In Fig.\ref{fig:dsig_dw13} we compare distribution 
in invariant mass of the forward produced $p \pi^{0}$ system
for the $\pi^{0}$-bremsstrahlung contribution and $y_{\pi^{0}}>0$.
The discussed here the $pp \to pp \pi^{0}$ process gives a sizable
contribution to the low mass $(M_{X} > m_{p} + m_{\pi^{0}})$ single diffractive cross section.
\begin{figure}[!ht]
(a)\includegraphics[width = 0.465\textwidth]{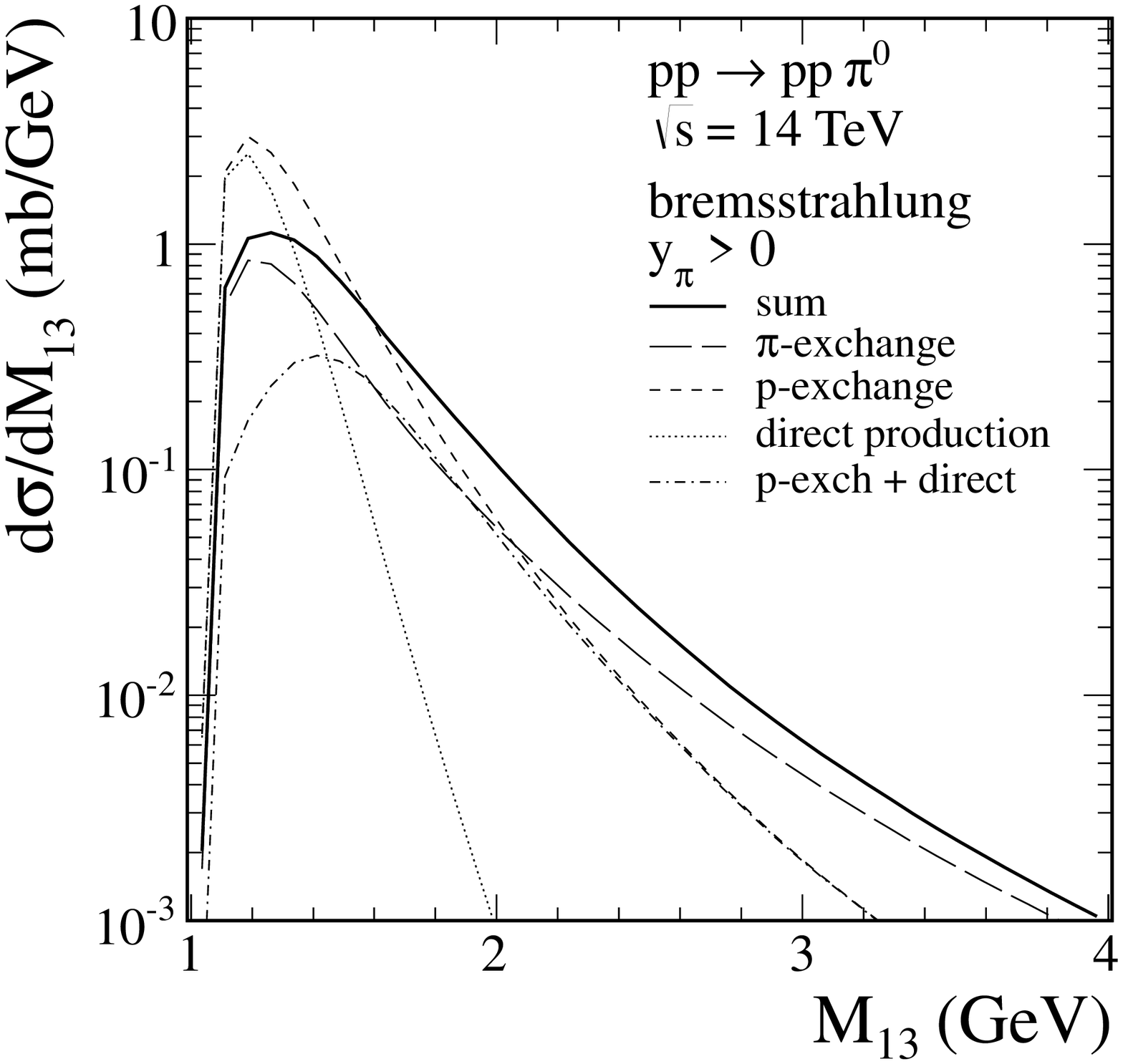}\\
(b)\includegraphics[width = 0.465\textwidth]{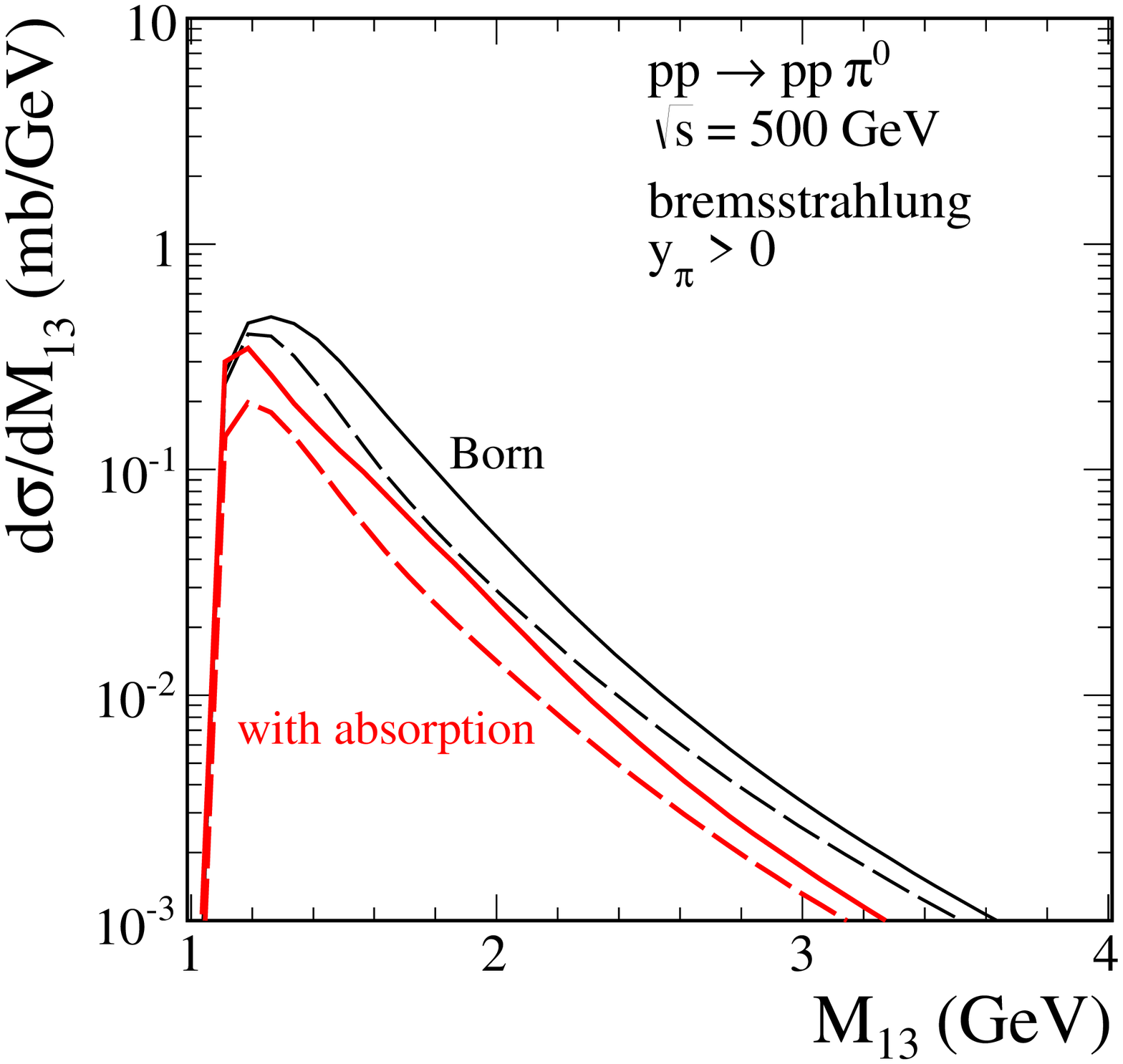}
(c)\includegraphics[width = 0.465\textwidth]{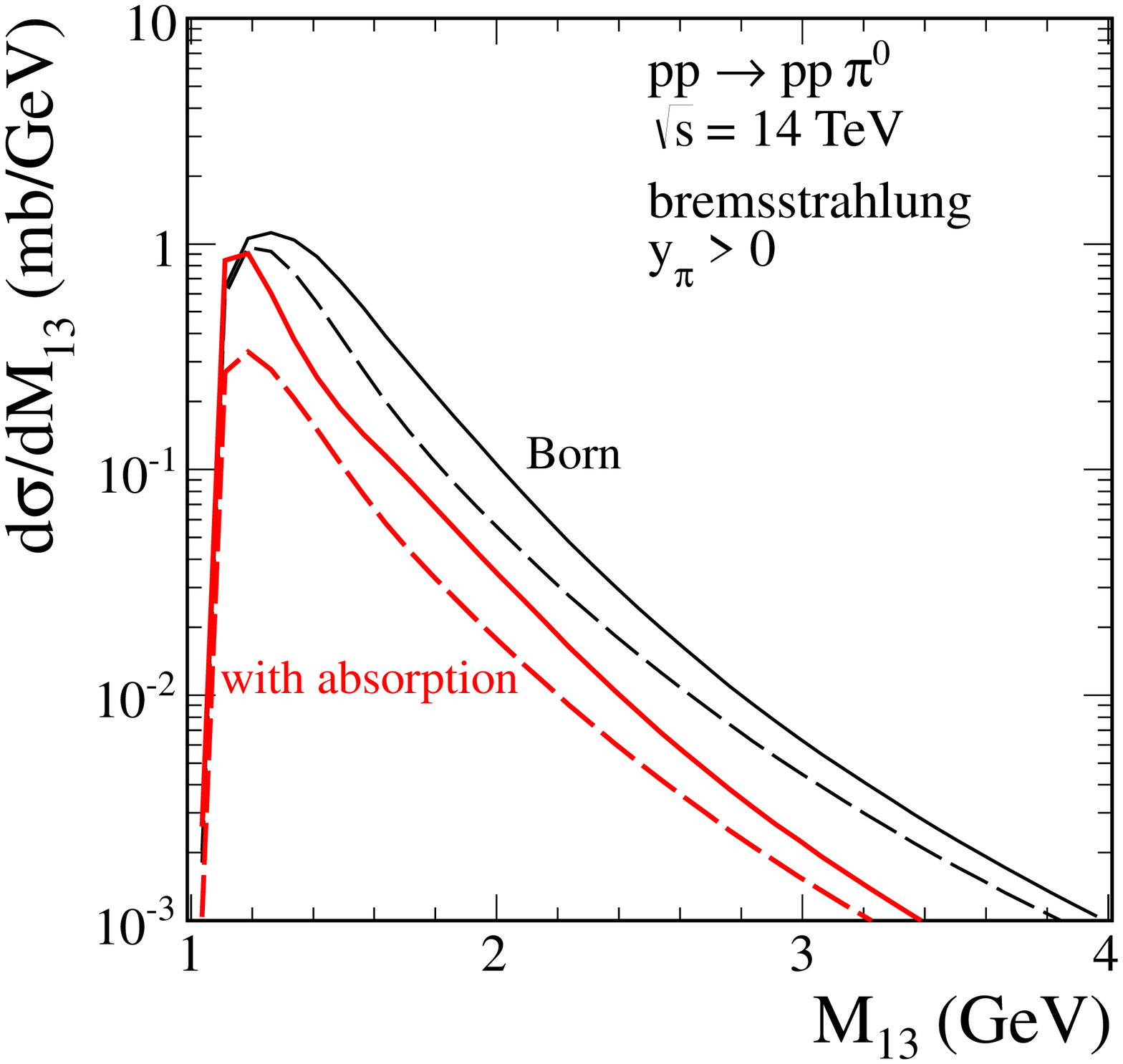}
  \caption{\label{fig:dsig_dw13}
  \small
Distribution in proton-pion invariant mass $M_{13}$
at $\sqrt{s} = 0.5, 14$~TeV and for $y_{\pi^{0}} > 0$.
In panel (a) we show individual contributions to the Born cross section.
Theoretical uncertainties are presented in panels (b) and (c).
Here $\Lambda_{N} = \Lambda_{\pi} = 1$~GeV (solid line)
or $\Lambda_{N} = 0.6$~GeV and $\Lambda_{\pi} = 1$~GeV (dashed line). 
}
\end{figure}

In Fig.\ref{fig:dsig_dphi12} we show correlation function 
in azimuthal angle between outgoing protons.
As can be seen in panel (a) the $\pi^{0}$-bremsstrahlung contribution
is peaked at back-to-back configuration ($\phi_{12} = \pi$).
For comparison, the contribution for other mechanism 
$\gamma \gamma$ fusion and $\gamma \omega$ fusion 
are peaked at $\phi_{12}=\pi/2$ and are much smaller.
We observe a strong cancellation between the initial and the final state radiation.
There [see panels (b) and (c)] is a sizable difference in shape between
the result obtained with the bare amplitude
and the result with inclusion of absorption effects.
We doubt if such a correlation can be measured at the LHC in the future.
\begin{figure}[!ht]    
(a)\includegraphics[width=0.465\textwidth]{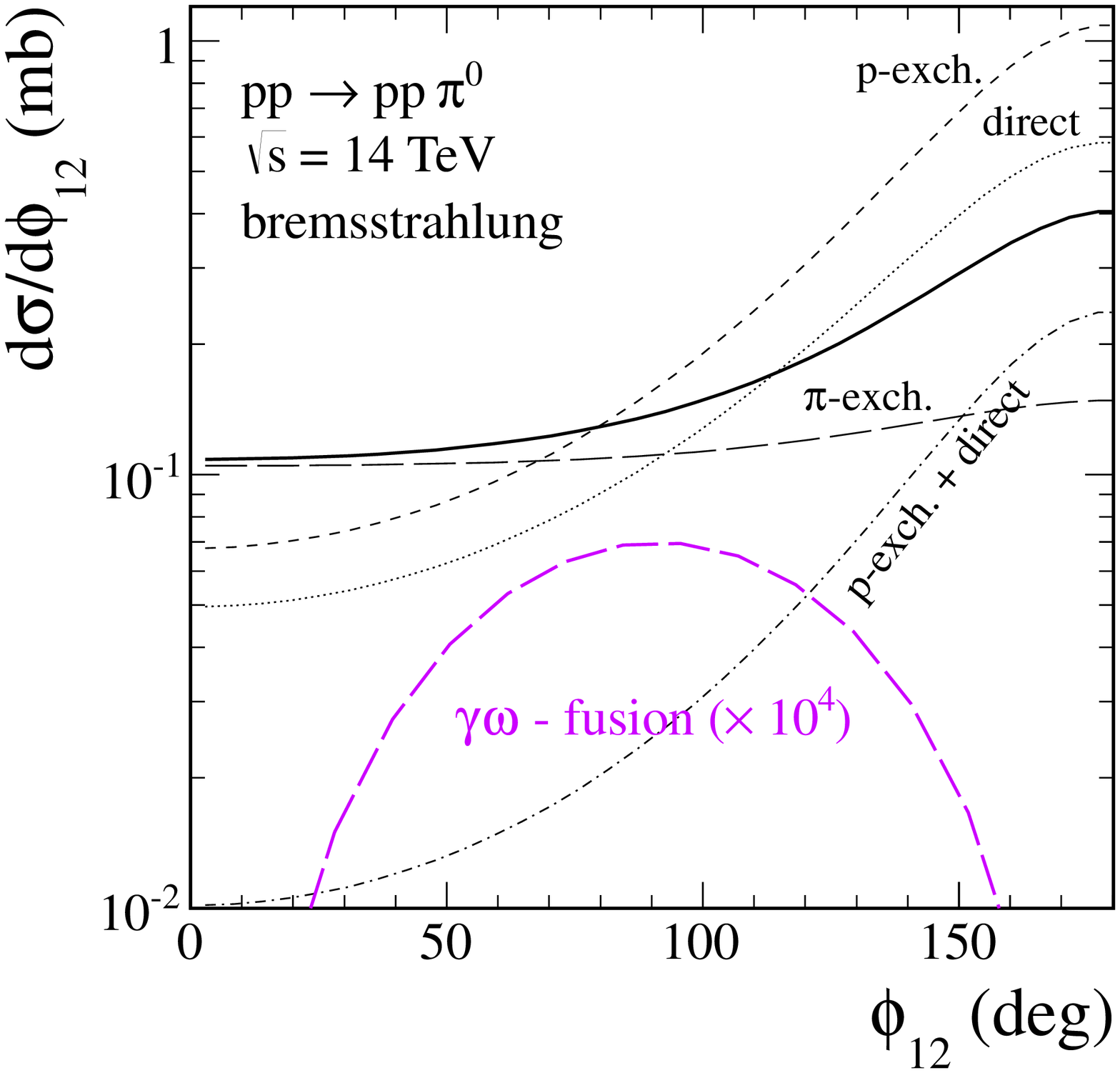}\\
(b)\includegraphics[width=0.465\textwidth]{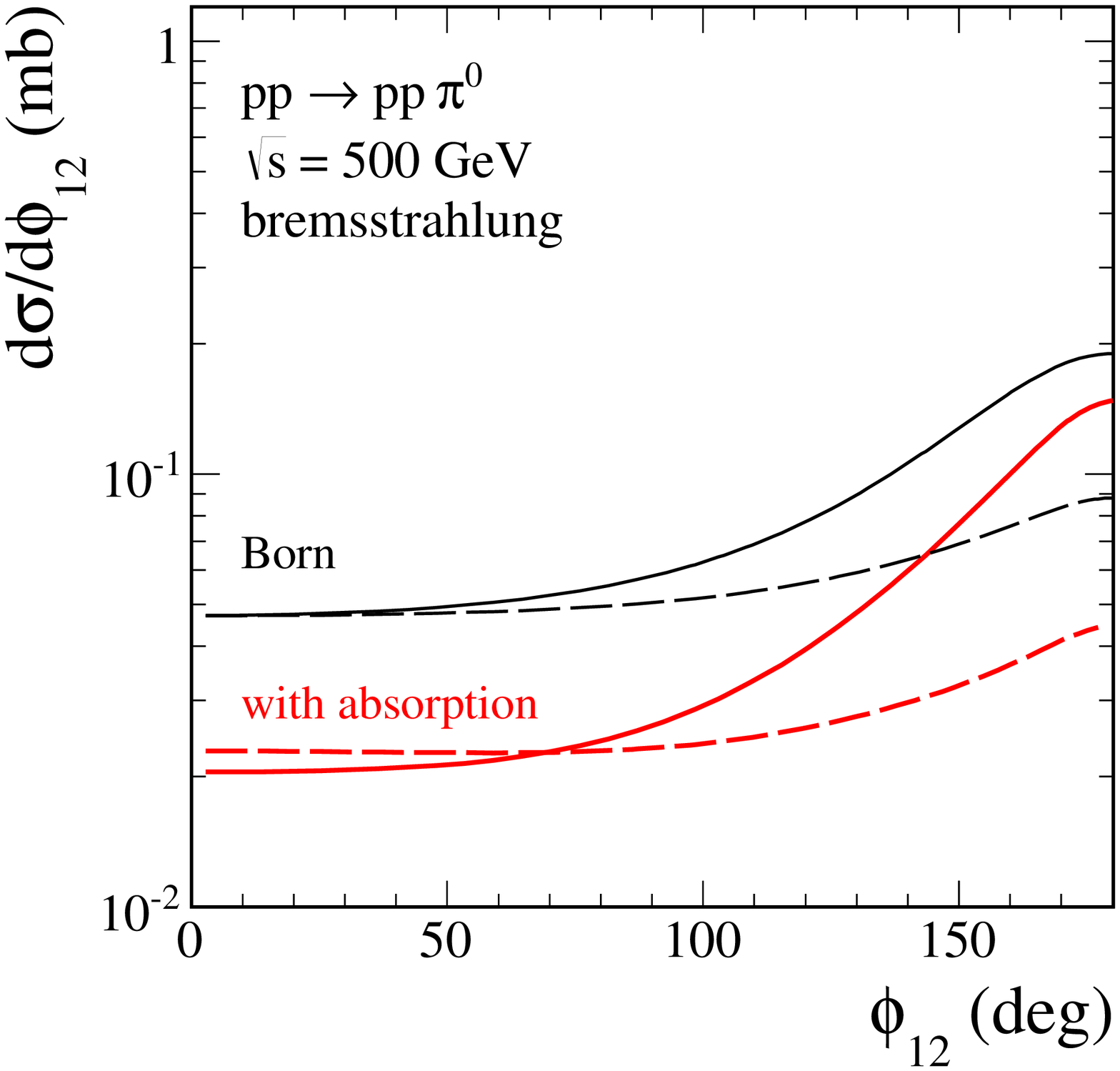}
(c)\includegraphics[width=0.465\textwidth]{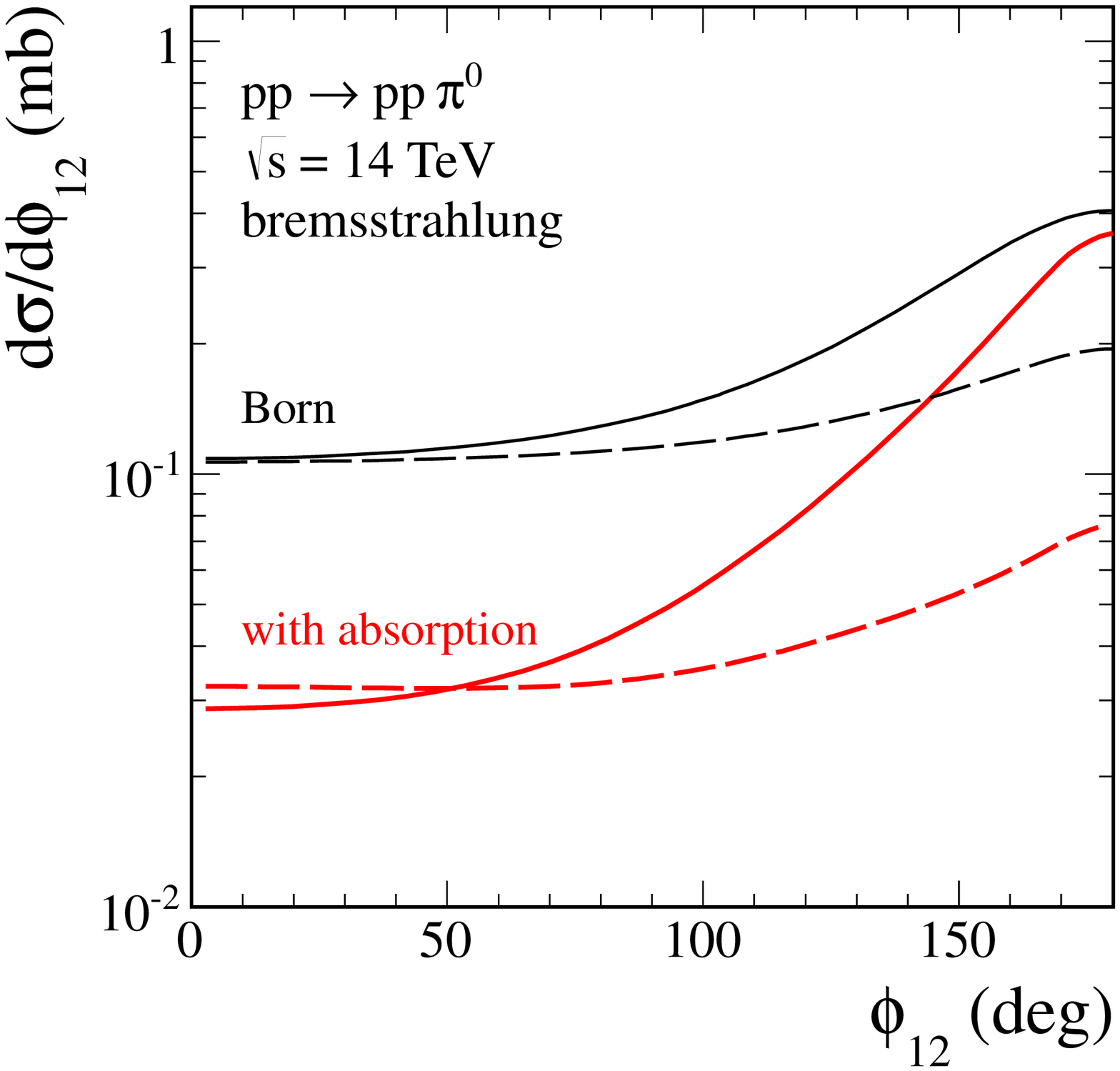}
\caption{\label{fig:dsig_dphi12}
\small  
The distribution in azimuthal angle between outgoing protons for $\sqrt{s}=0.5, 14$~TeV.
In panel (a) we show individual contributions to the Born cross section. 
Theoretical uncertainties are presented in panels (b) and (c).
Here $\Lambda_{N} = \Lambda_{\pi} = 1$~GeV (solid line)
or $\Lambda_{N} = 0.6$~GeV and $\Lambda_{\pi} = 1$~GeV (dashed line).
}
\end{figure}

In Fig.\ref{fig:map_t2M13} we show distribution in two-dimensional space ($t_{2}, M_{13}$).
One can observe different behavior of slope in four-momentum transfer squared $t_{2}$ 
for different masses of the $p \pi^{0}$ system.
A similar effect was observed for $pp \to p(n \pi^{+})$ \cite{Kerret1976}
and $np \to (p \pi^{-})p$ \cite{Biel} reactions at much lower energies.
As can be seen in Figs.\ref{fig:dsig_dt1_14000_absf} and \ref{fig:dsig_dw13}
the large contribution comes from the $\pi$-exchange diagram
and the baryon-exchange terms are suppressed due to amplitude cancellations.
The differential cross section peaks for invariant masses close to threshold
and disappears rapidly with increasing invariant mass, 
giving an approximately exponential behavior for large masses.
The absorptive effects could be partially responsible 
for the irregular structure in two-dimensional space ($t_{2}, M_{13}$)
at small $|t_{2}|$ and $M_{13} \sim 1.3$~GeV.
\begin{figure}[!ht]
(a)\includegraphics[width = 0.465\textwidth]{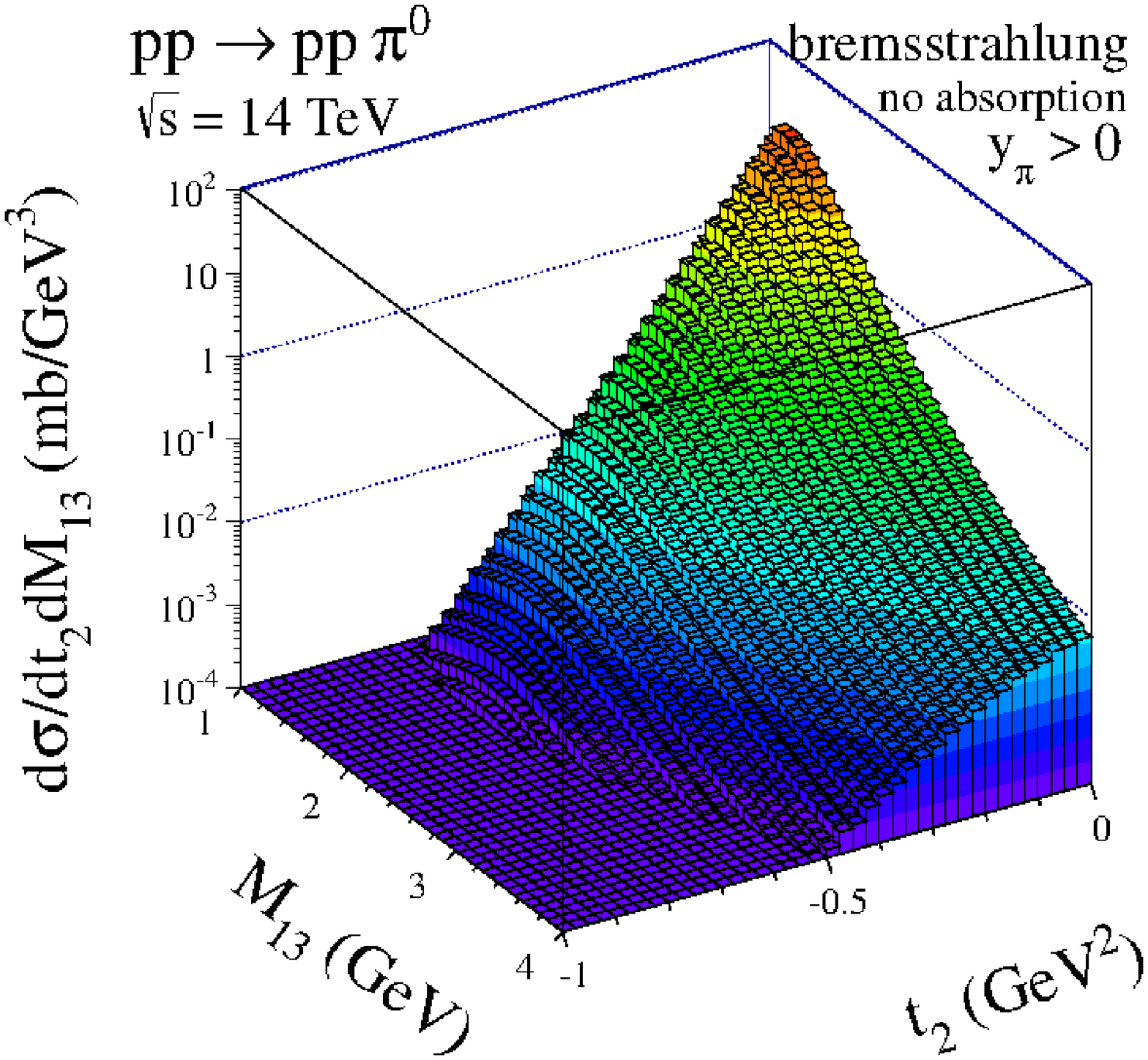}
(b)\includegraphics[width = 0.465\textwidth]{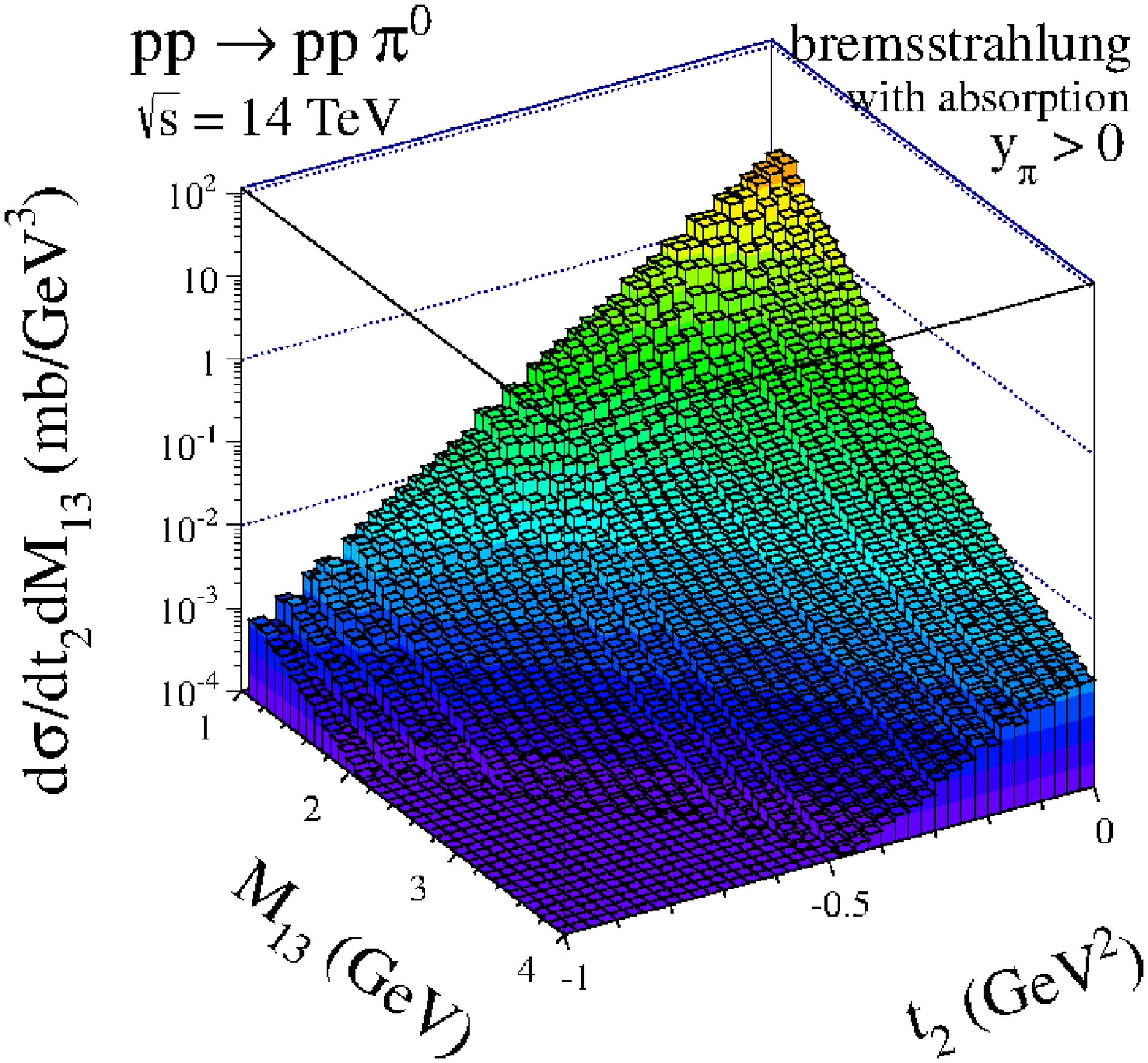}\\
(c)\includegraphics[width = 0.465\textwidth]{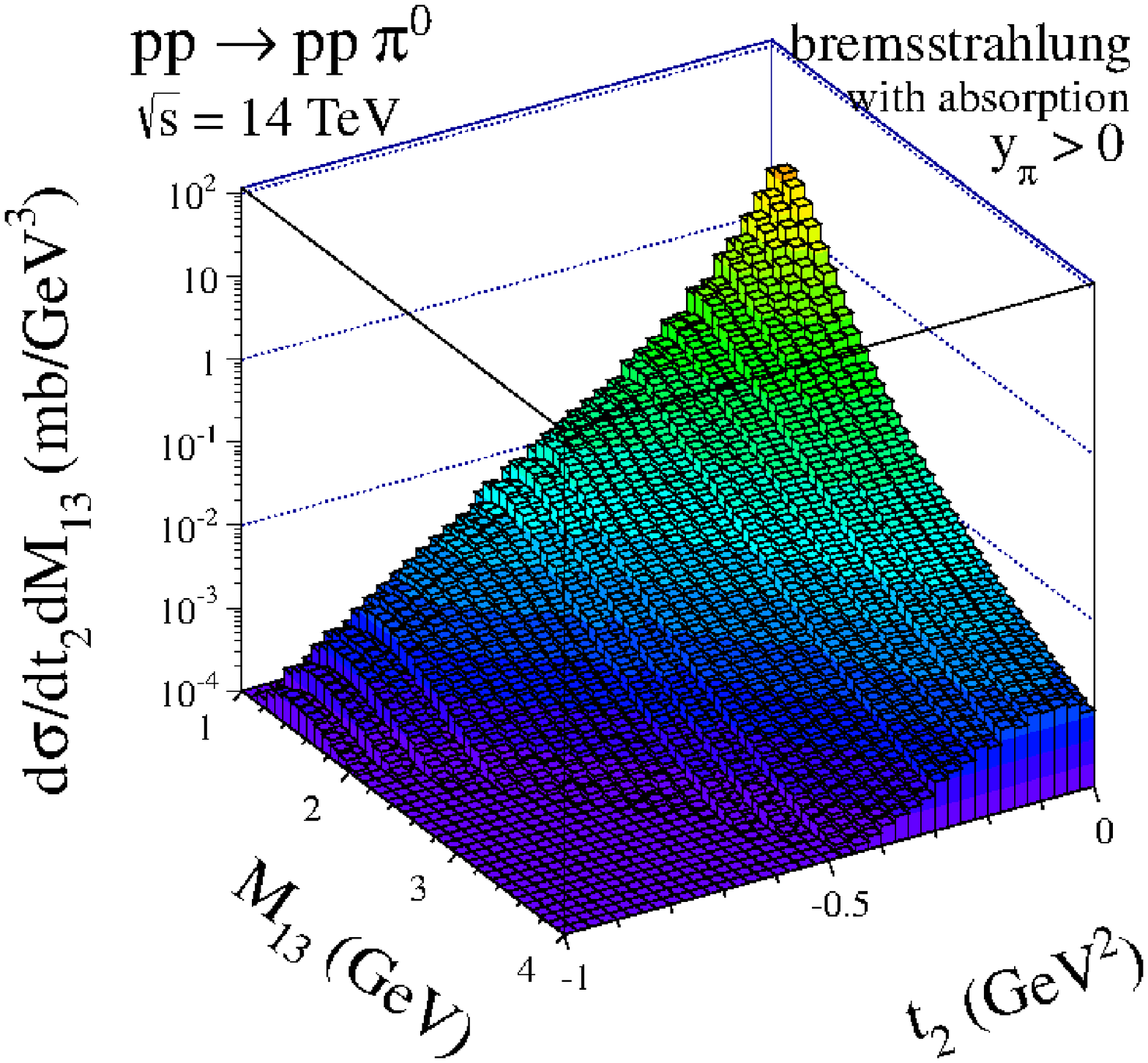}
(d)\includegraphics[width = 0.465\textwidth]{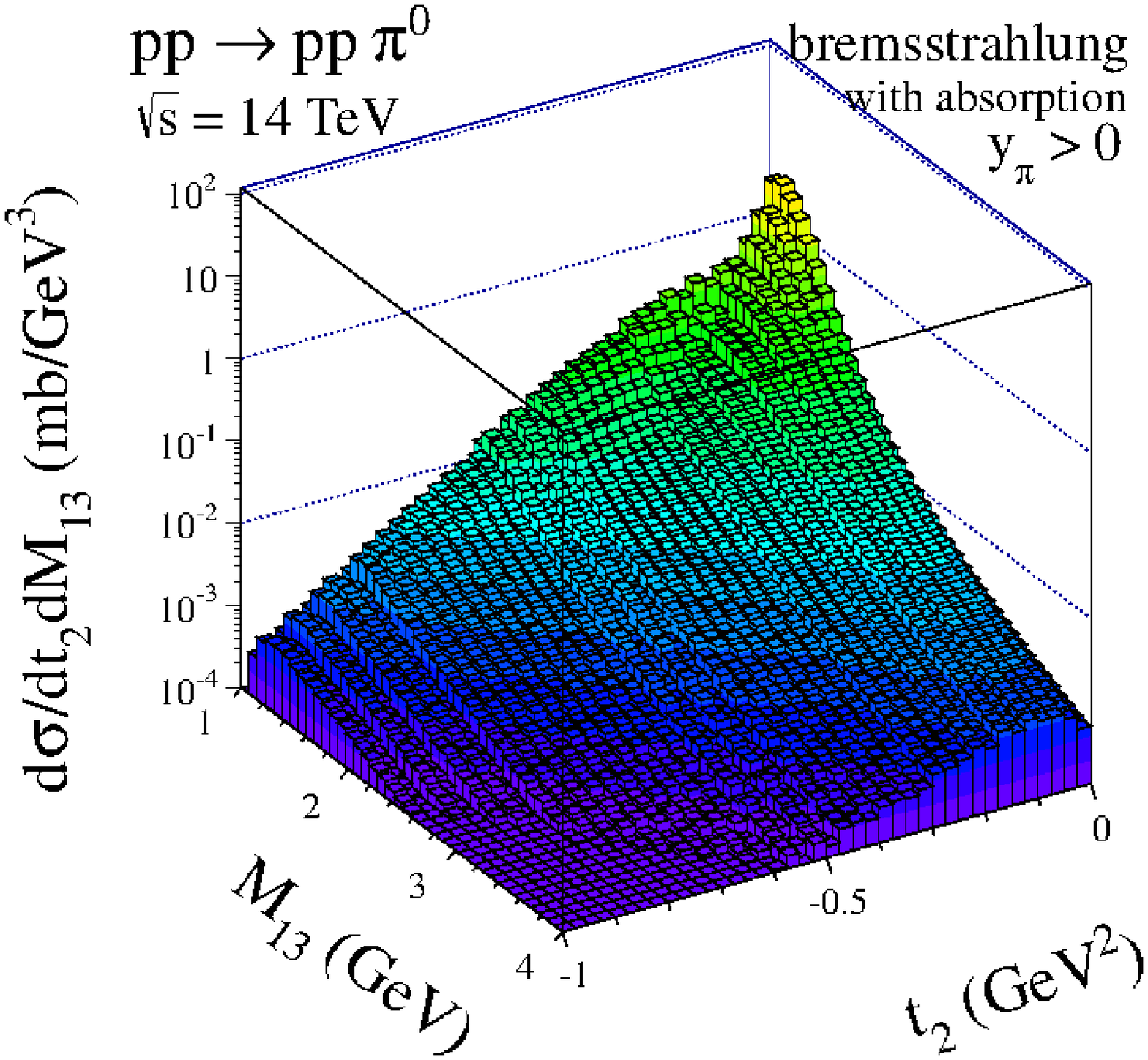}
  \caption{\label{fig:map_t2M13}
  \small
Distribution in ($t_{2}, M_{13}$) for the $\pi^{0}$-bremsstrahlung contribution
at $\sqrt{s} = 14$~TeV and for $y_{\pi^{0}}>0$
without [panel (a)] and with absorption effects in the final state only [panels (b) and (c)]
and absorption effects in the initial state only [panel (d)].
Here $\Lambda_{N} = \Lambda_{\pi} = 1$~GeV [panels (a) and (b)]
and $\Lambda_{N} = 0.6$~GeV, $\Lambda_{\pi} = 1$~GeV [panels (c) and (d)].
}
\end{figure}

In Fig.\ref{fig:map_y3pt3} we show corresponding two-dimensional
distributions in ($y_{\pi^{0}}, p_{t,\pi^{0}}$) for $y_{\pi^{0}}>0$.
Sizable correlations between pion rapidity and transverse momentum
can be observed which is partially due to interference of different components (amplitudes).
\begin{figure}[!ht]
\includegraphics[width = 0.465\textwidth]{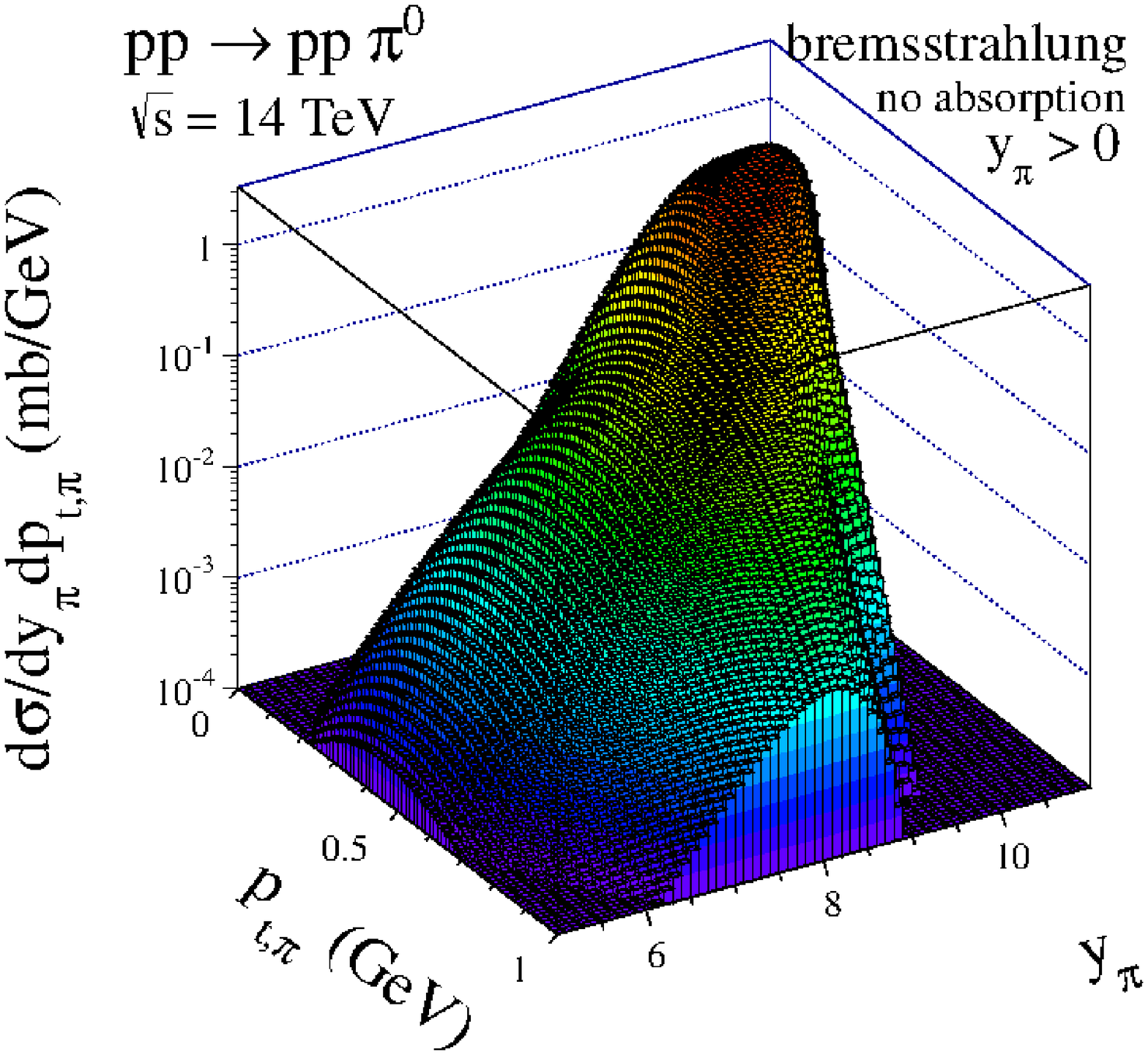}
\includegraphics[width = 0.465\textwidth]{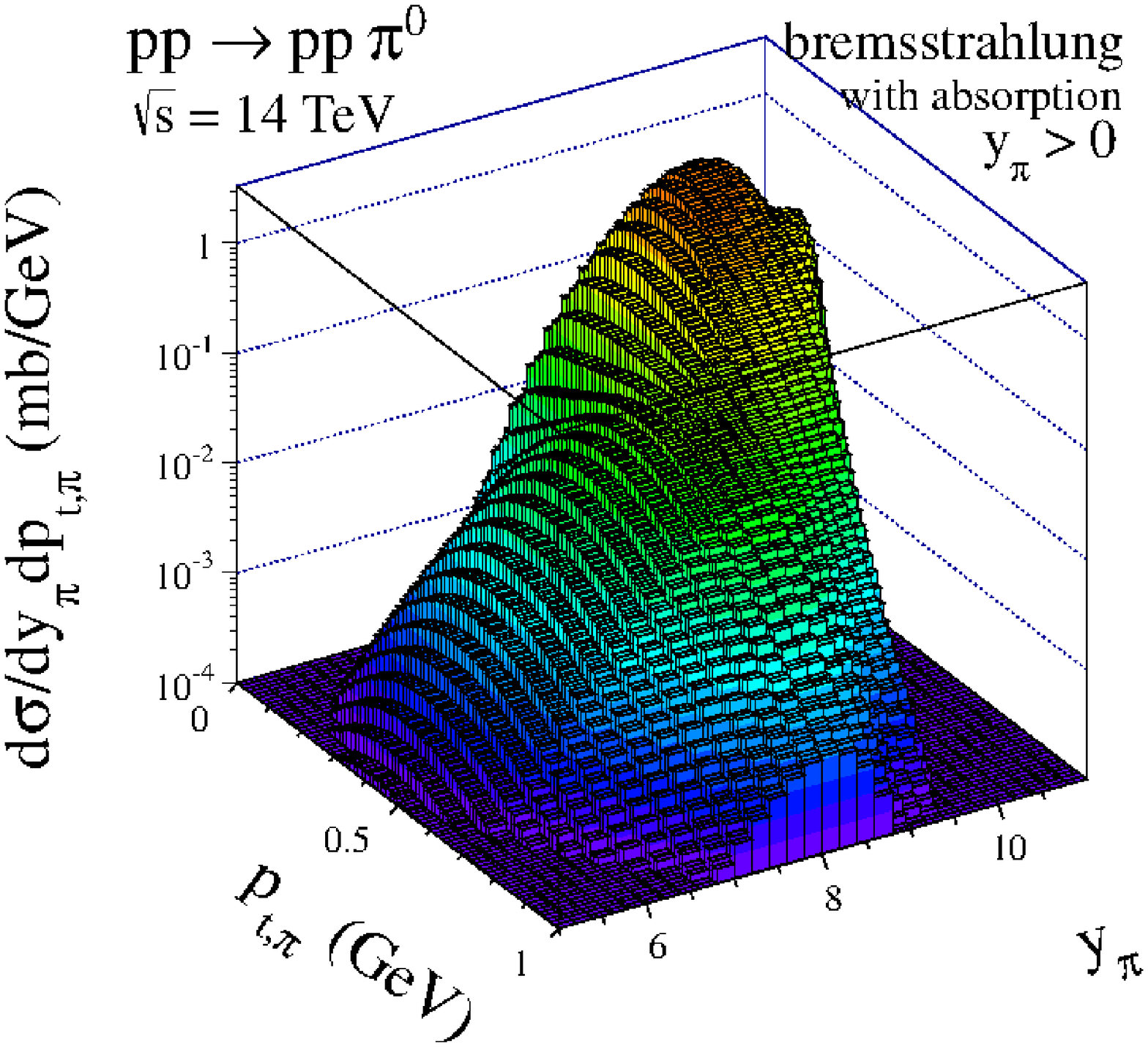}
  \caption{\label{fig:map_y3pt3}
  \small
Distribution in ($y_{\pi^{0}}, p_{t,\pi^{0}}$) at $\sqrt{s} = 14$~TeV
for the $\pi^{0}$-bremsstrahlung contribution and $y_{\pi^{0}}>0$
without (left panel) and with (right panel) absorption effects.
Here $\Lambda_{N} = \Lambda_{\pi} = 1$~GeV.
}
\end{figure}

In Table~\ref{tab:sig_tot_pppi0} we have collected 
numerical values of the integrated cross section $\sigma_{p p \to p p \pi^0}^{DHD}$
after taking the forward region ($y_{\pi^{0}} > 0$) into account 
for exclusive production of $\pi^{0}$ at different c.m.~energies $\sqrt{s}$.
Our results depend on the $\Lambda$ parameters of the hadronic form factors.
The cross section obtained from ISR experiments (see e.g., Ref.\cite{Kerret1976}) are roughly reproduced.
\begin{table}
\caption{The integrated value of cross sections in $\mu b$
for the $pp \to pp \pi^{0}$ reaction
at $\sqrt{s}=45$~GeV (ISR), 500~GeV (RHIC), and 14~TeV (LHC)
and $y_{\pi^{0}} >0$ is taken into account only.
The lower limit corresponds to the result
when $\lambda_{N} = 0.6$~GeV and $\lambda_{\pi} = 1$~GeV are imposed in calculations
and the upper limit when $\lambda_{N} = \lambda_{\pi} = 1$~GeV.
}
\label{tab:sig_tot_pppi0}
\begin{center}
\begin{tabular}{|l|c|c|c|}
\hline
Model & $\sqrt{s} = 45$~GeV & $\sqrt{s} = 500$~GeV & $\sqrt{s} = 14$~TeV \\
\hline
No absorption         & $103-146$ & $177-251$ & $402-575$ \\
Absorption in initial state & $\;\,46- \;\,76$ &  $\;\,62-125$ &  $\;\,94-357$ \\
Absorption in final state   & $\;\,60- \;\,91$ &  $\;\,84-139$ & $128-290$ \\
\hline
\end{tabular}
\end{center}
\end{table}

\begin{figure}[!ht]    
\includegraphics[width=0.465\textwidth]{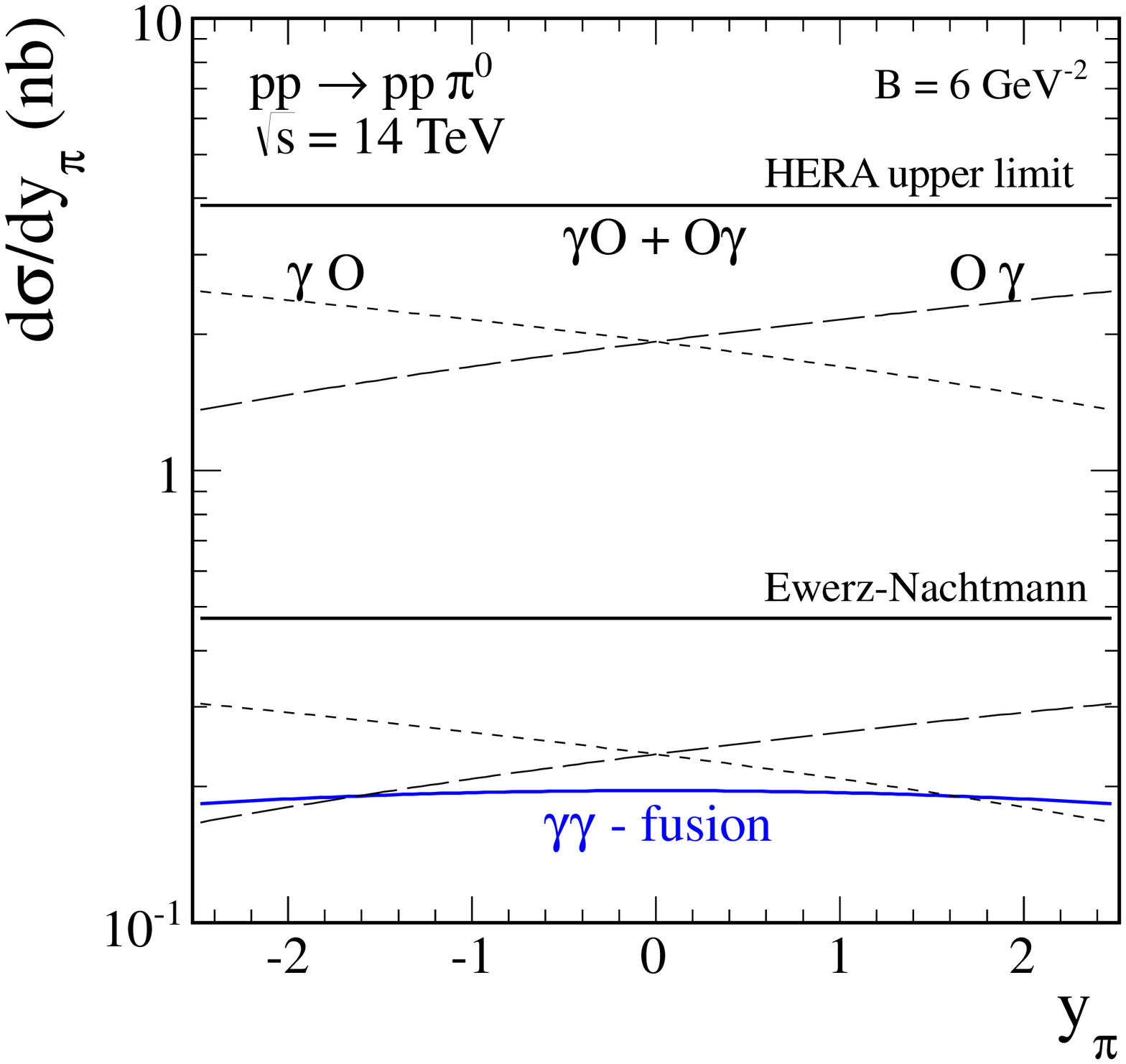}
\includegraphics[width=0.465\textwidth]{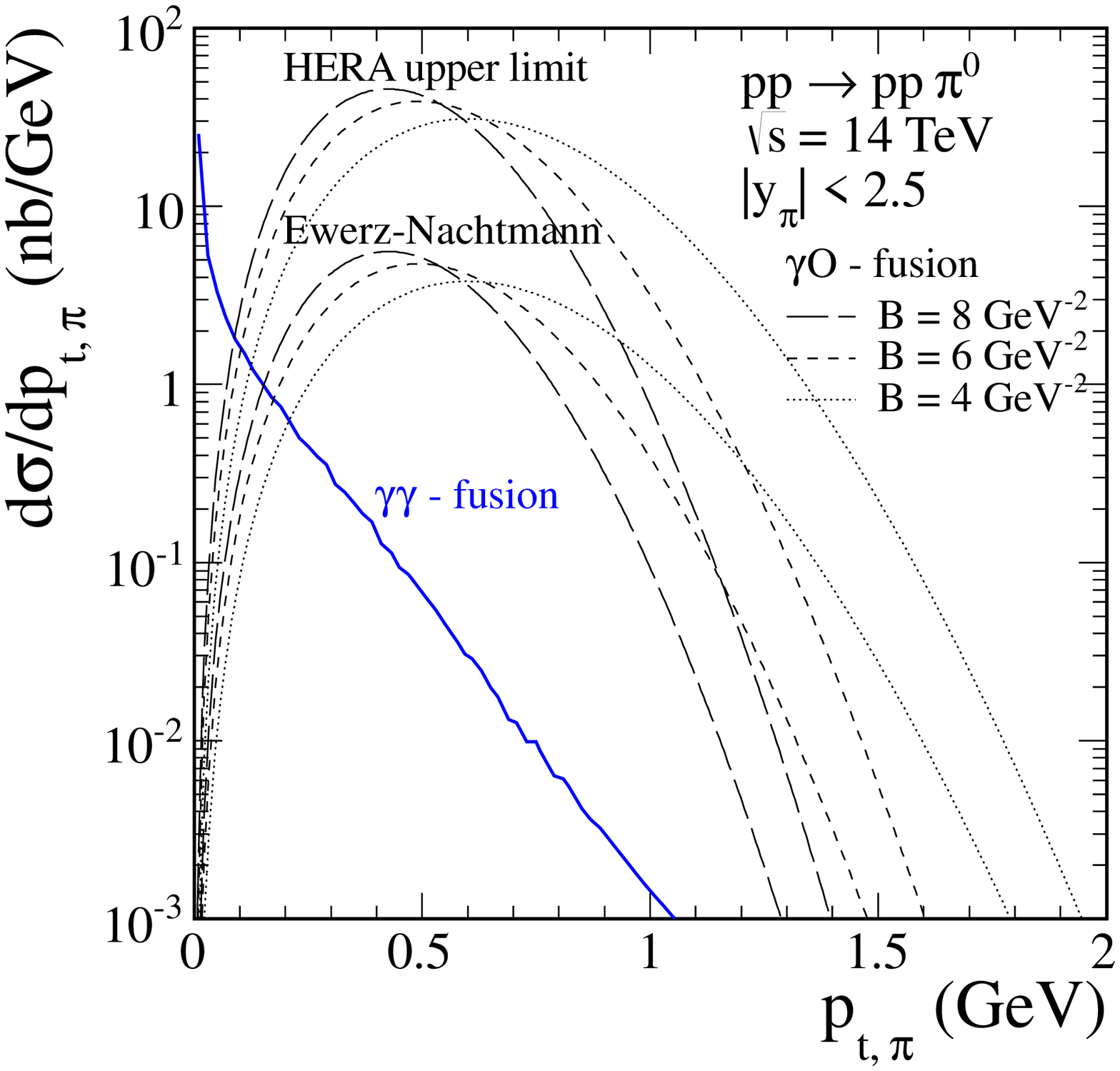}
\caption{\label{fig:fig_odderon}
\small
Rapidity distribution of neutral pions (left panel) 
produced in $\gamma \mathcal{O}$-fusion and $\mathcal{O} \gamma$-fusion 
(black upper solid lines) compared to the $\gamma\gamma$ contribution (blue lowest solid line) 
for $\sqrt{s}=14$~TeV.
Individual contributions of photon-odderon (short dashed line)
and odderon-photon (long dashed line) are shown separately.
We show predictions for the HERA upper limit ($\sigma_{\gamma p \to \pi^{0} p} = 49$~nb)
and for the Ewerz-Nachtmann estimate ($\sigma_{\gamma p \to \pi^{0} p} = 6$~nb).
In the right panel we make similar comparison of contributions of the two mechanisms 
for transverse momentum distribution of $\pi^{0}$'s
in the rapidity region $-2.5 < y_{\pi^{0}} < 2.5$.
For the odderon contributions we have used different values of slope parameters $B = 4, 6, 8$~GeV$^{-2}$.
}
\end{figure}

For completeness, in Fig.\ref{fig:fig_odderon}~(left panel) we compare the photon-odderon
and odderon-photon contributions with the $\gamma \gamma$ contribution.
We show results for $B = 6$~GeV$^{-2}$ and two different estimates
of the $\gamma p \to \pi^0 p$ cross section (energy independent) as 
specified in the figure caption. 
The total cross section for the odderon
contributions, corresponding to the HERA upper limit, 
is less than 20~nb in the rapidity region $|y_{\pi^{0}}| < 2.5$.
The corresponding curve is  more than an order
of magnitude larger than the photon-photon contribution.

Finally, in Fig.\ref{fig:fig_odderon}~(right panel) we make similar comparison of
contributions of the two mechanisms for transverse momentum distribution
of neutral pions for different slope parameters and $|y_{\pi^{0}}| < 2.5$. 
The curve corresponding to the HERA upper limit is considerably
larger than the photon-photon contribution starting from $p_{\perp,\pi^{0}} > 0.2$~GeV.
Even with the Ewerz and Nachtmann limit, one can observe deviations 
from the $\gamma \gamma$ curve at transverse momenta $p_{\perp,\pi^{0}} > 0.3$~GeV.
The cut on meson $p_{\perp,\pi^{0}}$ should enhance relative odderon contribution.
In principle, the ALICE collaboration could try to measure the
transverse momentum distribution of exclusively produced neutral pions.

\section{A comment on single diffractive cross section at low proton excitations}

The measurement of an inelastic proton-proton cross section is one of
the standard and obligatory measurements at each collision energy.
At the LHC single diffraction (SD) and double diffraction (DD) processes 
constitute a large contribution to the inelastic cross section (about a half).
Unfortunately it is very difficult to truly measure the cross section
for the low mass excitation at the LHC and often educated extrapolations are
required. Usually $1/M^{2}$ triple-Regge fit is used for this purpose.
Do we have the expertise on the very low mass excitations?
This issue was critically discussed recently \cite{JKOS2012}.
The authors presented predictions of a dual-Regge model 
with a nonlinear proton Regge trajectory \cite{JKLMO11}
with parameters fitted to the single diffractive cross section
measured at low energies (for a review of the low energy SD data see e.g.~\cite{Alberi_Goggi,Goulianos1983}).
In their fit the low mass excitation is dominated by the excitation
of the proton resonances
$N^{*}(1440)$ with $J^{P}=\frac{1}{2}^{+}$ and 
$N^{*}(1680)$ with $J^{P}=\frac{5}{2}^{+}$.
While the presence of the latter is rather natural
-- it is a member of the same Regge trajectory as proton --
the huge contribution of the Roper resonance is not too clear to us.
The low-energy experimental SD data \cite{Goulianos1983} 
show up a huge peak at the nominal position of the Roper resonance. 
This is the region where the absorbed Drell-Hiida-Deck mechanism
(the nonresonant background model) predicts an enhancement (see Fig.\ref{fig:dsig_dw13}).
The arguments against large Roper contribution in single diffraction at high energies 
were exposed in Ref.\cite{Tsarev2}.
We wish to emphasize that the DHD contribution was not included in 
the analysis of the SD mass spectrum in \cite{JKOS2012} 
where only a purely mathematical fit was used.
The fitted background seems to have quite different properties than 
the discussed here DHD mechanism with absorption (different both in $M_{X}$ and in $t$). 
In our opinion inclusion of a realistic absorbed DHD contribution could
dramatically change, or even eliminate, the contribution of the Roper resonance. 
This issue requires further studies.

The resonances contributing to the SD cross section
discussed in \cite{JKOS2012} naturally contribute
also to the $p p \to p p \pi^0$ channel and the corresponding cross section is
\begin{equation}
\sigma_{p p \to p p \pi^0}^{N^*} =
\sigma_{SD}^{N^*} \times BR(N^* \to N \pi) \times \frac{1}{3} \; .
\label{resonance_contribution}
\end{equation}
The last factor comes from the fact that the considered diffractively
excited baryon resonances have isospin $I = \frac{1}{2}$.
The branching fractions $BR(N^{*} \to N \pi)$ have been measured
\cite{PDG} and are about 65\% for both discussed states.
The same situation occurs in the $p p \to p (n \pi^+)$ and
$n p \to (p \pi^-) p$ reactions (a factor 2 larger cross section), 
where no clear signal of the Roper $N^{*}(1440)$ resonance was identified 
(see e.g., \cite{Kerret1976,Mantovani1976,Babaev1976}) 
while the $N^{*}(1680)$ resonance was observed.
\footnote{In Ref.\cite{Kerret1976} results on diffractive dissociation
of protons into ($n \pi^{+}$) in $pp$ collisions at the CERN ISR $\sqrt{s} = 45$~GeV energy
and the cross sections $\sigma_{pp \to p (n \pi^+)} = (400 \pm 110)$~$\mu$b,
$\sigma_{pp \to p N^{*}(1680)} = (170 \pm 60)$~$\mu$b have been reported.}
The situation should be better clarified in the future.
The discussed there resonances were not included in our analysis 
but could be included in principle.

Our DHD mechanism contributes to 
the single diffraction cross section as
\begin{equation}
\sigma_{SD}^{DHD} = 3 \, \sigma_{p p \to p p \pi^0}^{DHD} \,.
\label{contribution_of_DDH_to_SD}
\end{equation}
The factor 3 comes from the isospin symmetry of the $NN\pi$ coupling constant. 
Taking our numbers from Table~\ref{tab:sig_tot_pppi0} we predict
certainly not a negligible contribution
to the total inelastic cross section at high energies
(for both-side SD the $\sigma_{p p \to p p \pi^0}^{DHD}$ should be multiplied by a factor 2).
To our knowledge the DHD contribution is not included in the existing
Monte Carlo codes simulating high-energy diffractive processes.

\section{Conclusions and discussion}

In the present analysis we have calculated differential cross sections 
for the exclusive $p p \to p p \pi^0$ reaction at high energies
relevant for RHIC and LHC.
We have included the $\pi^{0}$-bremsstrahlung 
from the initial and final state, diffractive $\pi^0$-rescattering,
photon-photon fusion and photon-omega (omega-photon) fusion processes.
The diffractive $\pi N$ and $N N$ rescattering amplitudes have been 
related to the total $\pi N$ and $N N$ cross sections. 
The Donnachie-Landshoff parametrization has been used for energy
dependence of the latter.
Absorptive effects have been included in addition.
They lower the cross section by a factor 2 to 3; see Table~\ref{tab:sig_tot_pppi0}.

We have found very large cross sections of the order of mb. 
The total (integrated over phase space) cross section is almost 
energy independent. 
The dominant contributions are placed at large rapidities. 
The larger c.m. energy, the larger rapidities are populated.
On the other hand, the diffractive contribution is absent at midrapidity ($y_{\pi}=0$).
The higher the collision energy the larger the unpopulated region. 
This opens a window for other mechanisms with much smaller cross section. 
For example at the LHC the two-photon fusion mechanism ``wins" with 
the diffractive mechanisms at $y_{\pi} \approx 0$,
where the diffractive contributions are very small.
However, the transverse momenta of neutral pions in this region
are very small and therefore such pions are very difficult to measure.
The $\gamma \omega$ or $\omega \gamma$ exchanges have been found to be 
significant only in backward or forward rapidities, respectively, 
and are small at midrapidities due to $\omega$ reggezation. 
In principle, also $a_{2}$-pomeron and pomeron-$a_{2}$
exchanges or $\rho^{0}$-odderon and odderon-$\rho^{0}$ exchanges could
play some role but not at midrapidities.
In addition, it is rather difficult to make for them realistic predictions.
A larger cross section than predicted here at midrapidities 
would be an interesting surprise.

We have shown several other differential distributions.
If one limits to separate regions of $y_{\pi^0} < 0$ or $y_{\pi^0} > 0$ 
(one-side excitation), then the distributions in proton transverse
momenta $p_{1t}$ and $p_{2t}$ are quite different -- one reflecting 
the pion/nucleon exchange and the second reflecting the pomeron exchange. 
The same is true for the $t_1$ and $t_2$ (transferred four-momentum
squares) distributions. 
Analysis of such details would be a useful test of the model. 
The distribution in the mass of the excited $\pi^0 p$ 
system peaks at small $M_{\pi p}$ and quickly drops when the mass increases.
Such a distribution reminds the spectral shape of the Roper resonance 
fitted recently to an old single-diffractive data.
We have obtained an interesting correlation between
the mass of the excited system and the slope of the $t$ distributions
well represented in a two-dimensional plot $\frac{d \sigma}{dt dM}(t,M)$.
Similar effects were observed in the past for the
$p p \to p (n \pi^+)$ and $n p \to (p \pi^-) p$ reactions
at the CERN ISR and Fermi National Accelerator Laboratory (Fermilab).

Particularly interesting is the distribution in azimuthal angle
between outgoing protons, not studied so far, including low-energy
$p p \to p (n \pi^+)$ and $n p \to (p \pi^-) p$ reactions measured
in the 1970's at ISR and Fermilab.
The distribution has a maximum at relative angle $\phi_{12} = \pi$.
The detailed shape of the distribution is, however, very sensitive 
to the relative contribution of different ingredients of the model. 
The sensitive nature of the cancellation between 
proton-exchange and direct production amplitudes 
leads to a situation where minor changes
in the parametrizations of these amplitudes can
have large effects on discussed distributions.
Experimental analysis of such a distribution would therefore help 
in fixing model parameters such as cutoff parameters of hadronic form factors, 
not known very precisely.

At the LHC the $\pi^0$ mesons could be measured with the help
of zero degree calorimeters. Such measurements are possible only
at rather large pseudorapidities $|\eta_{\pi^0}| > 8-9$. 
The pions at midrapidities 
have rather small transverse momenta so their registration is probably
very difficult. On the other hand, protons could be measured
with the ALFA detector of ATLAS or the TOTEM detector 
associated with the CMS main detector.
The latter may be more difficult from an organization point of view. 

The reaction discussed here is interesting also in a much broader context.
First of all, it may constitute a sizable fraction of 
the pion inclusive cross section at very forward/backward (pseudo)rapidities. 
A comparison with nondiffractive Monte Carlo code would be therefore very valuable. 
Second, it leads to a production of very energetic photons 
($\sim 0.5-2$~TeV) from the decay of the forward $\pi^0$'s.
These two issues will be a subject of future investigations.
Third, the DHD mechanism sizably contributes to the single diffractive
cross section and as a consequence to the total inelastic cross section.
This contribution is not included in any existing Monte Carlo code.
Needless to say, these codes are used when extrapolating the
measured high-mass SD cross section down to the $\pi N$ threshold, 
which obviously leads to an underestimation of the extracted 
(measured and extrapolated) cross section for single diffraction
and/or inelastic processes.
Finally, because the cross section for the discussed 
reaction is large, detailed studies could help to 
test model(s) of soft absorption, so important in 
the context of more fundamental searches such as, e.g., exclusive
production of the Higgs boson in diffractive processes.

In the present paper we have calculated only
contributions with intermediate protons
in the ground state to the $p p \to p p \pi^0$ reaction. 
There are also resonance contributions, due to 
diffractive excitation of some nucleon resonances
and their subsequent decays into the $p + \pi^0$ ($\bar p + \pi^0$) channels.
The dominant contributions are due to $N^{*}$ resonant states being
members of the nucleon trajectory. 
The $N^{*}(1680)$~5/2$^+$ state is the best candidate. 
Although a huge contribution of the Roper resonance $N^{*}(1440)$
was suggested recently \cite{JKOS2012}, as discussed in our paper,
their contribution may be to some extent an artifact of a fit 
which does not include the DHD mechanism discussed in our paper,
neither in the $p p \to p p \pi^0$ nor in the $p p \to p n \pi^+$ channel.

In the present analysis we have considered single exclusive $\pi^0$ production.
One could think about immediate extension of the present study to
double diffractive, double DHD mechanism producing two exclusive $\pi^0$'s, 
not considered so far in the literature and not included in any Monte Carlo code.
Again we expect a rather large cross section for such an inelastic process.
This would be a competitive mechanism to the central $\pi^0 \pi^0$
production via pomeron-pomeron fusion considered by us for $\pi^+ \pi^-$ production; 
see \cite{LS10,SLTCS11}.
 

We have presented first estimates of the photon-odderon and
odderon-photon contributions based on the upper limit of
the $\gamma p \to \pi^0 p$ cross section obtained at the HERA
as well as estimates based on a nonperturbative approach of Ewerz 
and Nachtmann which makes use of chiral symmetry and PCAC. 
Based on the HERA upper limit we conclude that the cross section for 
the contribution to the $p p \to p p \pi^0$ reaction
is smaller than 20~nb in the rapidity region $|y_{\pi^{0}}| < 2.5$.

Any deviation from the $\gamma \gamma \to \pi^0$ contribution
to transverse momentum distribution of neutral pions at midrapidity 
would be a potential signal of photon-odderon (odderon-photon) contributions. 
One can expect potential deviations from the photon-photon
contribution at $p_{\perp,\pi^{0}} \sim 0.5$~GeV. 
This requires dedicated studies
if the considered process could be measured by, e.g., the ALICE Collaboration at the LHC.

\vspace{0.5cm}

{\bf Acknowledgments}

This work was partially supported by the Polish grant No. PRO-2011/01/N/ST2/04116.
A part of the present calculations has been carried out 
with the Institute of Nuclear Physics (PAN) cloud computing system.

\end{document}